\documentclass[twocolumn,english,aps,pra,floatfix,tightenlines,superscriptaddress,nofootinbib]{revtex4-1}
\usepackage[T1]{fontenc}
\usepackage[utf8]{inputenc}
\usepackage{color}
\usepackage[english]{babel}
\usepackage{amsmath}
\usepackage{amsthm}
\usepackage{amssymb}
\usepackage{graphicx,subfigure}
\usepackage{esint}
\usepackage{bbold}
\usepackage{amsfonts}
\usepackage{bbold}
\usepackage{mathtools}
\usepackage{commath}
\usepackage[unicode=true,
 bookmarks=false,
 breaklinks=false,pdfborder={0 0 1},backref=false,colorlinks=true]
 {hyperref}
\hypersetup{
 linkcolor=cyan,citecolor=magenta,filecolor=magenta,urlcolor=cyan,runcolor=cyan}
 \usepackage[normalem]{ulem}
 \usepackage{cleveref}
 \usepackage{braket}%Dirac Notation in QM

\makeatletter

\begin{document}

\date{\today}

\title{Geometric-Arithmetic Master Equation in Large and Fast Open Quantum Systems}

\author{Dragomir Davidovi\'c}
\affiliation{Georgia Institute of Technology, Atlanta, Georgia, United States}
\email{dragomir.davidovic@physics.gatech.edu}
\begin{abstract}

Understanding nonsecular dynamics in open quantum systems is addressed here, with emphasis on systems with large numbers of Bohr frequencies, zero temperature, and fast driving.
We employ the master equation, which replaces arithmetic averages of the decay rates in the open system, with their geometric averages, and find that it can improve
the second order perturbation
theory, known as the Redfield equation, while enforcing complete positivity on quantum dynamics. The characteristic frequency scale that governs the approximation is the minimax frequency:
the minimum of the maximum system oscillation frequency and the bath relaxation rate; this needs to be larger than the dissipation rate for it to be valid.
The concepts are illustrated on the Heisenberg ferromagnetic spin-chain model.
To study the accuracy of the approximation, a Hamiltonian is drawn from the Gaussian unitary ensemble, for which we calculate the fourth order time-convolutionless master equation,
in the Ohmic bath at zero temperature.
Enforcing the geometric average, decreases the trace distance to the exact solution. Dynamical decoupling of a qubit is examined by applying the Redfield and the geometric-arithmetic master equations, in the interaction picture
of the time dependent system Hamiltonian,
and the results are compared to the exact path integral solution. The geometric-arithmetic approach is significantly simpler and
can be super-exponentially faster compared to the Redfield approach.
\end{abstract}
\maketitle

An interesting form of a quantum master equation (ME) has emerged in recent years,
which approximates an arithmetic mean of the decay rates between energy eigenstates with their geometric mean. The ideas about this equation vary, from purely phenomenological Lindblad master equations capable of describing nonsecular dynamics~\cite{Benoit,perlind}, to improvements of the first principles secular approximation~\cite{Massimiliano,Kleinherbers},
to a highly accurate approximation of the first-principles Redfield equation~\cite{Davidovic2020}, culminating with the claim of an universal Lindblad equation, with a range of validity determined solely by the heat bath, independent of the system~\cite{Nathan}.

The need for such an equation is to model nonsecular quantum dynamics, where the environmental relaxation rate is larger than at least one nonzero system oscillation frequency.
This is relevant in quantum information science~\cite{Nielsen},
in the problems of negative probabilities
of observables and instabilities of reduced states~\cite{dumcke1979proper,Spohn,Suarez,Dunn}, to quantum phase transitions driven by the
environmental coupling~\cite{Leggett,ulrich,HUR20082208}. Lastly, there is
also evidence that non-secular
open quantum dynamics plays a large role in
the efficient light-to-chemical energy transfer that takes
place in the early stages of photosynthesis~\cite{Engel,THORWART2009234}. %Understanding
%this mechanism may allow for the development
%of efficient artificial photosynthetic light harvesting systems,
%with the potential to provide an accessible source
%of clean, sustainable energy globally~\cite{barber}.

In many-body open quantum systems, closely spaced levels can develop strong correlations due to the coupling to the bath. The unitary influence of the bath can renormalize two such  states into a nearly dark state and a state with twice the brightness~\cite{Davidovic2020}. In a weakly detuned two qubit system,
this interaction mechanism with the bath can induce non-local qubit correlations~\cite{Benatti_2009,Benatti,Hartmann}.

The Redfield equation, or the second order perturbative quantum master equation, has a well-known ability to describe such nonsecular effects.
Unfortunately, the Redfield equation has a stability issue: a state error slowly accrues in time and can even diverge if the coupling to the bath is strong. The dynamics is not bounded.

A master equation without the instability is usually sought in the Gorini-Kossakowski-Sudarshan-Lindblad form
(GKSL)~\cite{Gorini,lindblad1976}. The first such equation, derived rigorously from first principles, was created by Davies~\cite{davies1974}. It is valid in the secular approximation, which of course means that it cannot capture nonsecular dynamics.
Davies master equation predicts nonlocal quantum correlations only
at exact degeneracy~\cite{Hartmann}.

Many schemes that address this shortcoming have been put forward, which all enforce complete positivity. This includes
phenomenological approaches~\cite{Benoit,perlind}, later microscopically derived~\cite{Massimiliano,Kleinherbers,Davidovic2020,Nathan};
the adiabatic ME~\cite{Albash_2012}, the coarse-graining ME~\cite{Schaller,Benatti,Benatti2,Majenz,Cresser,Giovannetti,Hartmann,mozgunov}, the partial secular
approximation~\cite{Vogt,Tscherbul} and its recent cousin with a rigorous bound~\cite{Trushechkin}.
There is also a truncated Redfield equation (RE)~\cite{Becker} and the refined weak coupling ME~\cite{Rivas,winczewski}.
The debates range between universality~\cite{Nathan} and
fallacies of MEs with an enforced complete positivity (CP)~\cite{tupkary2021fundamental}.

In this paper we calculate an upper bound on a distance between the maps solved by the Redfield and the further approximated geometric-arithmetic master
equation (GAME)~\cite{Davidovic2020}, which enforces complete positivity on the former. %To simplify the terminology,
%here we will refer to the further approximated equation as the geometric master equation, or an equation in the geometric form.
Such bound can never be rigorous, because this would contradict the situations where the Redfield equation has diverging solutions,
while a CP map cannot have such solutions.

\begin{figure}
\centering
\includegraphics[width=0.42\textwidth]{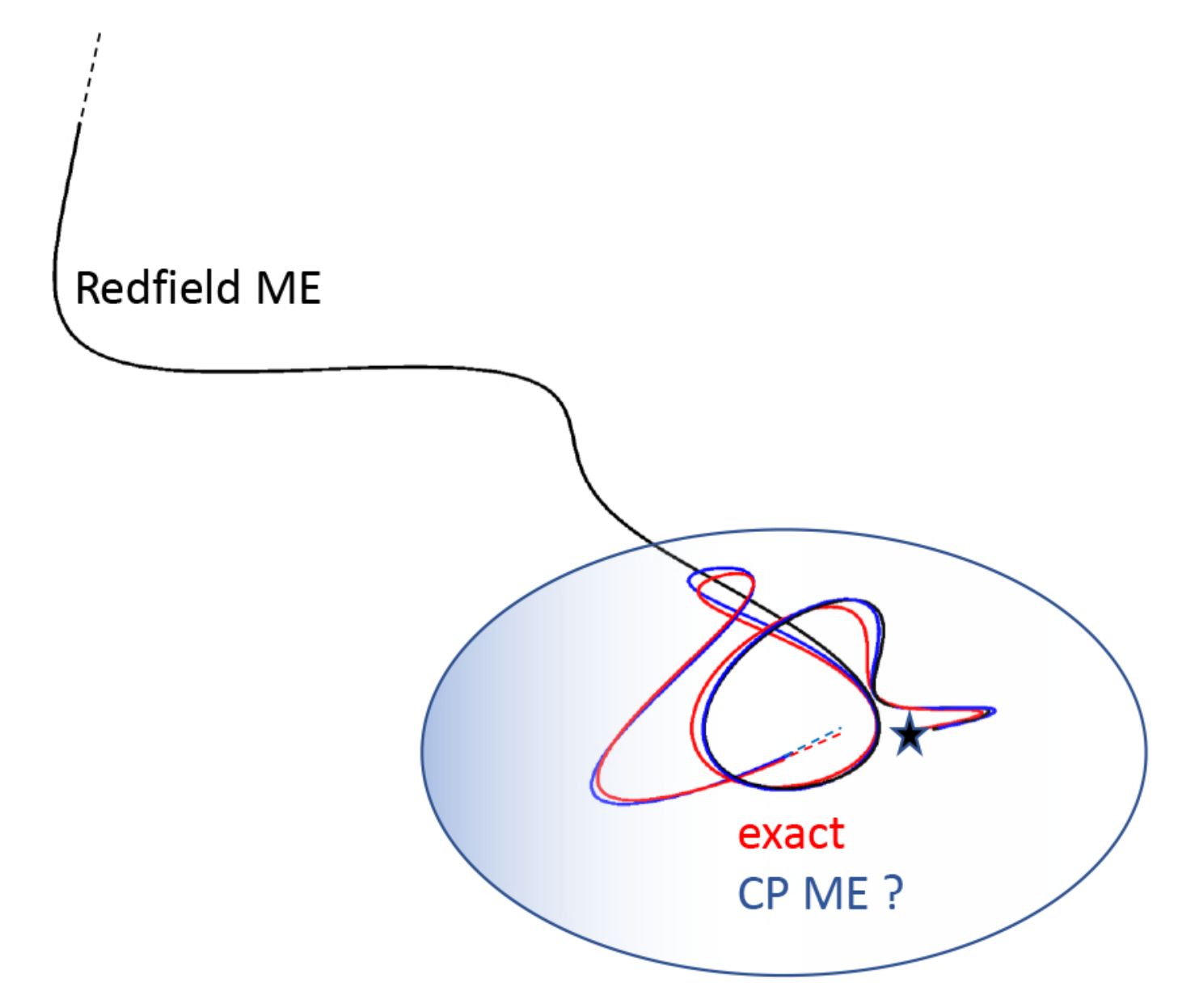}
  \caption{\label{Fig:divergence} Convex set of quantum states of the open quantum system. The initial state is indicated by the star. The three trajectories represent the solutions of the Redfield (black) and a completely positive master equation (blue), and the exact quantum dynamics (red).}
\end{figure}

In Fig.~\ref{Fig:divergence},  quantum states form a convex set in the Hilbert-Schmidt space. The three sketched trajectories
are the exact, the perturbative master equation (e.g., the Redfield),
and the further approximated CP master equation, all initialized into the same state indicated by the star.
Initially, the trajectories are closely intertwined, while the distance
between them increases with time.
On a time scale comparable to the relaxation time, the distance between the Redfield and the exact trajectory
slowly increases, and it can start to diverge.

Then, what will be the attractor of the CP master equation?
Will the strong constraint of complete positivity  stabilize the distance to the exact quantum trajectory,
or will the attraction to the Redfield solution drifting away from the boundary, prevail?

Until now, the answer to this question has been that the exact state can be the attractor, but under very strict conditions.
As we wrote above, a rigorous solution maintaining distance to the exact trajectory
was determined by Davies, in the secular approximation~\cite{davies1974}.
It was only in the last year, that the nearly 40-year old bound of the secular approximation was extended, by Trushechkin~\cite{Trushechkin}, who determined a rigorous bound between a
partially non-secular CP dynamics and the exact quantum dynamics.

Now that this issue has been resolved, the question becomes not if, but how well nonsecular dynamics can be accounted by a quantum master equation.
This brings us to the idea of the universal Lindblad equation (ULE),~\cite{Nathan}
which purports description of nonsecular open quantum dynamics,
with accuracy comparable to that of the Redfield equation and independent of the frequencies of the system,
and an added bonus of complete positivity.

The ULE and the GAME were developed approximately at the same time and independently.
Although the geometric averages in a quantum master equation have been utilized prior to this development~\cite{Massimiliano,Kleinherbers}, as ad hoc interventions, it was the GAME~\cite{Davidovic2020} and the ULE~\cite{Nathan}
where the accuracy of the approximation were assessed, including the microscopic origins and the assumptions.

The only difference between the GAME~\cite{Davidovic2020} and the ULE~\cite{Nathan}, besides the derivation, is in the Lamb-shift, which does not affect the accuracy significantly, as we showed by
an example in Ref.~\cite{Davidovic2020}. Thus, in further discussion we will ignore the differences in the Lamb-shift and focus on the dissipator.
The GAME derivation maps the set of initial states into the displacement between the master equations, and uses a coarse-graining argument.~\cite{Davidovic2020}
Coarse-graining in general is not a rigorous process, but empirically justified, although it can be done rigorously as in the derivation of the completely-positive
coarse-grained master equation~\cite{mozgunov}.
In this work, we put forward an alternative approach based on the properties of kernel induced matrices and the decay rates of their singular values,
which bypasses coarse-graining.

An advantage of the Redfield equation, on the other hand, is that the lack of the bound can be seen as an advantage, because unphysical solutions are a red flag
that the master equation is not valid in approach~\cite{Hartmann}.
A nonnecessity of the GKSL form of a Markovian ME was similarly discussed by Whitney~\cite{Whitney}.
In this paper we study examples showing that the accuracy of the Redfield map is not a prerequisite to
the accuracy of the ULE or the GAME.
That is, the situation depicted in Fig.~\ref{Fig:divergence} is possible. The Redfield equation
instability can be a false red flag, fixable by enforcing complete positivity.

We also study the effect of the system frequencies on open quantum dynamics described by the GAME.
 In relation to Fig.~\ref{Fig:divergence}, there is an initial pulse of
displacement between the geometric master equation solution and {\it both} the perturbative and the exact reduced quantum state.
During and soon after the pulse, the Redfield equation is more accurate than the GAME.
But the displacement in the former creeps up in time due to the lack of a bound,
while the CP constraint maintains the displacement of the geometric-arithmetic master equation.
As a result, the solution of the latter equation becomes more accurate after some time.

%We will numerically examine a random six-by-six Hamiltonian and a 3-qubit system with high concentration of frequencies smaller than the coupling rate. By solving the fourth order perturbative master equation, we display how the trace distance of the geometric master equation rapidly saturates while the Redfield solution drifts, and becomes less accurate on a longer time scale. The more non-secular the quantum system, the improvement in the accuracy seems to be more significant, due to enforcing the geometric form.
%However, if the Redfield equation is failing, e.g., because of the strong coupling or any other reason at weak coupling, we leave out the possibility that the convex bound may be the only bound remaining for the accuracy of the geometric master equation.
%We will also study dynamical decoupling
%in a qubit immersed in an Ohmic environment at zero temperature, as an example where the solution of the geometric master equation maintains the distance to the exact quantum state, while the perturbative master equation is unstable.
%Finally, we will investigate a Heisenberg spin-chain, with a very large number of system frequencies, to support the arguments by illustration. We also briefly discuss adding a gauge to the approximation, and show that it can further enhance the accuracy.

The outline of the paper is as follows. In Sec.~\ref{sec:RE} we rederive the second order perturbative master equation and introduce the mathematical framework and the notation,
and discuss the relevant frequency scales. A reader familiar with this derivation may skip to Sec.~\ref{sec:CPimposition},
where complete positivity is enforced on the Redfield equation. Next, in Sec.~\ref{Sec:Range1}, we develop a
 model of the state displacement responding to the CP-enforcement.
In subsection~\ref{Sec:RegionOfGame} we  conjecture the region of validity of the GAME. An alternative approach, based on kernel induced matrices, is presented
in Sec.~\ref{Sec:Range2}
The concepts are illustrated in a ferromagnetic spin chain in Sec.~\ref{sec:heisenberg}.
In section~\ref{sec:GUE} we discuss a sample from the Gaussian unitary ensemble, and show that the GAME can be more accurate than the Redfield equation.
In~\ref{Sec:DD}, we demonstrate very good agreement between the geometric-arithmetic approach and exact quantum dynamics, in a dynamical decoupling example,
in the regime where the Redfield equation is much more difficult to use correctly.

\section{Perturbative Quantum Master Equations~\label{sec:RE}}

We start by an overview of the derivation of the Redfield equation (RE) and explaining the notation.
Consider a quantum system $S$ weakly coupled to an environment $B$ described by the Hamiltonian
\begin{equation}
H=H_S(t)+H_B+H_I,
\label{Eq:Hamiltonian}
\end{equation}
where $H_S(t)$ and $H_B$ are the Hamiltonians of the system and the environment, respectively, and $H_I$
is the interaction Hamiltonian between the two. In the following the words environment and bath will mean the same.
In the Schr\"odinger picture, $H_S(t)$ is time-dependent, while $H_I$ and $H_B$ are assumed to be time-independent.

The transformation of an operator (X) from the Schr\"odinger picture to the interaction picture is performed according to $X\to U(t)^\dagger X U(t)$,
where $U(t)$ is the free system-bath propagator,
\begin{equation}
U(t)=T_\leftarrow e^{-i\int_0^td\tau H_S(\tau)}\otimes e^{-iH_Bt},
\end{equation}
where $T_\leftarrow$ is the chronological time ordering operator.

The von Neumann equation of motion for the density matrix of the total system
in the interaction picture is given as
\begin{equation}
\frac{d\varrho}{dt}=-i[H_I(t),\varrho].
\label{eq:rtotal}
\end{equation}
This equation can be formally solved as
 \begin{equation}
\varrho(t)=T_\leftarrow e^{-i\int_0^td\tau [H_I(\tau),\bullet]}\varrho(0).
\label{eq:rtotal}
\end{equation}

If we assume factorized
initial condition of the system environment hybrid, [$\varrho_S(0)\otimes\varrho_B(0)$], expand the exponential into time-ordered series, take the partial trace over the environment, and assume
that the odd powers of $H_I$ are partially traceless, we will arrive at the exact expression
\begin{widetext}
\begin{eqnarray}
\nonumber
\varrho_S(t)&=&\varrho_S(0)-\int\limits_0^t dt_1\int\limits_0^{t_1}dt_2\,\text{tr}_B[H_I(t_1),[H_I(t_2),\varrho_S(0)\otimes\varrho_B(0)]]\\
\label{eq:Cumulant}
&+&\int\limits_0^t dt_1\int\limits_0^{t_1}dt_2\int\limits_0^{t_2} dt_3\int\limits_0^{t_3}dt_4\,\text{tr}_B[H_I(t_1),[H_I(t_2),[H_I(t_3),[H_I(t_4),\varrho_S(0)\otimes\varrho_B(0)]]]]-...
\end{eqnarray}
\end{widetext}
The time-dependence in the system Hamiltonian is implicit, and involves the
time-ordered exponential in the system unitary propagator. If the system Hamiltonian is time-independent, this will be
 the same as~\cite[Eq.~(9.65)]{BreuerHeinz-Peter1961-2007TToO}.

For the initial reduced state of the environment, $\varrho_B(0)$, which from now on we
write as $\varrho_B$, we assume that
$[\varrho_B,H_B]=0$, e.g., $\varrho_B$ is diagonal in the eigenbasis of $H_B$, which will also be the preferred basis in the environment.

The Redfield equation is obtained, by taking the time derivative of Eq.~\ref{eq:Cumulant}, dropping the quartic commutator, and substituting $\varrho_S(0)$ on the right hand side (RHS)
with $\varrho_S(t)$.
This leads to the master equation
\begin{equation}
\frac{d\varrho_S}{dt}=-\text{tr}_B\int\limits_{0}^{t}ds[H_I(t),[H_I(s),\varrho_S(t)\otimes\varrho_B]],
\label{eq:rtotal1}
\end{equation}
which is formally the same as the RE in Ref.~\cite[Eq.~(3.117)]{BreuerHeinz-Peter1961-2007TToO}, except for the implicit time ordered exponential in the system unitary propagator,
due to the driving.
Changing the integration variable from $s$ to $s'=t-s$, changes the equation to
\begin{equation}
\frac{d\varrho_S}{dt}=-\text{tr}_B\int\limits_{0}^{t}ds'[H_I(t),[H_I(t-s'),\varrho_S(t)\otimes\varrho_B]].
\label{eq:rtotal1}
\end{equation}

This form of the Redfield equation, with implicit time ordered exponential as system propagator, is utilized in Hamiltonian
open quantum systems toolkit~\cite{ChenHuo}. In Refs.~\cite{Nathan} and~\cite{Dann}, this form is used in the limit $t\to\infty$. The error
induced by the Born-Markov approximation is assessed in Ref.~\cite{Nathan}, and agrees with that for the time-independent system Hamiltonian.

We assume for the interaction Hamiltonian in the Schr\"odinger picture
\begin{equation}
H_I=\sum\limits_{\alpha}A_\alpha\otimes B_\alpha,
\end{equation}
where $A_\alpha$ and $B_\alpha$ are time independent hermitian operators of the system and environment, respectively.
The system operators are dimensionless in physical units, but the system-environment coupling constant, which has unit of frequency, is embedded in the bath operators $B_\alpha$ (which also have physical units of frequency).
Without loss in generality, we assume that $\Vert A_\alpha\Vert_F=1$, where $F$ is the Frobenius norm.

Due to the time ordered exponential
in $U(t)$, the bath spectral properties can be incorporated only indirectly, or by making further assumptions such as the adiabatic approximation~\cite{Davies1978OpenQS,Albash_2012}
or periodicity of the
driving~\cite{AlickiPer,Szczygielski,szczygielski2020howland}.
For example, in the approach in Ref.~\cite{Dann}, $A_{\alpha}(t)$ is expanded in a complete basis of the system’s Lie algebra.

In this paper we take a purely formal approach. We utilize the Fourier transform of $A_{\alpha}(t)$,
defined as
\begin{equation}
\mathcal{A}_{\alpha}(\omega)=\int_{-\infty}^\infty dt A_\alpha(t)e^{i\omega t},
\label{Eq:ft}
\end{equation}
while $A_{\alpha}(t)$ is given by the inverse Fourier transform
\begin{equation}
A_\alpha(t)=\int_{-\infty}^{\infty}\frac{d\omega}{2\pi} \mathcal{A_{\alpha}(\omega)}e^{-i\omega t}.
\end{equation}

After some algebra, Eq.~\ref{eq:rtotal1} is equivalent to
\begin{eqnarray}
\nonumber
\frac{d\varrho_S}{dt}&=&\sum\limits_{\alpha\beta}\iint\limits_{-\infty}^{\infty}\frac{d\omega d\omega'}{(2\pi)^2} e^{i(\omega'-\omega)t} \Gamma_{\alpha\beta}(\omega,t)
\big[ {\mathcal A}_\beta(\omega)\varrho_S {\mathcal A}_\alpha^\dagger(\omega')\\
\label{eq:REbreuerInt}
&-&{\mathcal A}_\alpha^\dagger(\omega'){\mathcal A}_\beta(\omega)\varrho_S\big]+\text{h.c.}
\end{eqnarray}

Here, $\Gamma_{\alpha\beta}(\omega, t)$ is the timed spectral density, which is calculated from the bath correlation function (BCF),
\begin{equation}
\Gamma_{\alpha\beta}(\omega, t)=\int_0^t d\tau\,e^{i\omega\tau}C_{\alpha\beta}(\tau).
\label{Eq:timedSD}
\end{equation}
The BCF is defined as $C_{\alpha\beta}(s)=\text{tr}_B[\varrho_B B_\alpha(s)B_\beta(0)]$. Since we assume $[\varrho_B,H_B]=0$, the condition
$C_{\alpha\beta}(s)=C_{\beta\alpha}^\star(-s)$ is valid. Under that condition, the following identities are valid:
\begin{eqnarray}
\Gamma_{\alpha\beta}(\omega, t)&=&J_{\alpha\beta}(\omega,t)+iS_{\alpha\beta}(\omega,t)\\
\label{Eq:SDT}
J_{\alpha\beta}(\omega,t)&=&\frac{1}{2}\int\limits_{-t}^t ds C_{\alpha\beta}(s)e^{i\omega s}\\
\label{Eq:PDT}
S_{\alpha\beta}(\omega,t)&=&\frac{1}{2i}\int\limits_{-t}^t dsC_{\alpha\beta}(s)e^{i\omega s}\text{sign}(s),
\end{eqnarray}
where both $J_{\alpha\beta}(\omega,t)$ and $S_{\alpha\beta}(\omega,t)$ are hermitian.

If $t\gg\tau_c$, where $\tau_c$ is the
environmental correlation time, then $\Gamma_{\alpha\beta}(\omega, t)$ approaches its asymptotic value
$\Gamma_{\alpha\beta}(\omega)=J_{\alpha\beta}(\omega)+iS_{\alpha\beta}(\omega)$.
By the Bochner's theorem, the matrix
$J(\omega)$ is positive semidefinite (PSD)~\cite{BreuerHeinz-Peter1961-2007TToO}.

The RHS of Eq.~\ref{eq:REbreuerInt} can be recast into the form used in Ref.~\cite{ChenHuo},
\begin{equation}\label{eq:useful}
\frac{d\varrho_S}{dt}=\sum\limits_{\alpha\beta}\big\{[\Lambda_{\alpha\beta}(t)\varrho_S,A_\alpha(t)]+[A_\alpha(t),\varrho_S\Lambda_{\alpha\beta}^\dagger(t)]\big\},
\end{equation}
where
\begin{equation}
\Lambda_{\alpha\beta}(t)=\int\limits_0^t d\tau C_{\alpha\beta}(\tau)A_\beta(t-\tau).
\label{eq:lambdat0int}
\end{equation}
An alternative expression in terms of the Fourier transform follows,
\begin{equation}
\Lambda_{\alpha\beta}(t)=\int\limits_{-\infty}^{\infty}\frac{d\omega}{2\pi}\Gamma_{\alpha\beta}(\omega,t)\mathcal{A}_{\beta}(\omega) e^{-i\omega t}.
\label{eq:lambdat1int}
\end{equation}

\subsection{Time Convolutionless Master Equation}

If $\varrho_S(0)\to\varrho_S(t)$
is invertible as a linear map,
the time convolutionless ME is derived
by taking the derivative of Eq.~\ref{eq:Cumulant},
followed by expressing $\varrho_S(0)$ in terms of $\varrho_S(t)$ by the inverse map.
We adopt the notation TCL, for the exact time convolutionless equation.
TCL equation can be derived perturbatively in a systematic way,
by utilizing the technique of time ordered cumulants, originally derived by van Kampen~\cite{VANKAMPEN1,VANKAMPEN2}, or by the projection operator technique from the
Nakajima-Zwanzig equation~\cite{Nakajima,Zwanzig}.
We use the notation TCL$2n$ if the equation is calculated at $2n$th order in the interaction with the bath.
The RE is the time convolutionless equation TCL$2$.
Further technical details about how to get the coefficients of the TCL$4$ equation
are given in appendix~\ref{appendix:tcl}.

\subsection{The Dissipator and the Lamb Shift~\label{sec:UnLamb}}

The generator of any trace preserving and hermiticity preserving semigroup can be uniquely divided between
the unitary and the dissipative dynamics, as
\begin{equation}
\frac{d\varrho_S}{dt}=-i[H_{ls}(t),\varrho_S]+\mathcal{D}(t)\varrho_S,
\label{Eq:dissipatorRE}
\end{equation}
where $H_{ls}(t)$ is the Lamb shift and $\mathcal{D}(t)$ is the dissipator.
For details see appendix~\ref{appendix:Dissipator}.
The dissipator can be expressed as
\begin{equation}
\mathcal{D}(t)=D(t)-\frac{1}{2}\{\text{Tr}_1[G(t)]\otimes \mathbb{1}+\mathbb{1}\otimes \text{Tr}_1[G^T(t)]\}.
\label{Eq:DisTensor}
\end{equation}
where $D(t)$ is the dynamical tensor, and $G(t)$ is the relaxation tensor obtained by the realignment $G_{ni,mj}(t)=D_{nm,ij}(t)$.
The partial trace is defined as $[\text{Tr}_1(X)]_{ij}=\sum_k X_{ki,kj}$. The superscript $T$ is transposition.

If the system coupling operators $A_\alpha$ are all traceless, then, $\Lambda_{\alpha\beta}$ are also traceless, (see Eq.~\ref{eq:lambdat0int}), and then we find:
\begin{enumerate}
  \item The traceless Hamiltonian, or the Lamb shift,
\begin{equation}
H_{ls}(t)=\frac{1}{2i}\sum\limits_{\alpha\beta}\big\{[A_\alpha(t)\Lambda_{\alpha\beta}(t)-\Lambda_{\alpha\beta}^\dagger(t)A_\alpha(t)].
\label{eq:ULS}
\end{equation}
  \item The dynamical tensor
\begin{equation}
D(t)=\sum\limits_{\alpha\beta}[A_\alpha^\star(t)\otimes\Lambda_{\alpha\beta}(t)+\Lambda_{\alpha\beta}^\star(t)\otimes A_{\alpha}(t)],
\label{Eq:Dmat}
\end{equation}
\end{enumerate}
where the superscript $^\star$ indicates elementwise complex conjugation.

Note the redundancy of tensors and superoperators, since we can recast the Redfield equation without them,
in terms of matrices only (two-index):
\begin{equation}
\frac{d\varrho_S}{dt}=
\sum_{\alpha\beta}\Lambda_{\alpha\beta}(t)\varrho_S A_\alpha (t)-A_\alpha(t)\Lambda_{\alpha\beta}(t)\varrho_S +\text{h.c.}
\label{Eq:classicRE}
\end{equation}
The extension to a nonzero trace of the coupling operator is given in appendix~\ref{appendix:Dissipator}.

Throughout the paper, we utilize column-major ordering of matrices, e.g., valid in Fortran/MATLAB, but not in Python/JULIA. Respectively, $(A \otimes B)\rho$ translates, into the operator notation, as $B\rho A^T$, not $A\rho B^T$.
See appendix~\ref{appendix:column} for further details, including the derivation of Eq.~\ref{Eq:classicRE} from Eqs.~\ref{Eq:dissipatorRE} and~\ref{Eq:Dmat}.

\subsection{Time-Independent System Hamiltonian\label{sec:RE_timeindependent}}
If $H_S$ is time-independent, the previous formalism can be reduced to the usual Redfield equation in the eigenbasis of $H_S$. The free system-environment propagator will be $U(t)=\exp[-i(H_S+H_B)t]$. While the analysis so far did not involve a preferred basis in the system Hilbert space, in the time-independent case the preferred basis is the eigenbasis of $H_S$. In terms of the discrete eigenenergies $E_n$ of $H_S$, and the projectors on the respective eigenspaces denoted by $\Pi_n$,
$A_\alpha(t)$ can be written as the eigenoperator decomposition~\cite{BreuerHeinz-Peter1961-2007TToO},
\begin{equation}
A_\alpha(t)=\sum\limits_k e^{-i\omega_k t}a_{\alpha}(\omega_k)=\sum\limits_k a_{\alpha}(\omega_k,t),
\label{Eq:FTAd}
\end{equation}
where $\omega_k$ are the discrete Bohr frequencies, and
\begin{equation}
a_{\alpha}(\omega_k)=\sum\limits_{nm,E_m-E_n=\omega_k}\Pi_nA_\alpha\Pi_m.
\label{eq:eigenprojectors}
\end{equation}

Due to the orthogonality of the projectors $\Pi_n$, the operator projections are orthogonal as well, satisfying
\begin{equation}
\text{Tr}[a_{\alpha}^\dagger(\omega_k,t)a_{\beta}(\omega_q,t)]=\delta_{kq}\text{Tr}[ a_{\alpha}(\omega_k)^\dagger a_{\beta}(\omega_k)],
\label{Eq:orthogonality}
\end{equation}
along with the normalization
\begin{equation}
\sum_k \text{Tr}[a_\alpha^\dagger (\omega_k)a_\beta (\omega_k)]=\text{Tr}(A_\alpha A_\beta).
\end{equation}

Now we calculate the Fourier transform by substituting Eq.~\ref{Eq:FTAd} into Eq.~\ref{Eq:ft}, and obtain
\begin{equation}
\mathcal{A}_\alpha(\omega)=2\pi\sum_k a_{\alpha}(\omega_k)\delta(\omega-\omega_k).
\label{eq:timeindep}
\end{equation}

Inserting Eq.~\ref{eq:timeindep} into Eq.~\ref{eq:REbreuerInt}, we arrive at the form of the RE similar to~\cite[Eq.~(3.132)]{BreuerHeinz-Peter1961-2007TToO},
\begin{eqnarray}
\nonumber
\frac{d\varrho_S}{dt}&=&\sum\limits_{kq\alpha\beta}e^{i(\omega_q-\omega_k)t} \Gamma_{\alpha\beta}(\omega_k,t)
\big[ a_\beta(\omega_k)\varrho_S a_\alpha^\dagger(\omega_q)\\
\label{eq:REbreuer}
&-&a_\alpha^\dagger(\omega_q)a_\beta(\omega_k)\varrho_S\big]+\text{h.c.}
\end{eqnarray}
Similarly, we find
\begin{equation}
\Lambda_{\alpha\beta}(t)=\sum_{k}\Gamma_{\alpha\beta}(\omega_k,t)a_{\beta}(\omega_k)e^{-i\omega_k t}.
\label{eq:lambdat0sum}
\end{equation}
{\it From now until section~\ref{Sec:DD}, we will assume that the system Hamiltonian is time-independent.}

\subsection{Characteristic Frequencies~\label{Sec:Freqs}}

To introduce various frequency and time scales that we  use throughout the paper, let us examine a quantum system coupled to a single bosonic Ohmic bath at zero temperature.
The bath operator is
\begin{equation}
B=\sum_\lambda g_\lambda(b_\lambda+b_\lambda^\dagger),
\end{equation}
where $b_\lambda$ is the canonical annihilation operator for bath mode $\lambda$ with energy $\omega_\lambda$, in units where $\hbar=1$.
For the Ohmic bath, the bath correlation function (BCF) is assumed to be
\begin{equation}
\langle B(t)B\rangle=C(t)=\sum_\lambda \vert g_\lambda\vert^2 e^{-i\omega_\lambda t}=\frac{\alpha\omega_c^2}{2(1+i\omega_ct)^{2}}.
\label{eq:OhmicBCF}
\end{equation}

There are two characteristic frequencies of the bath~\cite{mozgunov,Nathan}: 1, the bath relaxation rate or
\begin{equation}
\omega_{bath}=\frac{\int_0^\infty dt\,\vert C(t)\vert}{\int_0^{T_c} dt\, \vert t C(t)\vert}=\pi\omega_c/\ln[1+(\omega_cT_c)^2],
\label{Eq:omegabath}
\end{equation}
and 2, the  maximum relaxation rate permitted by the bath on any system,
\begin{equation}
\max(\omega_r)=4\int_0^\infty\vert C(t)\vert dt=\alpha\omega_c\pi.
\label{Eq:omegarMax}
\end{equation} The corresponding time scales are $\tau_{bath}=1/\omega_{bath}$ and $\tau_{r}=1/\omega_{r}$.
Here $T_c$ is the time of the dynamics and is applied at zero temperature, to cut off the divergence in the limit $T_c\to\infty$.
In terms of these frequencies, the weak-coupling constant is defined as $\alpha_{sb}=\max(\omega_r)/\omega_{bath}$, or
\begin{equation}
\alpha_{sb}=\alpha\ln[1+(\omega_cT_c)^2).
\label{Eq:alphaBath}
\end{equation}
{\it In all order of magnitude estimates in this paper, we will not worry about the cut-off and use:
$\omega_{bath}=\omega_c$ and $\max(\omega_r)=\alpha\omega_c$.}

The third frequency governs the geometric-arithmetic approximation, which is the system-bath frequency spread, defined as
\begin{equation}
\omega_s\equiv\sqrt{\langle \vert\Gamma\vert^2\rangle-\vert\langle\Gamma\rangle\vert^2}/\alpha.
\label{Eq:compressionFreq}
\end{equation}
Here we introduced the frequency average as  $\langle X\rangle=(1/N_f)\sum_{\omega}X(\omega)$,
where $N_f$ is the number of Bohr frequencies of the system. This frequency is not uniquely defined.
For example, $\max[J(\omega)]/\alpha$ would work similarly well, and in the Ohmic bath at $T=0$ one easily verifies that this is $J(\omega_{\text{minimax}})/\alpha$,
where $\omega_{\text{minimax}}=\min[\max(\omega),\omega_c]$ and $\max(\omega)$ is the maximum system oscillation frequency.
The effects specific to the system-bath coupling operator are not studied here, except that only active Bohr frequencies are included,
those with $\Vert a_\alpha(\omega)\Vert>0$.

In the limit $t\to\infty$, the spectral density of the single Ohmic bath is
\begin{eqnarray}
\nonumber
J(\omega)&=&\pi\sum\limits_\lambda \vert g_\lambda\vert^2\delta(\omega-\omega_\lambda)\\
&=&\frac{\pi}{2}\alpha\omega
\Theta(\omega)e^{-\frac{\omega}{\omega_c}},
\label{eq:spdensity}
\end{eqnarray}
where $\Theta(\omega)$ is the Heaviside step function.
Similarly, the principal value defined by Eq.~\ref{Eq:PDT} in the limit $t\to\infty$ becomes
\begin{equation}
S(\omega)=-\frac{\alpha\omega_c}{2}\left[1-\frac{\omega}{\omega_c}e^{-\frac{\omega}{\omega_c}}ei(\frac{\omega}{\omega_c})\right],
\label{Eq:ei}
\end{equation}
where $ei(x)$ is the exponential integral. Note that $dS/d\omega=(\alpha/2)[x+(1-x)e^{-x}ei(x)]$, where $x=\omega/\omega_c$, diverges as $\text{ln}(x)$ at $x\to 0$.
Also, a continuation of $ei(x)$ into the complex plane has a branch cut on the real axis interval $(-\infty,0]$.

Finally, the timed spectral density is $\Gamma(\omega,t)=\int_0^t d\tau\,C(\tau)e^{i\omega\tau}$:
\begin{eqnarray}
\nonumber
\Gamma(\omega,t)=-i\frac{\alpha\omega_c}{2}\Bigg\{
1-\frac{e^{i\omega t}}{1+i\omega_ct}-\frac{\omega}{\omega_c}e^{-\frac{\omega}{\omega_c}}\\
\bigg[
ei(\frac{\omega}{\omega_c})-ei(\frac{\omega}{\omega_c}+i\omega t)-i\pi\Theta(-\frac{\omega}{\omega_c})
\bigg]
\Bigg\}.
\label{eq:sdtOhmic}
\end{eqnarray}

\section{Geometric-Arithmetic Approximation~\label{sec:CPimposition}}

In this section we enforce the GKSL form on the dissipator in Eq.~\ref{Eq:DisTensor}.
In  step one, we insert Eqs.~\ref{Eq:FTAd} and~\ref{eq:lambdat0sum} into Eq.~\ref{Eq:Dmat},
to make the  dynamical and the relaxation tensors explicit in the interaction picture,
\begin{eqnarray}
\nonumber
            D_{r}(t)&&=\sum\limits_{\alpha\beta kq}e^{i(\omega_q-\omega_k)t}
\\
\label{Eq:DmatD}
&&[\Gamma_{\alpha\beta}(\omega_k,t)+
\Gamma_{\beta\alpha}^\star(\omega_q,t)]
            a_{\alpha}^\star(\omega_q)\otimes a_{\beta}(\omega_k)
\end{eqnarray}
and
\begin{eqnarray}
\nonumber
            G_{r}(t)&&=\sum\limits_{\alpha\beta kq}e^{i(\omega_q-\omega_k)t}
\\
\label{Eq:DmatG}
&&[\Gamma_{\alpha\beta}(\omega_k,t)+
\Gamma_{\beta\alpha}^\star(\omega_q,t)]
           \vert a_{\beta}(\omega_k)\rangle\langle a_{\alpha}(\omega_q)\vert,
\end{eqnarray}
respectively.
Here the subscript $r$ is for the Redfield and $\vert a_{\alpha}(\omega_k)\rangle$
represents the vector format of $a_{\alpha}(\omega_k)$ in column major order. The realignment between the tensors swaps out the tensor and the outer products.

In the next step, we show that the dissipator~\ref{Eq:DisTensor} is separable sum on the bath and the system variables.
For the system, we introduce the superoperator quadratic in the system operator, e.g.,
\begin{eqnarray}
\nonumber
&&\mathbb{A}_{\alpha\beta}(\omega,\omega')=a_{\alpha}^\star(\omega')\otimes a_{\beta}(\omega)\\
&&-\frac{1}{2}\{[a_\alpha(\omega')^\dagger a_\beta(\omega)]^T\otimes\mathbb{1}
+\mathbb{1}\otimes a_\alpha^\dagger(\omega')a_\beta(\omega) \}.
\label{Eq:quadratic}
\end{eqnarray}
Inserting Eqs.~\ref{Eq:DmatD} and~\ref{Eq:DmatG} in Eq.~\ref{Eq:DisTensor}, applying the identity
$\text{Tr}_1\vert X\rangle\langle Y\vert=[Y^\dagger X]^T$, and organizing the system operators,
we recast the dissipator into the sum,
\begin{equation}
\mathcal{D}_r(t)=
\sum\limits_{\alpha\beta kq}
e^{i(\omega_q-\omega_k)t}[\Gamma(\omega_k)+
\Gamma(\omega_q)^\dagger]_{\alpha\beta}\mathbb{A}_{\alpha\beta}(\omega_k,\omega_q),
\label{Eq:dissipRE}
\end{equation}
which is the separable sum.

Now, we apply the geometric-arithmetic mean approximation on the bath part,
to enforce complete positivity, leaving the system part intact. Let us first study the Markovian limit of the dynamics, i.e.,  $\Gamma(\omega,t)\to\Gamma(\omega)$,
and perform the ansatz:
\begin{eqnarray}
\label{eq:ansatz}
\Gamma(\omega)+\Gamma(\omega')^\dagger &\to& 2\sqrt{Z(\omega')^\dagger} \sqrt{Z(\omega)},\\
\label{eq:constraint}
J(\omega)^\dagger J(\omega)&=&Z(\omega)^\dagger Z(\omega).
\end{eqnarray}
On the RHS of Eq.~\ref{eq:ansatz} there is a frequency order and the square root is the usual matrix root.
$Z(\omega)$ is the impedance matrix.
The constraint~\ref{eq:constraint} assures that the norm  distance $\Vert$RHS$-$LHS$\Vert$ of~\ref{eq:ansatz} scales linearly
versus $\omega-\omega'$, at small $\vert\omega-\omega'\vert$. The LHS and the RHS are
Hermitian [e.g., $X(\omega,\omega')^\dagger=X(\omega',\omega)^\star$]. More details about the geometric-arithmetic approximation of matrices are presented in appendix~\ref{appendix:MGAM}.

After applying the ansatz, we obtain the dissipator of the GAME,
\begin{equation}
\mathcal{D}_g(t)=\sum\limits_{\alpha\beta kq}
e^{i(\omega_q-\omega_k)t}2[\sqrt{Z(\omega')^\dagger}\sqrt{Z(\omega)}]_{\alpha\beta}
\mathbb{A}_{\alpha\beta}(\omega_k,\omega_q).
\label{Eq:dissipGA}
\end{equation}

In our previous work and in this paper, except in Sec.~\ref{sec:GUE}, we will assume $Z(\omega)=J(\omega)$ for simplicity.
Further improvements await to be done. Then,
Eqs.~\ref{Eq:DmatD} and~\ref{Eq:DmatG} are converted to the geometric form
\begin{eqnarray}
\nonumber
            D_{g}(t)&=&\sum\limits_{\alpha\beta kq}e^{i(\omega_q-\omega_k)t} 2\big[\sqrt{J(\omega_q)^\dagger}\sqrt{J(\omega_k)}\big]_{\alpha\beta}\\
            &&a_{\alpha}^\star(\omega_q)\otimes a_{\beta}(\omega_k)=\sum_\gamma M_{\gamma}^\star(t)\otimes M_{\gamma}(t)
 %           \vert\mathcal{A}_{\beta}(\omega)\rangle\langle\mathcal{A}_{\alpha}(\omega')\vert
\label{Eq:DmatDG}
\end{eqnarray}
and
\begin{eqnarray}
\nonumber
            G_{g}(t)&=&\sum\limits_{\alpha\beta}e^{i(\omega_q-\omega_k)t}2\big[\sqrt{J(\omega_q)^\dagger}\sqrt{J(\omega_k)}\big]_{\alpha\beta}\\
          && \vert a_{\beta}(\omega_k)\rangle\langle a_{\alpha}(\omega_q)\vert=\sum_\gamma\vert M_{\gamma}(t)\rangle\langle M_{\gamma}(t)\vert.
\label{Eq:DmatGG}
\end{eqnarray}
respectively. Here,
 $M_{\gamma}$ are the  jump operators in the interaction picture, defined as
\begin{eqnarray}
M_{\gamma}(t)&=&\sum\limits_{\beta k}\big[\sqrt{2J(\omega_k)}\big]_{\gamma\beta}e^{-i\omega_k t}a_\beta(\omega_k)\\
&=&\sum\limits_\beta\int\limits_{-\infty}^{\infty}d\tau\,g_{\gamma\beta}(\tau)A_\beta(t-\tau).
\label{eq:MofT}
\end{eqnarray}
The jump correlator is defined as the Fourier transform
\begin{equation}
g(t)=\int\limits_{-\infty}^{\infty}\frac{d\omega}{2\pi}\,e^{-i\omega t}\sqrt{2J(\omega)}.
\label{eq:jumpf}
\end{equation}

After some algebra, we arrive at the GKSL master equation
\begin{eqnarray}
\nonumber
&&\frac{d\varrho_S}{dt}=-i[H_{ls}(t),\varrho_S]\\
\label{Eq:lindblad}
+&&\sum\limits_\alpha \big[M_\alpha(t)\varrho_S M_\alpha^\dagger(t)-\frac{1}{2}\{M_\alpha^\dag(t)M_\alpha(t),\varrho_S\}\big],
\end{eqnarray}
which is essentially the same as the ULE, except for the Lamb-shift difference.

By direct integration, for the Ohmic bath at $T=0$, we find
\begin{equation}
g(t)=\sqrt{\frac{\alpha\omega_c^3}{2}}\frac{1}{(1+2i\omega_ct)^{3/2}}.
\label{eq:ohmicjump}
\end{equation}
The weak coupling constant corresponding to the jump correlator is the maximum relaxation rate~\cite{Nathan},
\begin{equation}
\max(\omega_r)=4\left(\int_0^\infty dt\vert g(t)\vert\right)^2=2\pi\alpha\omega_c\frac{\Gamma^2(\frac{5}{4})}{\Gamma^2(\frac{3}{4})},
\label{eq:gamma34}
\end{equation}
divided by the bath-relaxation rate, given by Eq.~\ref{Eq:omegabath} after swapping out $C(t)$ and $g(t)$. In Eq.~\ref{eq:gamma34}, $\Gamma(\frac{5}{4})$ and $\Gamma(\frac{3}{4})$
is the gamma function.
We obtain,
\begin{equation}
\alpha_{sj}=\alpha \frac{\sqrt{2\pi}\Gamma(\frac{5}{4})}{\Gamma(\frac{3}{4})}\sqrt{\omega_cT_c}\approx 1.86\alpha\sqrt{\omega_cT_c},
\end{equation}
which is very different from Eq.~\ref{Eq:alphaBath}. Thus, the bath correlation function and the jump correlator are not equivalent at zero temperature. In the previous work, we showed that
the trace distance between the Redfield and the GAME solutions is comparable to $\alpha$ independent of the evolution time,
numerically demonstrating the approximation's validity at zero temperature. Thus, it is not the approximation, but the ULE bound, that is not applicable at zero temperature~\cite{Nathan}.

\section{Time Dependent Fluctuations of Displaced States - I~\label{Sec:Range1}}

In this section we study the differences between the dynamics of the Redfield and the geometric-arithmetic master equations.
The ansatz~\ref{eq:ansatz} introduces a state displacement,
which needs representation, before the range of applicability of the approximation can be discussed.

\subsection{The Position-Displacement Superoperators~\label{Sec:DisplacementSuperoperator}}

\begin{figure}
\centering
\includegraphics[width=0.3\textwidth]{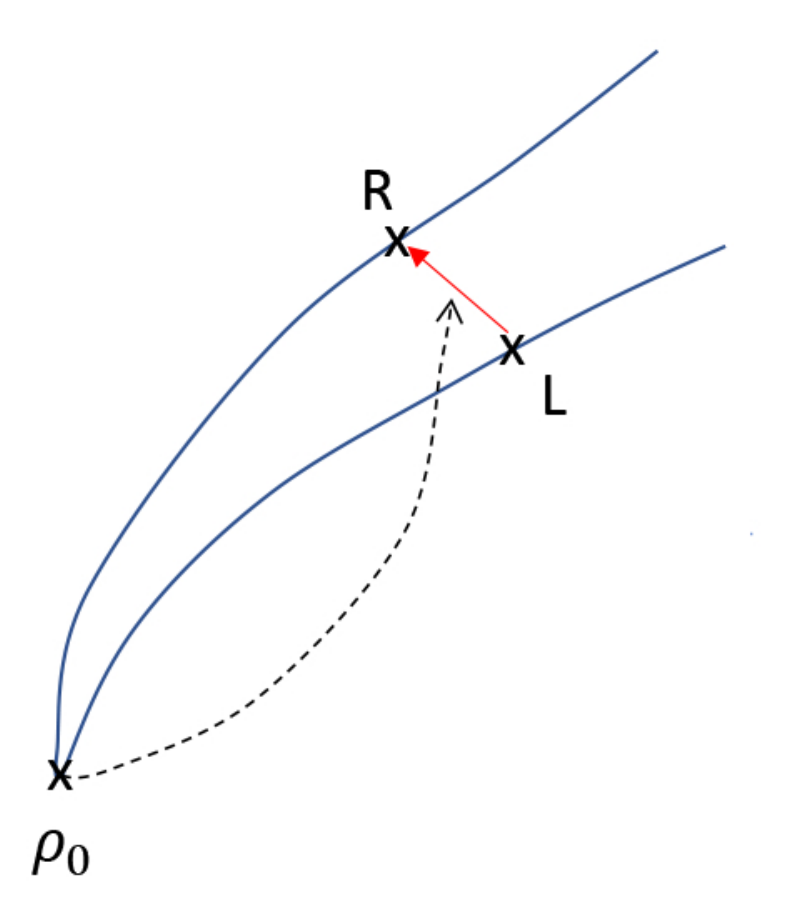}
  \caption{\label{Fig:frames} Displacement superoperator maps the initial set of states into the displacements.}
\end{figure}

The displacement superoperator, $\Delta_t\mathfrak{X}$, is introduced as follows.
The dynamics starts at time $t=0$.  $\Delta_t\mathfrak{X}$ maps the initial state $\varrho_0$ into the
displacement between the Redfield and the Lindblad state at time $t$, e.g.,
\begin{equation}
\varrho_r(t)-\varrho_g(t)\equiv\Delta_t\mathfrak{X}\varrho_0 .
\label{Eq:brut}
\end{equation}
This superoperation is illustrated  in Fig.~\ref{Fig:frames}.

In Liouvillian terms, the Redfield and the geometric-arithmetic master equations are
\begin{eqnarray}
\frac{d\varrho_{r}}{dt}&=&\mathcal{L}_r(t)\varrho_{r}(t),\\
\frac{d\varrho_{g}}{dt}&=&\mathcal{L}_g(t)\varrho_{g}(t),
\end{eqnarray}
where $\mathcal{L}_{r/g}(t)=-i[H_{ls}(t),.]+\mathcal{D}_{r/g}(t)$,
both in the interaction picture.
After integrating in time, there is a time ordered series,
\begin{eqnarray}
\nonumber
\varrho_r(t)&=&\varrho_0+\int_{0}^{t}dt_1\mathcal{L}_r(t_1)\varrho_0\\
&+&\int_{0}^{t}dt_1\int_{0}^{t_1}dt_2\mathcal{L}_r(t_1)\mathcal{L}_r(t_2)\varrho_0+...\\
\nonumber
\varrho_g(t)&=&\varrho_0+\int_{0}^{t}dt_1\mathcal{L}_g(t_1)\varrho_0\\
&+&\int_{0}^{t}dt_1\int_{0}^{t_1}dt_2\mathcal{L}_g(t_1)\mathcal{L}_g(t_2)\varrho_0+...
\end{eqnarray}
Taking the difference, we have
\begin{eqnarray}
\label{Eq:varDIF}
&&\Delta\varrho(t)=\varrho_r(t)-\varrho_g(t)=\int_{0}^{t}dt_1[\mathcal{L}_r(t_1)-\mathcal{L}_g(t_1)]\varrho_0\\
\nonumber
&&+\int_{0}^{t}dt_1\int_{0}^{t_1}dt_2[\mathcal{L}_r(t_1)\mathcal{L}_r(t_2)-\mathcal{L}_g(t_1)\mathcal{L}_g(t_2)]\varrho_0+....
\end{eqnarray}
As defined above, we obtain
\begin{equation}
\label{eq:stateDISPL}
\Delta_t\mathfrak{X}=\int_{0}^{t}dt_1[\mathcal{D}_r(t_1)-\mathcal{D}_g(t_1)]+O(\omega_rt)^2,
\end{equation}
where the higher order integrals are embedded in $O(\omega_rt)^2$.

Inserting Eqs.~\ref{Eq:dissipRE} and~\ref{Eq:dissipGA}
into Eq.~\ref{eq:stateDISPL} and
replacing $Z(\omega)$ with $J(\omega)$, we get
\begin{equation}
\label{Eq:displacedLiuv}
\Delta_t\mathfrak{X}=\mathfrak{X}_t-\mathfrak{X}+O(\omega_r t)^2,
\end{equation}
where
\begin{equation}
\mathfrak{X}=\sum\limits_{\alpha\beta qk}
X_{\alpha\beta}(\omega_k,\omega_q)
\mathbb{A}_{\alpha\beta}(\omega_k,\omega_q),
\label{Eq:position}
\end{equation}
and
\begin{equation}
\mathfrak{X}_t-\mathfrak{X}=\sum\limits_{\alpha\beta qk}
\left[e^{i(\omega_q-\omega_k)t}-1\right]X_{\alpha\beta}(\omega_k,\omega_q)
\mathbb{A}_{\alpha\beta}(\omega_k,\omega_q).
\label{Eq:positiont}
\end{equation}

The last equation suggests that we call $\mathfrak{X}$ the position superoperator,
so that the displacement superoperator is the change in the position. While $\mathfrak{X}_t-\mathfrak{X}$ is unique,
$\mathfrak{X}$ is not uniquely defined.
The latter can be seen as induced by the
$X$-{\it kernel},
\begin{equation}
X(\omega,\omega') = i\frac{\Gamma(\omega)+
\Gamma^\dagger(\omega')
-2\sqrt{J^\dagger(\omega')}\sqrt{J(\omega)}}{\omega-\omega'}
\label{Eq:PsiS}
\end{equation}
while $\mathfrak{X}_t-\mathfrak{X}$ is induced by the corresponding, unique displacement kernel,
\begin{equation}
X_t(\omega,\omega')-X(\omega,\omega')=\left[e^{i(\omega'-\omega)t}-1\right]X(\omega,\omega').
\label{Eq:XtminusX}
\end{equation}
We also introduce the unique velocity kernel,
\begin{eqnarray}
\nonumber
V(\omega,\omega')&=&\dot{X}_t(\omega,\omega')=i(\omega'-\omega)X(\omega,\omega')\\
&=&\Gamma(\omega)+
\Gamma(\omega')^\dagger
-2\sqrt{J(\omega')^\dagger}\sqrt{J(\omega)}.
\label{Eq:velocityGA}
\end{eqnarray}
which is hermitian [$V(\omega,\omega')=V^\star(\omega',\omega)$] and zero on the diagonal [$V(\omega,\omega)=0$]. $V$ is uniformly continuous on $[-\max(\omega),\max(\omega)]^2$, assuming that the spectral density is also
uniformly continuous. In the secular approximation, on the other hand, the corresponding velocity matrix can be similarly written down, but it cannot be kerneled with a continuous kernel.

While the displacement kernel matrix is unique, the diagonal elements of the $X$-matrix, $X(\omega_k,\omega_k)$, are
not uniquely defined. Thus, there is a degree of freedom in the $X$-kernel, that we use at our convenience. Namely, we define
\begin{equation}
X(\omega,\omega)\equiv-\frac{\partial S}{\partial\omega},
\end{equation}
which ensures local continuity of $X(\omega,\omega')$, in the noncompact region that excludes the point $\omega=\omega'=0$.
We want the continuity in the $X$-kernel, to increase the decay rate of its singular values, which are sensitive to the smoothness of the kernel
as will be discussed in Sec.~\ref{Sec:Xdecompose}.

Now, $X(\omega,\omega')$ diverges at $\omega=\omega'=0$ at $T=0$.
We regularize this by enforcing the Lipschitz continuity on the principal density function $S(\omega)$, which means $\forall \omega,\omega'\in[-\max(\omega),\max(\omega)]^2$, $\exists L>0$,
so that
\begin{equation}
\vert S(\omega)-S(\omega')\vert<L\vert \omega-\omega'\vert.
\label{Eq:LipschitzS}
\end{equation}
In practice, we floor the derivative as
\begin{equation}
\left.\frac{\partial S}{\partial\omega}\right\vert_{\omega=0}=\frac{1}{2}\left.\frac{\partial S}{\partial\omega}\right\vert_{-\delta_m}+\frac{1}{2}\left.\frac{\partial S}{\partial\omega}\right\vert_{\delta_m},
\end{equation}
where
$\delta_m$ is the frequency spacing at $\omega=0$. Then $L=\max_{\pm\delta_m}\vert\partial S/\partial\omega\vert$.
As a result of~\ref{Eq:LipschitzS}, the velocity kernel is Lipschitz continuous function of frequency $\omega$, for any $\omega'$.
This construction makes the $X$-matrix, as well as $\mathfrak{X}$, bounded, without any effect on the matrix elements of $X_t-X$, $V$, and $\mathfrak{X}_t-\mathfrak{X}$.

With the bound on $\mathfrak{X}$ in place, we readily find that the state displacement is
bounded according to:
\begin{eqnarray}
\label{eq:bound}
\Vert\Delta\varrho (t)\Vert_F
&\leq&
2\Vert\mathfrak{X}\Vert_\infty\Vert\varrho_0\Vert_F+O(\omega_r t)^2\\
\label{Eq:finalBound}
&\leq &
2\Vert\mathfrak{X}\Vert_\infty+O(\omega_r t)^2,
\end{eqnarray}
where $\Vert\mathfrak{X}\Vert_\infty$ is the operator norm,
\begin{equation}
\Vert\mathfrak{X}\Vert_\infty=\sup\limits_{\Vert x\Vert_F=1}\Vert \mathfrak{X}x\Vert_F,
\label{eq:supremum}
\end{equation}
equal to the highest singular value (s-value) of $\mathfrak{X}$.
To prove line~\ref{eq:bound}, take the Frobenius norm of Eq.~\ref{Eq:brut}, invoke the operator norm on $\Delta_t\mathfrak{X}$, apply the triangle inequality and
unitary invariance. The next line follows because the Frobenius norm is $\leq$ than the trace norm, which is equal to one in $\varrho_0$.

{\bf Inequality 1.} There is a useful inequality, that connects the superoperator norm to the spectral properties of its kernel matrix:
For any superoperator $\mathbb{Y}$,
\begin{equation}
\mathbb{Y}=\sum\limits_{\alpha\beta qk}
Y_{\alpha\beta}(\omega_k,\omega_q)
\mathbb{A}_{\alpha\beta}(\omega_k,\omega_q),
\label{Eq:Y-operator}
\end{equation}
there is an upper bound
\begin{equation}
\Vert\mathbb{Y}\Vert_\infty\leq 2\sum_{n,k\alpha\beta}\vert y_n\vert U_{\alpha k,n}U_{\beta k,n}^\star\text{Tr}[ a_\alpha^\dagger(\omega_k)a_\beta(\omega_k)],
\label{Eq:boundExplBY}
\end{equation}
where $y_n$ and $U_{\alpha k,n}$ are the eigenvalues and the eigenvectors
of the hermitian kernel matrix,
\begin{equation}
\sum_{\beta}Y_{\alpha\beta}(\omega_k,\omega_q)U_{\beta q,n}=y_nU_{\alpha k,n}.
\label{eq:decompY}
\end{equation}
In single bath case, the inequality~\ref{Eq:boundExplBY} simplifies to
\begin{eqnarray}
\label{Eq:boundS}
\Vert\mathbb{Y}\Vert_\infty&\leq& 2\sum_{n,k}\vert y_n\vert
\vert U_{k,n}\vert^2\Vert a(\omega_k)\Vert_F^2\\
\label{Eq:boundSB}
&\leq& 2\sum_{n}\vert y_n\vert\max_{\omega_k}\vert U_{k,n}\vert^2
%&\leq& 2\Vert Y\Vert_1.
%\label{Eq:boundSBtn}
\end{eqnarray}
\ref{Eq:boundSB} follows from~\ref{Eq:boundS}, due to the normalization condition $\Vert A\Vert_F^2=\sum_k \Vert a(\omega_k)\Vert_F^2=1$.
The derivation of the inequality is given in appendix~\ref{appendix:normIneq}.

Everywhere in this paper,
the eigenvalues are sorted in descending order of their magnitudes, i.e., $\vert y_1\vert\geq\vert y_2\vert\geq...\geq\vert y_{N_f}\vert$.
Since these correspond to hermitian matrices, the s-values are the same as the eigenvalue magnitudes.
The following analysis will be done for the single bath case, but the results can be generalized to
multiple baths.

The inequality~\ref{Eq:boundS} relates the accuracy of the geometric-arithmetic approximation of the spectral density mean,  to the decay rate of the s-values
of the $X$-matrix, $\vert x_n\vert$. Assume that $\vert x_n\vert$  decays geometrically. Then, the state displacement induced by the approximation will be of order $\alpha$,
as follows.
Apply the inequality~\ref{Eq:boundSB} on the position superoperator. The eigenvectors of $X$
can be assumed to be delocalized and have amplitudes $\sim 1/\sqrt{N_f}$. This is because they inherit
the Lipschitz constant of $X$~\cite{braun2005spectral}.
So, replace $\max_{\omega_k}\vert U_{k,n}\vert^2$ with $1/N_f$ in Eq.~\ref{Eq:boundSB}.
Then, if $\vert x_n\vert$ decays geometrically,
the sum can be truncated to the highest s-value, which is proportional to $N_f\alpha$. The resultant state displacement, Eq.~\ref{Eq:finalBound},  is
$\lessapprox 4\vert x_1\vert/N_f+O(\omega_rt)^2 \approx 4\alpha+O(\omega_rt)^2$.

Let us examine an especially simple, but instructive situation where $\Vert a(\omega_k)\Vert_F=1/\sqrt{N_f}$, $\forall k$. Then, inequality~\ref{Eq:boundS} simplifies to
\begin{equation}
\Vert\mathbb{Y}\Vert_\infty \leq 2\Vert Y\Vert_1/N_f.
\label{Eq:tracenorm}
\end{equation}
where $\Vert Y\Vert_1$ is the trace-norm (sum of the s-values). Thus, the operator-norm of the superoperator
is related to the trace-norm of its kernel matrix. This notion will be relevant for estimating the state displacement, from the spectral properties of the kernel.

\subsection{Introductory Remarks on the Spectrum of the $X$-Matrix\label{Sec:Xdecompose}}

Determining the eigenvalues and eigenvectors of the $X$-matrix is very difficult, because its kernel
is neither separable, nor differentiable. This is due to the
notorious, yet fundamental, singularity in the spectral density at zero temperature and zero frequency,
i.e., the Heaviside and the branch cut, in Eqs.~\ref{eq:spdensity} and~\ref{Eq:ei}, respectively.

To deal with the issues caused by the singularity, from now on we assume that
the number of Bohr frequencies is much larger than 1. Most systems in condensed matter physics satisfy this condition.
Further analysis is also restricted to quantum systems with
measurable sets of Bohr frequencies, with the
measure denoted by $d\mu$. This maps
the matrix eigenvalue problem into the integral operator problem,
\begin{equation}
\int X(\omega,\omega') U_n(\omega') d\mu'=x_n U_n(\omega).
\label{Eq:Fredholm}
\end{equation}

Compact integral operators such as~\ref{Eq:Fredholm} have countable (discrete) spectra.
The distribution of eigenvalues of integral operators is a long studied subject, dating back to Weyl~\cite{Weyl}.
The  smoothness of the kernel is essential:
the smoother kernel, the faster decay of the eigenvalues.
If the m-th order derivative of the kernel exists and is continuous, then the eigenvalues decay as $\vert x_n\vert=O(n^{-m-1/2})$~\cite{Wathen}.
If there is an analytic continuation inside the so called Bernstein's ellipse,~\cite{Trefethen,Wathen}
the  decay will be exponential~\cite{little}.

In the open quantum system, these functional analysis results are very powerful at a temperature not much smaller than the frequency spread $\omega_s$.
Finite temperature lifts the singularity from the real axis into the Matsubarra poles, thereby making the kernel analytically continuable to
an open region in the complex plane. Thus, the functional analysis theorems
predict exponential decay of the s-values, which limits the state displacement to be within $\sim\alpha$, as shown in the previous section.
Since the focus of this paper is on the limit at low temperatures, we will not review these results further.
The error bound of the ULE, is also only valid at not too low of a temperature.

\subsection{Region of Validity of the GAME\label{Sec:RegionOfGame}}

At zero temperature, the general results do not predict geometric decay of the s-values of the $X$-kernel, because the spectral density is not analytic.
However, in the previous paper on the geometric-arithmetic approximation~\cite{Davidovic2020}, a simple coarse-graining argument,
independent of the singular properties of the spectral density, led to the conclusion that the
state displacement induced by the approximation is comparable to the weak coupling constant. We now repeat the essentials of the argument and expand it.

Here, the region of validity of a master equation is based on how good an approximation it makes of the Redfield equation.
This is not a rigorous region of validity, as determined by Davies~\cite{davies1974} in the secular approximation, or more recently by Trushechkin~\cite{Trushechkin}
in the partial secular approximation. Rigorous regions
reference exact quantum dynamics and do not use the Redfield equation as an in-between.

Although the Redfield equation convergence is not guaranteed, it should not be dismissed as an intermediary, because the secular and the partial secular approximations, which only have converging solutions, can be deduced from the Redfield equation.
If the relaxation rate is much smaller than the smallest oscillation frequency of the system,
the Davies master equation will arise from the Redfield equation, on coarse-grained time scale~\cite{mozgunov,Davidovic2020}. Similar can be said about the partial secular approximation~\cite{Vogt,Tscherbul}.
Thus, the coarse-graining process can be correct, even though it has been referred to as dubious~\cite{davies1974}.

It is important to recollect that our derivation of the displacement norm has the $O(\omega_r t)^2$ term. Thus, the results discussed here extrapolate from the limited time scale
to time scale comparable to the relaxation time. This is one reason we perform the following analysis on both the secular and the geometric-arithmetic mean approximations.
If we find that the extrapolation agrees with the rigorous result in the secular approximations, it will show that the methodology can be correct.
Another help is that the ULE bound, which also uses the Redfield as an intermediary, as us,
seems not to have the $O(\omega_r t)^2$ term. In the range of its validity, the ULE bound agrees with the inference we make below.

The displacement kernel matrices vary in time. After a unitary transformation $X_t\to X_t'=\exp[-i(\omega-\omega')t/2]X_t$
in Eq.~\ref{Eq:XtminusX},
the following expression for the coarse-grained derivative of the displacement kernel matrices follow:
\begin{eqnarray}
\nonumber
&&\frac{X_t'(\omega_k,\omega_q)-X'(\omega_k,\omega_q)}{t}= \text{sinc}\frac{(\omega_k-\omega_q)t}{2}\\
&&\times\left\{
                    \begin{array}{ll}
                      \Gamma(\omega_k)+\Gamma(\omega_q)^\star-2J(\omega_k)\delta_{k,q}, & \hbox{Davies;} \\
                      \Gamma(\omega_k)+\Gamma(\omega_q)^\star-2\sqrt{J(\omega_k)J(\omega_q)}, & \hbox{GAME.}
                    \end{array}
                  \right.
\label{eq:cgsdg}
\end{eqnarray}
Here $\text{sinc}(x)=\sin(x)/x$.
The origin of the sinc function is in the integral over the interval $[0,t]$ in Eq.~\ref{eq:stateDISPL}.
The time-limit has a bandwidth suppressing effect on the velocity kernel matrices, onto the right of the curly bracket. That is,
the multiplication is like a band-pass filter into the frequency region $\pm 1/t$ around the diagonal. On the diagonal itself the matrix elements
of the velocity kernels are zero to begin with.

Now, recall that the state displacement is related to the trace norm of the kernel, Eq~\ref{Eq:tracenorm}. Then, due to the Lipschitz continuity of $V$ in the GAME, the trace norm of the matrix on the RHS is suppressed in proportion to $1/t$, at the time-scale of the kernel.
This suppresses the trace norm of the coarse-grained derivative on the LHS,
so $\Vert X_t'-X'\Vert_1$, and the state displacement, stop growing.

Since the velocity kernel in the secular approximation is discontinuous, the time scale to suppress $\Vert X_t'-X'\Vert_1/t$ corresponds to
the frequency spacing at $\omega=\omega'$. This time scale is unrelated to the kernel.

\begin{figure}
\centering
\includegraphics[width=0.45\textwidth]{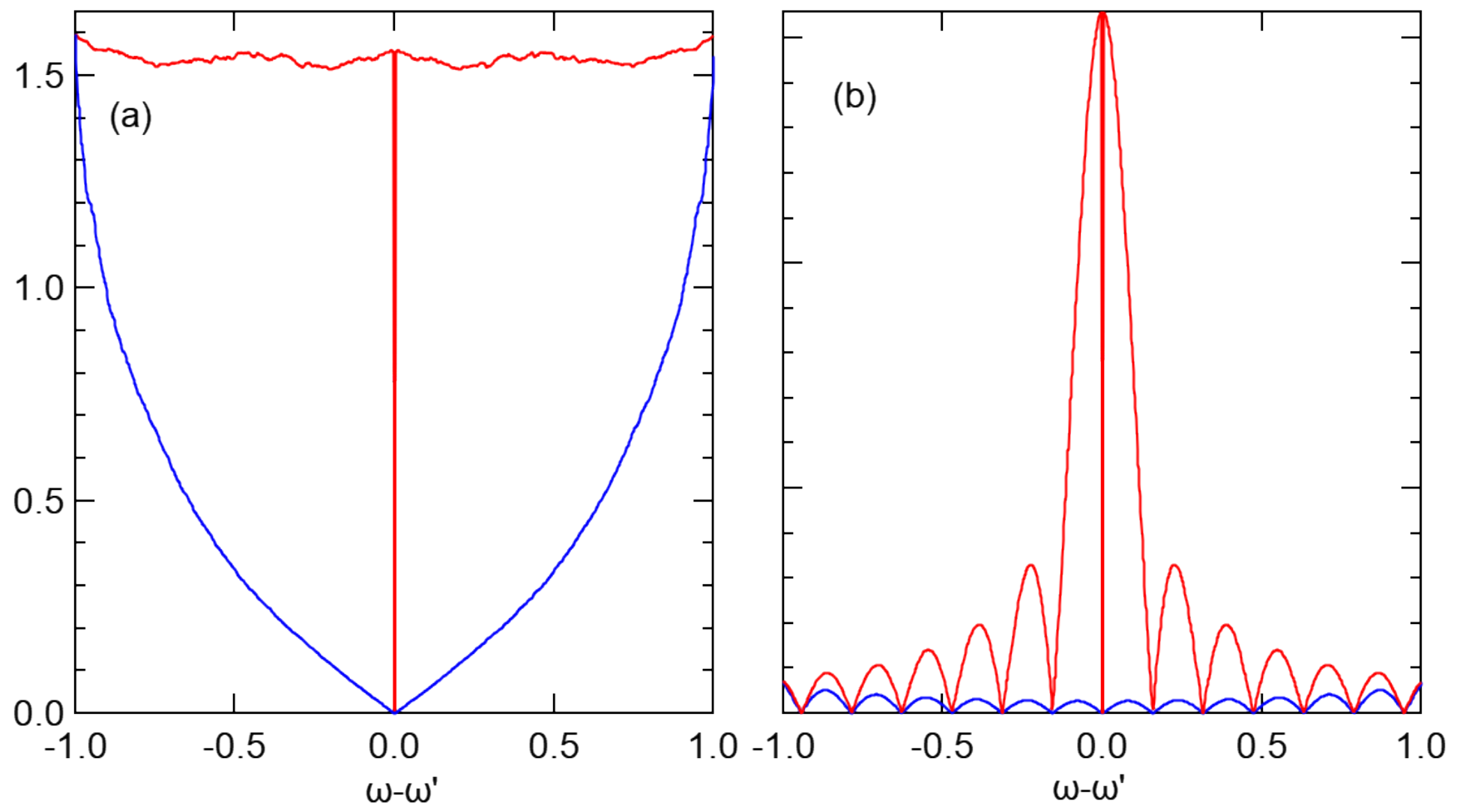}
  \caption{\label{Fig:sinc} Magnitude of the velocity kernel versus frequency difference, at the average frequency $\max(\omega)/2$,
in the secular approach (red) and the geometric-arithmetic approach (blue).
(a) Time $t=0$. (b) Time $t=40/\max(\omega)$.}
\end{figure}

\begin{figure}
\centering
\includegraphics[width=0.45\textwidth]{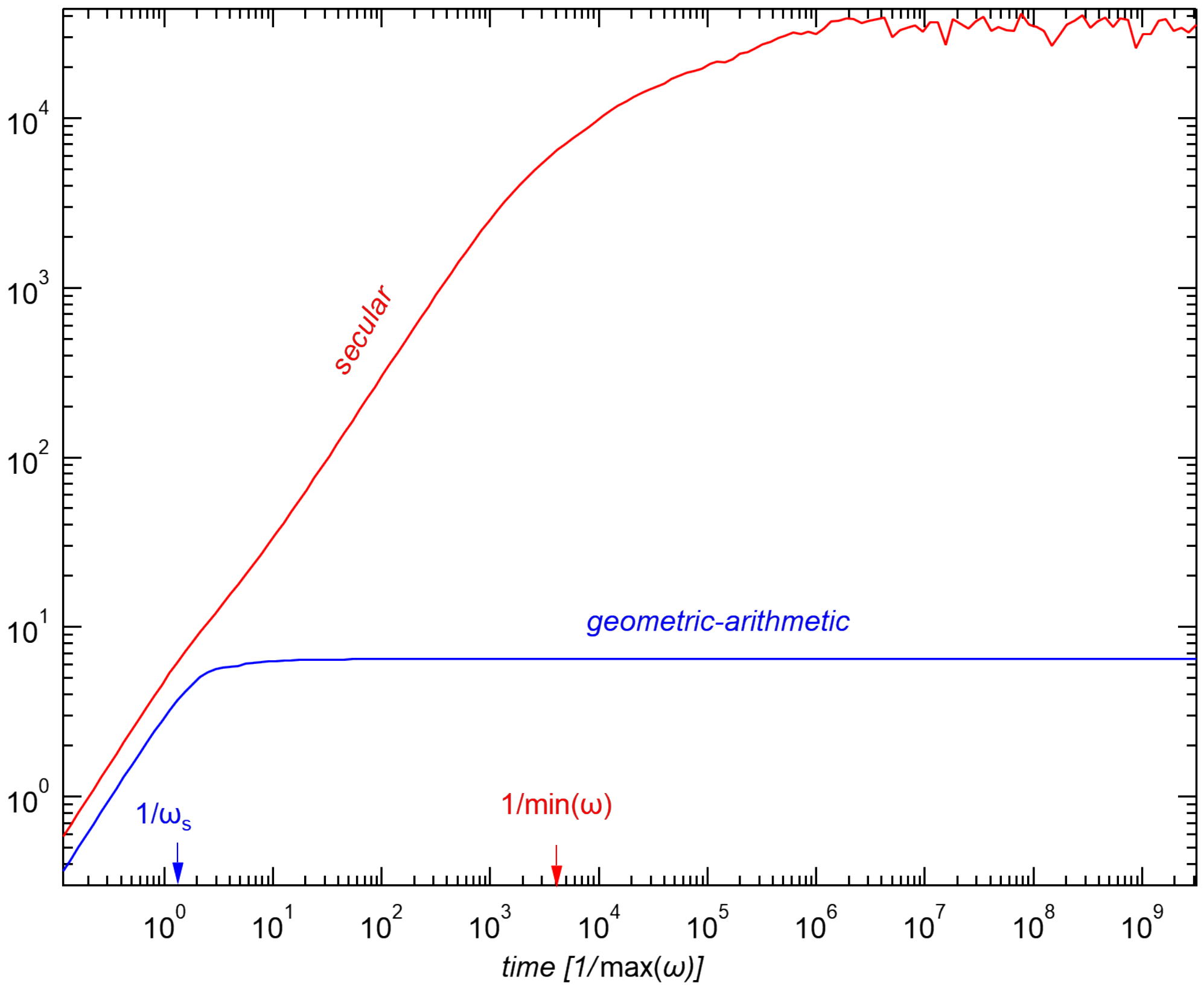}
  \caption{\label{Fig:normative} Operator norm of the displacement superoperator, $\Vert\Delta_t\mathfrak{X}\Vert_\infty/\alpha$, versus system evolution time.
 Blue and red correspond to the geometric-arithmetic and the secular approximations.
In order for the master equation to be a valid approximation of the Redfield equation, the plateau value multiplied by $\alpha$ should be less than one. Thus,
the relaxation rate needs to be smaller than the frequency spread and the minimum oscillation frequency, in the geometric-arithmetic and the secular approximations, respectively.}
\end{figure}

{\bf Example:} Assume a random sample of 1000 Bohr frequencies, with uniform distribution on $(0,1]$,
the Ohmic spectral density at $T=0$, and $\omega_c=10$. After adding the negative frequency image and zero frequency, the velocity kernels are calculated.
In this sample, $\min(\omega)=3.4\times 10^{-4}$ and $\omega_s=0.77$ (Eq.~\ref{Eq:compressionFreq}).

Fig.~\ref{Fig:sinc} displays $\vert V[\frac{\max(\omega)}{2}+\frac{\Delta\omega}{2},\frac{\max(\omega)}{2}-\frac{\Delta\omega}{2}]\vert$
in the secular and the geometric-arithmetic approximation.
In the former case, there is an apparent discontinuity in the matrix element
at $\Delta\omega=0$.
This originates from the discontinuity in the secular kernel at $\omega=\omega'$. In the geometric-arithmetic approximation, the kernel is continuous.

The effect of multiplying by the sinc function at $t=40/\max(\omega)$
is shown in Fig.~\ref{Fig:sinc}(b), showing a strong suppression of the $V$-kernel in the geometric-arithmetic approximation.
We calculate the bound on the operator norm $\Vert\Delta_t\mathfrak X\Vert_\infty$, which limits the state displacement via Eq.~\ref{Eq:brut}. See appendix~\ref{appendix:normIneq}
for details. Fig.~\ref{Fig:normative} shows that the displacement norm
saturates versus time, near $t=1/\omega_s$ and $t=1/\min(\omega)$, in the geometric-arithmetic and the secular approximations, respectively.
The validity threshold of the master equation is determined by the condition that the displacement norm be less than one.
In the secular approximation, we need $\alpha<2.5\times 10^{-5}$.  The operator norm of the dissipator~\ref{Eq:dissipRE} can be either directly calculated
or estimated by the inequality~\ref{Eq:boundS}: $\Vert\mathcal{D}_r\Vert_\infty=3.43\alpha$.
At the threshold $\alpha=2.5\times 10^{-5}$, this is equal to $2.5\min(\omega)$.
Thus, the weak coupling
regime will be valid, if the relaxation rate is near or smaller than the minimum system oscillation frequency.
This condition is rigorously established~\cite{davies1974}.

In comparison, we need $\alpha<0.15$ for the GAME to be valid. The threshold relaxation rate is $\Vert\mathcal{D}_r\Vert_\infty=0.67\omega_s$.
Now, the region of validity of the weak coupling regime, at any temperature,
is that the relaxation rate be smaller than the frequency spread. If that spread is the bath relaxation rate, then the RE and the GAME will be on equivalent level of accuracy.

\subsection{Non-Markovian Extension}
Next we consider the dynamics at time scale $t\ll\tau_c$,
where the GKSL form naturally emerges, as in Ref.~\cite{Whitney}.
Eq.~\ref{Eq:timedSD} can be approximated as,
\begin{equation}
 \Gamma_{\alpha\beta}(\omega,t)\approx C_{\alpha\beta}(0)t.
 \end{equation}
It is easily shown that matrix $C_{\alpha\beta}(0)$ is also PSD, and after some algebra, similar to that in the main part of Sec.~\ref{sec:CPimposition}, we find
\begin{equation}
D(t)=\sum\limits_{\gamma}M_{\gamma}^\star(t)\otimes M_{\gamma}(t),
\label{Eq:Dmatd}
\end{equation}
 where
 \begin{equation}
M_{\gamma}(t)=\sqrt{2t}\sum\limits_{\beta k}\big[\sqrt{C(0)}\big]_{\gamma\beta}e^{-i\omega_k t}a_\beta(\omega_k)
~\label{eq:dosta}
 \end{equation}
and the ME assumes the GKSL form once again.
\begin{figure}
\centering
\includegraphics[width=0.5\textwidth]{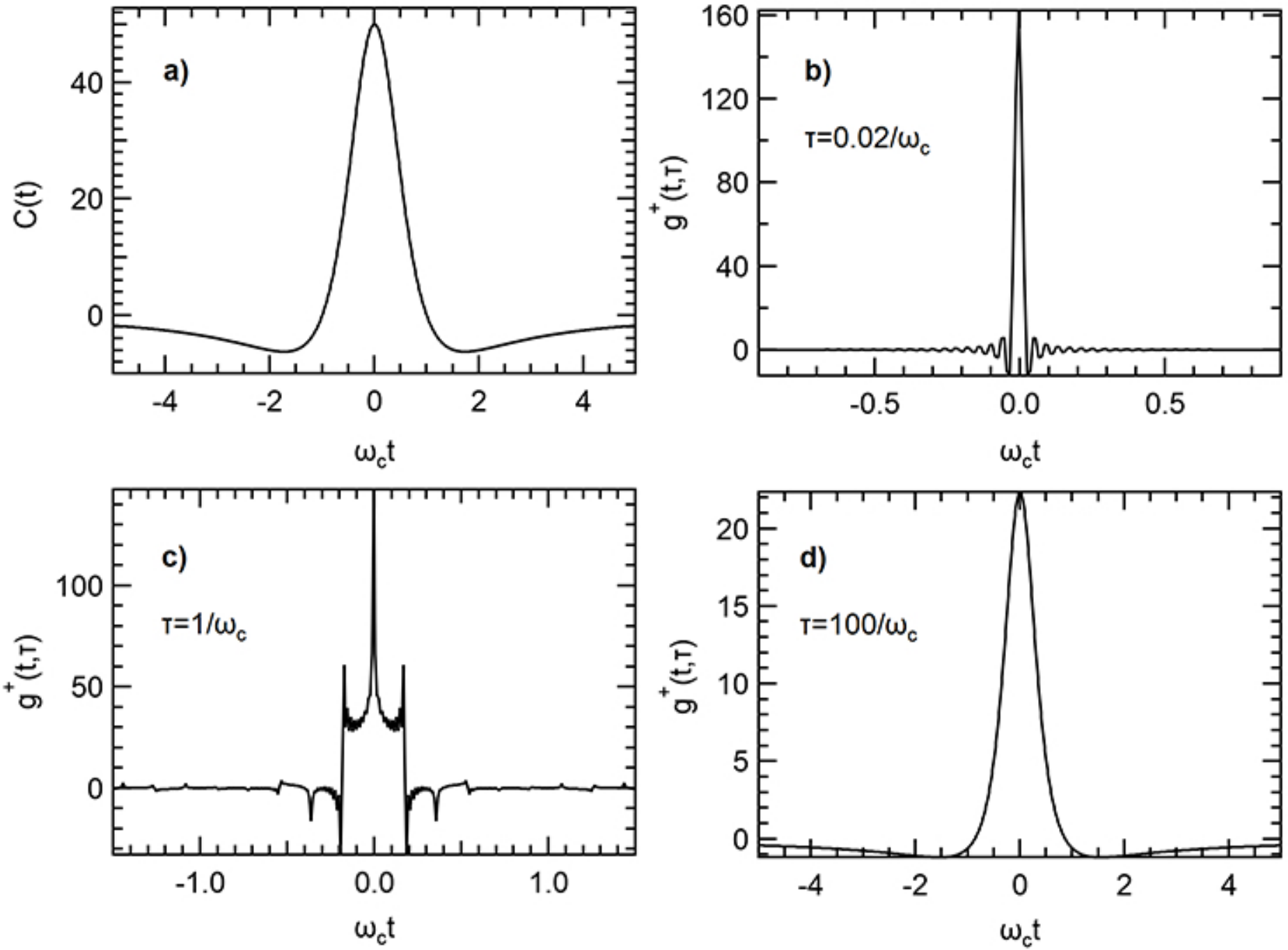}
  \caption{\label{Fig:BCF} Real part of correlation functions for the Ohmic bath at zero temperature. a) Bath correlation function. b)-d) Jump correlator at $\tau=
0.02/\omega_c$, $1/\omega_c$, and $100/\omega_c$, respectively.}
\end{figure}

At the problematic intermediate time scale, $\tau_c<t<\tau_s$, the real part of the time-dependent spectral density can be negative introducing a problem of how to take the square-root.
Introducing a complex impedance $Z(\omega)$ from Sec.~\ref{sec:CPimposition} may lead to a fix of some of the issues.

However, at this time, we utilize the Jordan decomposition
$J(\omega,t)=J^+(\omega,t)-J^-(\omega,t)$, where both  $J^+(\omega,t)$ and $J^-(\omega,t)$ are PSD, and interpolate. Namely, since per the above $J^-(\omega,t)$ approaches zero at $t\ll\tau_c$ and
$t\gg \tau_s$,
 dropping $J^-(\omega,t)$ will interpolate between the short and long time scale and guarantee complete positivity of the entire dynamics. We have
\begin{eqnarray}
\nonumber
M_\gamma(t)&=&\sum_{\beta k}\big[\sqrt{2J^+(\omega_k,t)}\big]_{\gamma\beta}e^{-i\omega_k t}a_\beta(\omega_k)\\
\label{Eq:jumpBCF}
&=&\sum\limits_\beta\int\limits_{-\infty}^{\infty}d\tau\,g_{\alpha\beta}^+(\tau,t)A_\beta(t-\tau).
\label{Eq:jumpBCFnonMarkov}
\end{eqnarray}
Here we introduced the timed jump correlator to be the matrix
\begin{equation}
g^+(\tau,t)=\int\limits_{-\infty}^{\infty}\frac{d\omega}{2\pi}\,e^{-i\omega \tau}\sqrt{2J^+(\omega,t)}.
\label{eq:timedjump}
\end{equation}
The dependence on two time arguments accounts for the memory in the bath.

Fig.~\ref{Fig:BCF}(a) displays the real part of the bath correlation function, while Figs.~\ref{Fig:BCF}(b-d)
show the real part of the jump bath correlator at different times.
The latter has a very complicated time dependence,
ranging from a delta at $\tau\to 0$ to the one in the long time limit, from Eq.~\ref{eq:timedjump}.
As seen in~\ref{Fig:BCF}(c), the time axis develops fractions of $t$, above and below $t$.
These ''time-fractals'' are due to taking the square root of an oscillating function around zero, and await further study. However irregular,
this non-markovian extension can be easily implemented when comparing non-markovian perturbative master equations with the geometric-arithmetic master equation.
This will be done in Sec.~\ref{sec:GUE}.

\section{Time Dependent Fluctuations of Displaced States - II~\label{Sec:Range2}}

In this section, we discuss the nuances in the spectra of the position-displacement kernel matrices.
Although the spectral approach is less intuitive than the coarse-graining argument, it opens ideas for future work as well as it relates to other research fields, such as
communications theory and machine learning. The spectral approach confirms the idea of the saturation of displacement induced by the geometric-arithmetic approximation.
At saturation, the displacement kernel becomes static, giving rise to
fluctuations analogous to how operators fluctuate in the Heisenberg picture.

\subsection{Initial Kinematics\label{subsec:kinematics}}

We proceed to study the displacement dynamics, by
expanding Eq.~\ref{Eq:XtminusX} into the Taylor series,
\begin{eqnarray}
\nonumber
X_t&=&X+\sum\limits_{kq}
[V(\omega_k,\omega_q)t+\frac{1}{2!}G(\omega_k,\omega_q)t^2\\
&+&\frac{1}{3!}W_3(\omega_k,\omega_q)t^3+...].
\label{Eq:LiuvDt}
\end{eqnarray}
The power series~\ref{Eq:LiuvDt} generates coefficient kernels, resembling the expansion of
cumulants. This series converges at any time, because it expands the exponential.
However, at large $t$, one needs to truncate into an impractically high
degree polynomial. Although it has this deficiency, the Taylor expansion exposes low ranks
of the coefficient kernels, which will suggest an alternative approach.

The velocity kernel was defined by Eq.~\ref{Eq:velocityGA}. Similarly, $G$ is a traceless and hermitian acceleration kernel
\begin{eqnarray}
&&G(\omega,\omega')=-(\omega'-\omega)^2X(\omega,\omega')\\
&&=i(\omega'-\omega)[\Gamma(\omega)+
\Gamma(\omega')^\dagger
-2\sqrt{J(\omega')^\dagger}\sqrt{J(\omega)]}.
\label{Eq:accelerationGA}
\end{eqnarray}
The nth order kernel can be written as $W_n(\omega,\omega')=[i(\omega'-\omega)]^{n-1}V(\omega,\omega')$.

These are all separable kernels. Let us solve the eigenvalue problem of the velocity kernel matrix,
\begin{equation}
\sum_{q}V(\omega_k,\omega_q)u(\omega_q)=v u(\omega_k),
\label{Eq:eigpro}
\end{equation}
where $v$ and $u(\omega_k)$ are the real eigenvelocities and the normalized eigenvector components, respectively.
Inserting from Eq.~\ref{Eq:velocityGA}, we have the solution
\begin{eqnarray}
\nonumber
\frac{v}{N_f} u(\omega_k)&=&\big[\Gamma(\omega_k)\langle u\rangle +\langle \Gamma^\star  u\rangle\\
&-&2\sqrt{J(\omega_k)}\langle \sqrt{J}u\rangle\big],
\label{eq:rank3}
\end{eqnarray}
where, recall from Sec.~\ref{Sec:Freqs}, $\langle ... \rangle=\sum_{\omega_k}(...)/N_f$.

Next, we recalculate the frequency averages based on the solution~\ref{eq:rank3}, and organize them into a matrix  problem
\begin{equation}
\left[
  \begin{array}{ccc}
    \langle \Gamma\rangle-\frac{v}{N_f} & 1 & -2\langle\sqrt{J}\rangle \\
    \langle\vert\Gamma\vert^2\rangle & \langle\Gamma^\star\rangle-\frac{v}{N_f}  & -2\langle  \Gamma^\star\sqrt{J}\rangle \\
    \langle\sqrt{J}\Gamma\rangle & \langle\sqrt{J}\rangle & -2\langle J\rangle-\frac{v}{N_f} \\
  \end{array}
\right]
\left[
  \begin{array}{c}
\langle u\rangle \\
\langle \Gamma^\star u\rangle\\
\langle \sqrt{J} u\rangle \\
  \end{array}
\right]
=0.
\label{eq:matrixproblem}
\end{equation}

All nonzero eigenvalues are the solutions of the cubic equation
\begin{eqnarray}
\nonumber &&
\left(\frac{v}{N_f}\right)^3-\left[\alpha^2\omega_s^2 -4\langle\sqrt{J}\,\rangle\left(\langle\sqrt{J^3\,}\rangle-\langle\sqrt{J}\,\rangle^3\right)\right]\frac{v}{N_f}\\
\nonumber &&
-2\langle J\rangle \alpha^2\omega_s^2+2\left(\langle\sqrt{J}\rangle^2\langle
\vert\Gamma\vert^2\rangle+\vert\langle\sqrt{J}\Gamma\rangle\vert^2\right)\\
&&
-2\langle\sqrt{J}\rangle
\left(\langle\sqrt{J}\Gamma\rangle\langle \Gamma^\star\rangle
+\langle\sqrt{J}\Gamma^\star\rangle\langle \Gamma\rangle\right)=0,
\label{Eq:cubic}
\end{eqnarray}
where $\omega_s$ is given by Eq.~\ref{Eq:compressionFreq}.

%This scaling neglects the singularity in the principal part at zero frequency, which is
%logarithmic and
%requires a cut-off. %The scaling cannot not be perfect, due to the singularity of the principal component
%at zero frequency, while the Heaviside step function of the dissipative component does not affect the scaling.
%In the Ohmic bath, there a logarithmic correction to scaling, while in sub-Ohmic baths at zero temperature, there is
%a power-law contribution and the scaling breaks down as the frequency spread of the system goes to zero, while in super-Ohmic phonon baths, the scaling will
%be nearly perfect. It is the scaling of the eigenvalues, in response to the frequency renormalization of teh system,
%that is at the essence why the geometric-arithmetic approximation works well.
Due to the tracelessness of $V$, the solutions satisfy $v_1+v_2+v_3=0$.
In our experience, $\vert v_3\vert$ is smaller than $\vert v_1\vert,\vert v_2\vert$, by factor of about 8.
If we neglect $v_3$, the dominant eigenvelocities, from the roots of Eq.~\ref{Eq:cubic}, become
\begin{equation}
v_{1,2}\approx\pm\alpha N_f\omega_s,
\label{eq:v12}
\end{equation}
hence our definition of the spread frequency in Eq.~\ref{Eq:compressionFreq}.
It can be shown that the acceleration matrix $G(\omega_k,\omega_q)$ is also separable and has rank 6.
Generally, in the $n$th order, we get a rank $3n$ kernel.

\subsubsection*{Mapping to Prolate Spheroidal Wave Functions}

Although the Taylor series~\ref{Eq:LiuvDt} converges at arbitrary time, and has separable kernels, it is very difficult to calculate the eigenvalues at long time scale.
At a moderately small time, comparable to $1/\omega_s$, prolate spheroidal wave functions (PSWF) introduced by Slepian and Pollak~\cite{Slepian1}
lead to a somewhat simpler approach.

We rewrite Eq~\ref{eq:cgsdg} as
\begin{equation}
X_t'-X'=2\pi K_{\frac{t}{2}}(\omega-\omega')V(\omega,\omega'),
\label{Eq:prolate}
\end{equation}
where
\begin{equation}
K_\tau(\omega-\omega')=\frac{\sin[\tau(\omega-\omega')]}{\pi(\omega-\omega')}
\label{Eq:sinc}
\end{equation}
is the band limiting kernel, which has been well studied in communications theory,
noise analysis, and machine learning. It is a PSD kernel with spectral properties known in great detail~\cite{Slepian1,Slepian5,BOYD2003168,LiLian,LiLian1}.
In our paper, we need to swap out time and frequency, to map to that literature.
The following table shows how our variables correspond to the variables in~\cite{Slepian1}:
\begin{center}
\begin{tabular}{|c|c|c|c|}
  \hline
  this paper & $t/2$ & $\omega$ & $2\max(\omega)$ \\
\hline
   Ref.~\cite{Slepian1} & $\Omega$ & $t$ & $T$\\
\hline
\end{tabular}
\end{center}

Next, we make the assumption that the kernel matrices have eigenvalues converging to those of their integral kernels.
In kernel methods in machine learning, it has been established that the eigenvalues of Lipschitz continuous kernel matrices
converge to the eigenvalues of
the respective integral operators, with robust relative errors,
even if the eigenvalue magnitudes decay exponentially~\cite{Trefethen,koltchinskii1998asymptotics,braun2005spectral}.

In addition, we  make the assumption of uniform measure: $d\mu=d\omega$.
The effect of the measure on the spectral properties of PSD kernel matrices has also been investigates in machine learning,
with the result that the measure did not significantly affect the
decay rates of the s-values~\cite{Belkin2018}.

If we transit between the kernel and the matrix, the product with $V(\omega,\omega')$ in~\ref{Eq:prolate} will transit to the Hadamard (elementwise) product.
Although this product is awkward to work with, it does
factor out the singularity in the spectral density, confining it to the rank-3, exactly diagonalizible, matrix.
To see how that works, we take advantage of the decomposition of the Hadamard product,
\begin{equation}
K_\tau\circ V=\sum_{s=1}^{3}\sum_{i=1}^{N_f} \mu_i v_s\vert u_s\rangle\circ\vert \psi_i\rangle \langle u_s\vert\circ\langle \psi_i\vert.
\label{Eq:Hadam}
\end{equation}
Here, $\mu_s$ and $v_s$ are the eigenvalues of $K_\tau$ and $V$, and $\vert \psi_s\rangle$ and $\vert u_s\rangle$ are the respective eigenvectors.
$1>\mu_1>\mu_2\geq...> \mu_{N_f}\geq 0$ and $\vert \psi_i\rangle$ converge to the PSWF.

The decomposition~\ref{Eq:Hadam} is useful to prove a well known inequality, $\text{rank}(K_\tau\circ V)\leq \text{rank}(K_\tau)\text{rank}(V)$.
In a similar vein, we can say that if $\mu_i $ decreases geometrically versus $i$, the eigenvalues of the Hadamard product will also decay geometrically.
This is due to the low rank of the velocity kernel.
One should note,~\ref{Eq:Hadam} is not the spectral decomposition, because the Hadamard product of two orthonormal sets, one complete $\{\vert \psi_i\rangle\}$ and
another not complete $\{\vert u_s\rangle\}$,
do not make a complete orthonormal set.

Although the decay of $\mu_n$ is supergeometric
at any time~\cite{LiLian1}, at a long time the prolate spectrum has a near, but finite, degeneracy at $\mu\approx 1$, making the mathematical analysis complex, which we leave for future studies.
Let us now examine the decay of the s-values of the displacement kernel at times $t_1=1/\max(\omega)$ and $t_2=2/\max(\omega)$. In Fig.~\ref{Fig:normative},
those times a slightly lower and higher than the spread time $1/\omega_s$.
Table~1 in Ref.~\cite{Slepian1} gives the following decays for $\mu$: $3.1\times 10^{-1}$, $8.6\times 10^{-3}$, $3.9\times 10^{-5}...$, at $t_1$ and
$5.7\times 10^{-1}$, $6.3\times 10^{-2}$, $1.23\times 10^{-3}...$, at $t_2$. The truncation to the highest eigenvalue can be done with good accuracy.
Since the velocity kernel has two dominant eigenvelocities, it follows that the eigenvalues of the Hadamard product can also be well truncated.
Thus, the displacement is of order $\alpha$.

No requirement of analyticity in the spectral density is required in the above, because the velocity kernel has rank 3 irrespective of the spectral density.
At long time scale, the analysis is much more complex due to the degeneracy discussed earlier. So, we take an alternative approach next.

\subsection{Saturation.\label{subsec:satkinematics}}

At time $t\gg\omega_s$,
the components of the displacement matrix~\ref{Eq:XtminusX},
$X(\omega_k,\omega_q)$ and $e^{i(\omega_q-\omega_k)t}X(\omega_k,\omega_q)$, are manifested by the smoothness of the former and the irregularity in
the latter. These matrices can be considered to be statistically uncorrelated. Their decoherence begins at time scale
$t\sim 1/\vert \omega_k-\omega_q\vert$, e.g., the
inverse of the frequency spread.

Let us apply the regular and the irregular kernel matrices, on the eigenvector
$\sum_q V_{q,n}\vert q\rangle$ of the regular matrix. In the limit $t\gg\tau_s$, we have
\begin{eqnarray}
\sum_{q}X(\omega_k,\omega_q)V_{q,n}&=&x_nV_{k,n}\\
\label{Eq:RHSfft}
\sum_{q}e^{i(\omega_q-\omega_k)t}X(\omega_k,\omega_q)V_{q,n}&\approx& 0.
\end{eqnarray}
The approximation in the second line works  like a reverse rotating wave approximation, where the oscillations average out in frequency at fixed time,
rather than over time at fixed frequency difference.
Similarly, the rapidly fluctuating vector $\sum_q e^{-i\omega_q t}V_{q,n}\vert q\rangle$ is an eigenvector of
both the irregular and the regular matrix, with the eigenvalues $x_n$ and approximately $0$, respectively.
The result is that the regular and irregular matrices
approximately commute. Their net contribution
displays time-independent eigenvalue doublets, $\pm x_n$.

By approximating the RHS in Eq.~\ref{Eq:RHSfft} with zero, we neglect the terms of order $n/(\omega_s t)$. At higher $n$, $V_{kn}$ varies faster versus $k$, (it has the number of nodes $\propto n$), so it takes a longer time
for the sum over $q$ in Eq.~\ref{Eq:RHSfft} to approach zero.

Thus, the eigenvalues of $X_t-X$ versus time saturate to $\pm x_n$,
with the saturation time comparable to $n/\tau_s$.
Above that time, the displacement superoperator $\mathfrak{X}_t-\mathfrak{X}$ continues to fluctuate in time, but on a time independent kernel. The dynamics of the displacement is
analogous to the dynamics of operators in the Heisenberg picture, in unitary quantum mechanics.

To conclude, in this section we showed that the decay of the s-values of the displacement kernel starts from the highest s-value and is rapid, at time $t=2/\max(\omega)$. This guarantees that
the state displacement from the Redfield equation solution is of order $\alpha$. We also explained how the displacement kernel saturates at the similar time scale, which implies that the
s-values
decay rapidly at any time. These conclusions are based on the spectral problem
of the displacement kernel, and do not involve any coarse-graining arguments that we utilized in the previous section.

These two pictures - the coarse-grained and the spectral - are in
agreement, both in terms of the time scale to saturate the displacement and the norm at saturation. Neither approach addresses the question if the distance between the solutions of the GAME
and the exact quantum
state approaches zero in the limit $\alpha\to 0$ {\it and} $N_f\to\infty$. This remains an unanswered fundamental question, regarding nonsecular dynamics in open quantum systems.

\section{\label{sec:heisenberg}Example 1. Heisenberg Spin-Chain}

\begin{figure}
\centering
\includegraphics[width=0.49\textwidth]{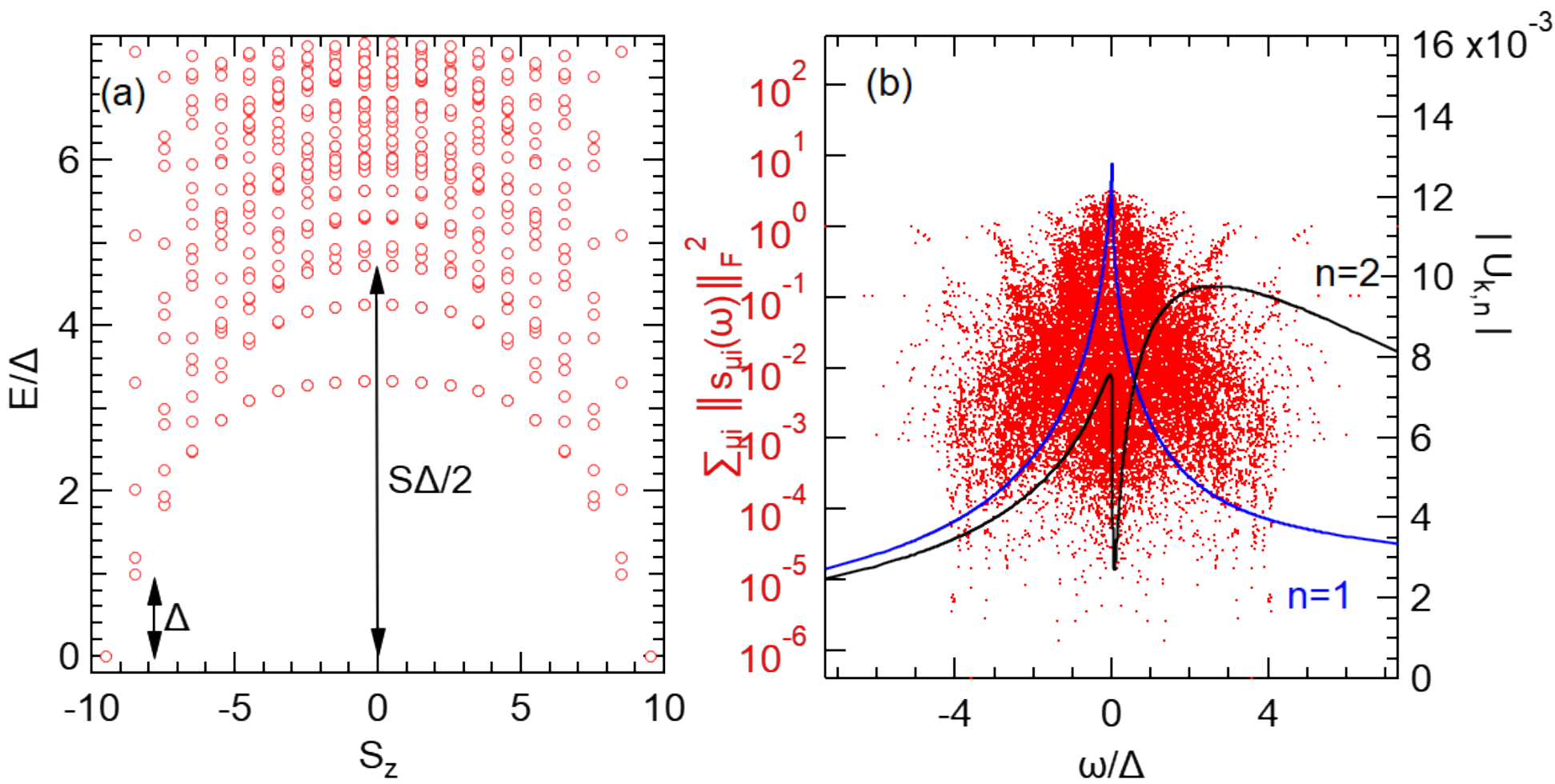}
  \caption{\label{Fig:spectrum} (a) Energy levels versus $S_z$ in a ferromagnetic spin chain with $S=19/2$ and uniaxial anisotropy.
(b) The net Frobenius norm versus Bohr frequency. Each red dot is a frequency.
Lines are the magnitudes of the lowest two eigenvectors of the $X$-kernel.}
\end{figure}

In this section we illustrate the concepts discussed in the previous sections, on a ferromagnetic spin-1/2 chain
with $N_s=19$ sites, interacting with nearest neighbour exchange interaction
and uniaxial anisotropy due to dipole-dipole coupling:
\begin{equation}
H_S=-J\sum_{i=1}^{N_s-1}\vec{s}_i\vec{s}_{i+1}-\epsilon_d\sum_{i=1}^{N_s-1}\sum_{j=i+1}^{N_s}\frac{3s_{zi}s_{zj}-\vec{s}_i\vec{s}_{j}}{(j-i)^3}.
\end{equation}
Here, $\vec{s}_i=(1/2)\vec{\sigma}_i$ is the spin operator at site $i$.
We introduced this model in our prior work~\cite{Davidovic2020}, to study the accuracy of the geometric arithmetic master equation. Note that $[H_S,S_z]=0$, where $S_z=\sum_{i=1}^{N_s} s_{zi}$ is the total spin z-component, so that $S_z$ is a good quantum number.
A characteristic frequency of the system is given by
the energy gap $\Delta$, which corresponds to the ferromagnetic resonance frequency.
$H_S$ is truncated to the lowest 472 eigenenergies, which depend on $S_z$  as displayed in Fig.~\ref{Fig:spectrum}(a). In this example, $J=25$, $\epsilon_d=0.375$, and $\Delta=1.242$,
while more generally,
$\Delta\approx 3.58\epsilon_d(1-1.29/N_s)$.

\begin{figure}
\centering
\includegraphics[width=0.49\textwidth]{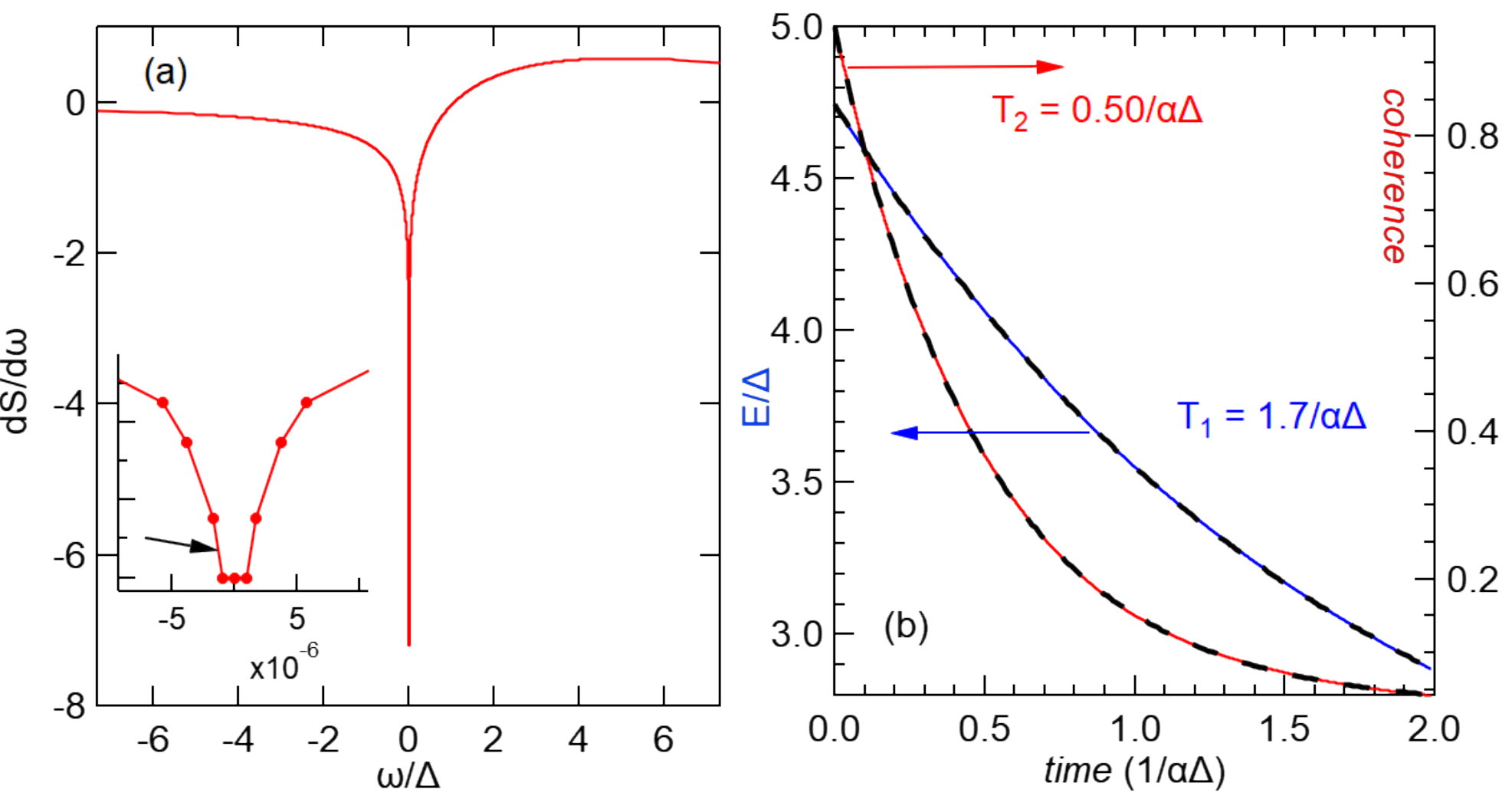}
  \caption{\label{Fig:cutoff} (a) $dS/d\omega$ versus frequency at $\alpha=1$. The inset displays the cut-off at zero frequency. The slope indicated by the arrow is the Lipschitz norm of $X$,
while magnitude of $X$ at the minimum is the Lipschitz norm of the velocity kernel of the GAME.
 (b) Redfield equation solution. Left and right axis: energy [$\text{tr}(\rho H_S)$] and the coherence ($\Vert \rho_c\Vert_1 /2$), respectively, versus scaled time. Dashed black lines are best fits to the exponential,
 from which the energy relaxation and decoherence time are extracted.
 The initial state $\vert S,S_x=S\rangle$.
$S=9/2$, $\omega_c=10\Delta$, and
$\alpha=0.001$.}
\end{figure}

\begin{figure}
\centering
\includegraphics[width=0.49\textwidth]{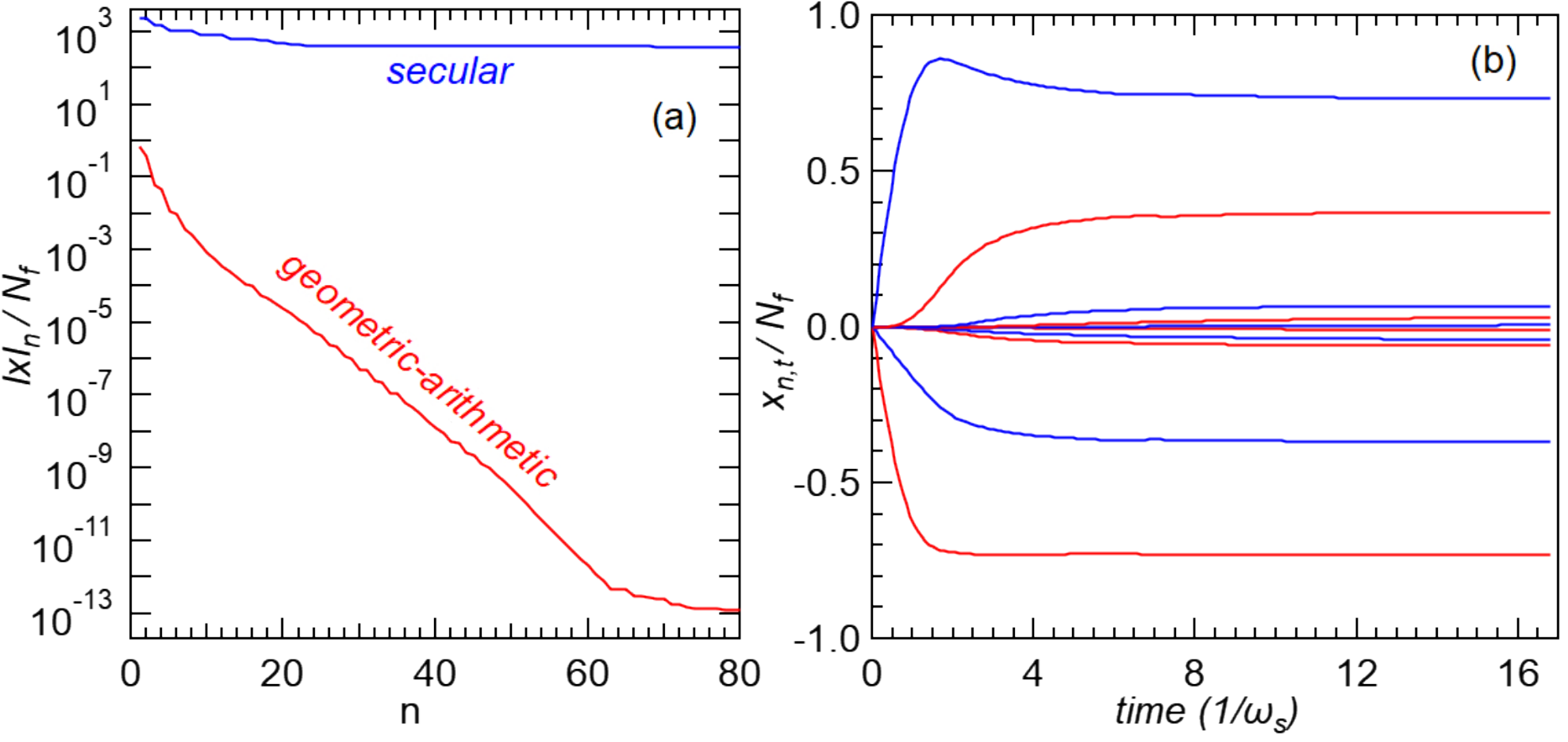}
  \caption{\label{Fig:filter} (a) S-values of $X$ vs their index $n$, in the secular (blue) and geometric-arithmetic (red) approximations.
(b) Eigenvalues of the displacement kernel versus time. At saturation, half of the eigenvalues match those of $X$ (red)
and another half match the same but opposite eigenvalues of $X$ (blue).}
\end{figure}

The spin chain is coupled to $3N_s$ independent Ohmic baths with the BCF given by Eq.~\ref{eq:OhmicBCF}
and system coupling operators $s_{xi}$, $s_{yi}$ and $s_{zi}$ at each site.
The coupling operators are not normalized in Frobenius
norm. To counteract that, when we plot the results, the norm will be explicitly shown.

The bath cut off frequency is set to $\omega_c=10\Delta$.
The model exhibits a range of frequencies, that can be associated with spin modes including ferromagnetic resonance $\Delta$, standing spin-waves,
and domain walls~\cite{Davidovic2020}. The spectrum has $N_f=19503$ Bohr frequencies with nonzero weight,
shown in Fig.~\ref{Fig:spectrum}(c). The minimum oscillation frequency is  min$|\omega|=9.27\times 10^{-7}\Delta$,
while the frequency spread is $\omega_s=\text{std}(\vert\Gamma\vert)/\alpha=1.360\Delta$, where std stands for standard deviation.
 This agrees well with our previous report where we observed that the trace distance between the Redfield and geometric-arithmetic equation
solutions saturated on time scale $1/\Delta$, longer than the bath correlation time.

$X(\omega,\omega')$ is singular at $\omega=\omega'=0$, because it has the derivative of the principal value on the diagonal, which is shown in Fig.~\ref{Fig:cutoff}(a).
The inset shows that at zero frequency, we take the nearest neighbour average around zero, as a cut-off.
The Lipschitz norms of $V$ and $X$ are the magnitude at minimum and the slope of the line
indicated by the arrow, respectively. Changing the cut-off value by factor of 10 has negligible effect near the largest five to ten s-values, but can have a significant
relative effect at very high $n$.

The
system is prepared into the initial state with magnetization along $x$ axis, perpendicular to the chain. That is, the initial state is approximately the spin eigenstate
with $S_x=9.5$ (within the span of the truncated Hilbert space).
We solve the markovian Redfield master equation numerically, and follow the relaxation of energy and of the norm of the offdiagonal density matrix elements, to determine the relaxation and
decoherence characteristic times, as shown in Fig.~\ref{Fig:cutoff}(b). $T_1$ and $T_2$ are obtained from the best fits to the exponential.

\begin{figure}
\centering
\includegraphics[width=0.45\textwidth]{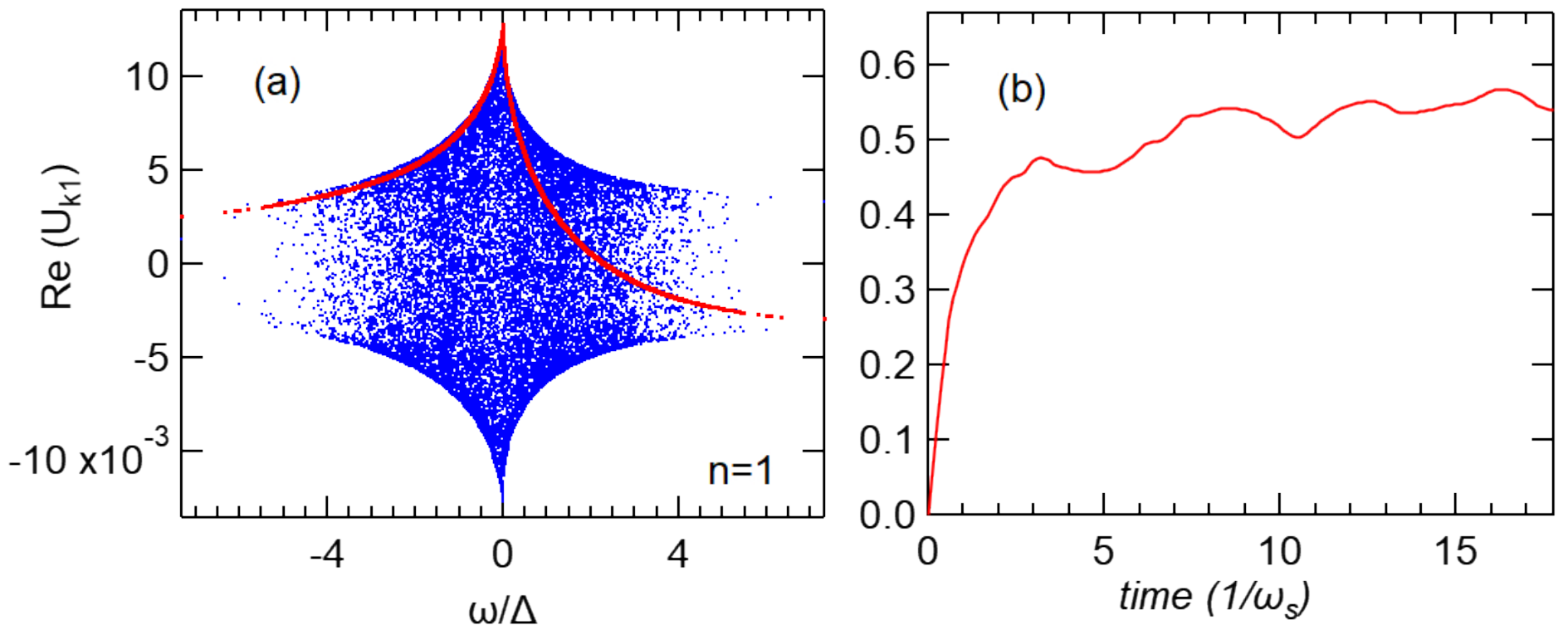}
  \caption{\label{Fig:saturation} (a) Real parts of the eigenvectors of the displacement kernel, corresponding to the highest magnitude eigenvalues of the $X$-kernel.
  Blue and red dots correspond to the top-most and the lowest-most eigenvalues at saturation in Fig.~\ref{Fig:filter}(b). (b) Displacement superoperator norm,
$\Vert \mathfrak{X}_t-\mathfrak{X}\Vert_\infty /\alpha\sum_{i\mu}\Vert s_{i\mu}\Vert_F^2$, versus time.}
\end{figure}

The velocity kernel has three nonzero eigenvalues, $v_1=1.01N_f\alpha\omega_s$, $v_2=-0.85N_f\alpha\omega_s$ and $v_3=-0.16N_f\alpha\omega_s$.
The corresponding eigenvectors are determined by Eq.~\ref{eq:rank3}, and span the frequency spread, with profiles determined by the spectral density of the bath.
The eigenvectors of $X$ also span the spread, as shown in Fig.~\ref{Fig:spectrum}(b) for $n=1,2$.

Fig.~\ref{Fig:filter}(a) displays the s-values of $X(\omega_k,\omega_q)$.
There is a rapid drop-off versus $n$. Also shown are the s-values in the secular approximation.
Not only they are much larger, but also they decay much more slowly with $n$ at high $n$, relative to the geometric-arithmetic case.

As a function of time, Fig.~\ref{Fig:filter}(b) shows the eigenvalues of the displacement kernel.
Only the highest three eigenvalues (in magnitude) are launched with nonzero velocity. As time increases,
eigenvalues approach saturation at time scale of the inverse spread frequency. At saturation, the eigenvalues display a split-doublet structure.
One half of the eigenvalues (red) saturate at the eigenvalues of the $X$-matrix.
The other half (blue) saturate at equal but opposite eigenvalues.

The eigenvectors of the highest doublet at saturation are shown in Fig.~\ref{Fig:saturation}(a).
The positive eigenvalue eigenvector has rapid oscillations and can be considered as uncorrelated with respect to the long-range eigenvector.
The result is that the displacement superoperator can be decomposed into two semi-orthogonal contributions, at saturation, illustrating the concepts discussed before.

We now calculate the operator norm of the displacement superoperator versus time, using the eigenvalues and the eigenvectors of the displacement kernel
plotted in Figs.~\ref{Fig:filter}(a,b) and~\ref{Fig:saturation}. The result is shown in Fig.~\ref{Fig:filter}(c), showing quasi-saturation on time scale $1/\omega_s$.
The normalized superoperator settles around $0.55\alpha$, or approximately at one half of the operator norm
$\Vert\mathfrak{D}\Vert_\infty$. But, in contrast to the kernel, the superoperator norm exhibits oscillations, which is why we refer to this as quasi-saturation.

\section{\label{sec:GUE} Example 2. Gaussian Unitary Sample}

\begin{figure}
\centering
\includegraphics[width=0.49\textwidth]{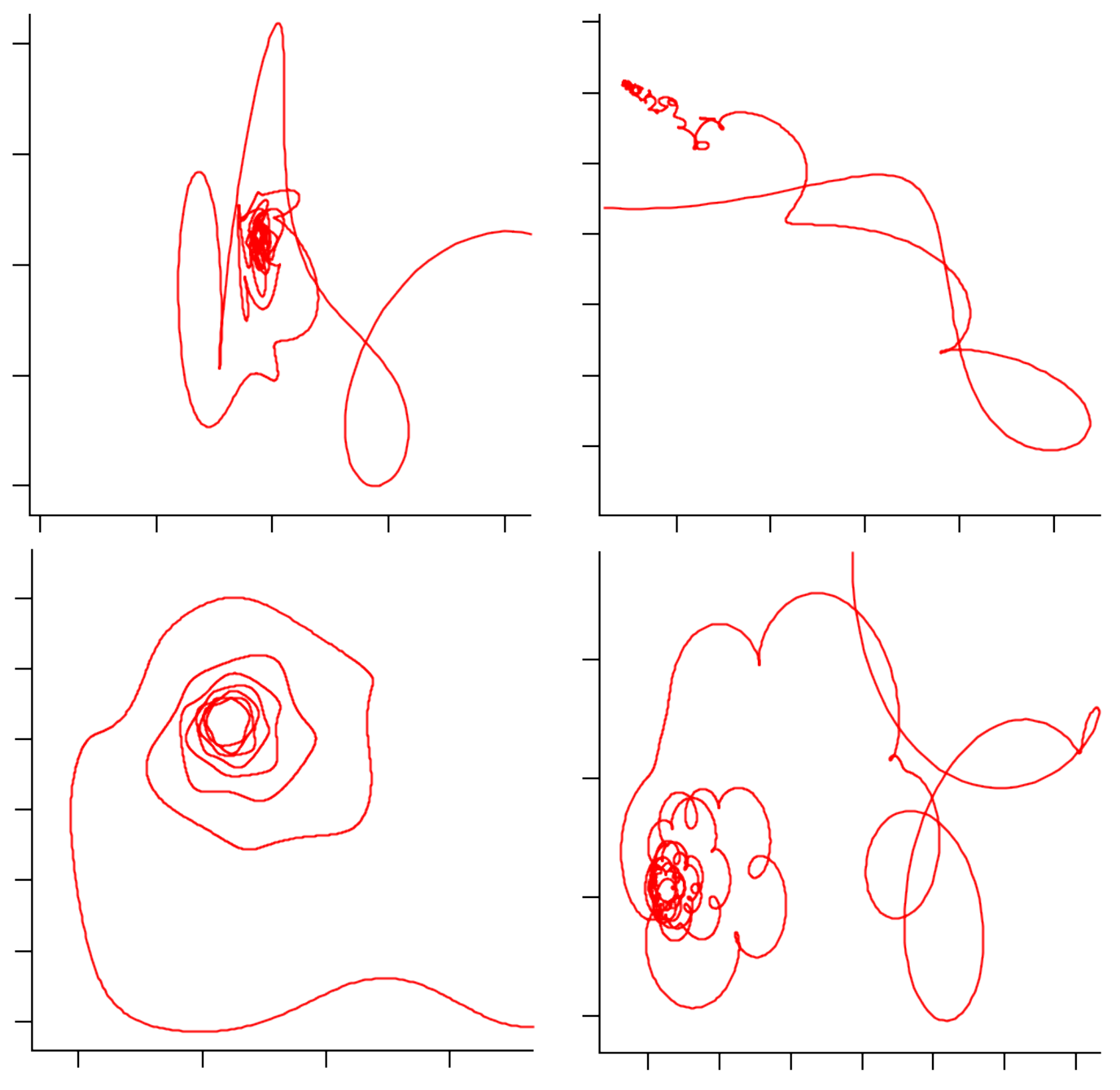}
  \caption{\label{Fig:circled} The imaginary versus the real part of four typical matrix elements of the fourth time ordered cumulant, for the random Hamiltonian sample.
  As time increases, the trajectories slowly approach the Markov limit while exhibiting complex dynamics.}
\end{figure}

To confirm that the completely positive GAME can be closer to exact quantum dynamics, than the less approximated Redfield equation, we draw a random Hamiltonian from the
Gaussian unitary ensemble (GUE) of $6\times 6$ matrices.
With 31 unique frequencies and 337 unique frequency differences, this system exhibits many oscillators that
we may expect in nonsecular dynamics. For example, Fig.~\ref{Fig:circled} displays 4 randomly selected matrix elements of the fourth order
cumulant in the TCL4 master equation. (We use the Ohmic spectral density in Eq.~\ref{eq:spdensity} and the bath cut-off frequency $\omega_c=10$.)
The plots show the imaginary versus the real part, as time increases.
The trajectories slowly approach the Markov limit, often circling the drain irregularly.
Yet, the system is sufficiently small, to let us calculate all of the coefficients of the TCL4 master equation, using
a 32 core AMD EPYC processor on a time scale of order one week.

\begin{figure}
\centering
\includegraphics[width=0.49\textwidth]{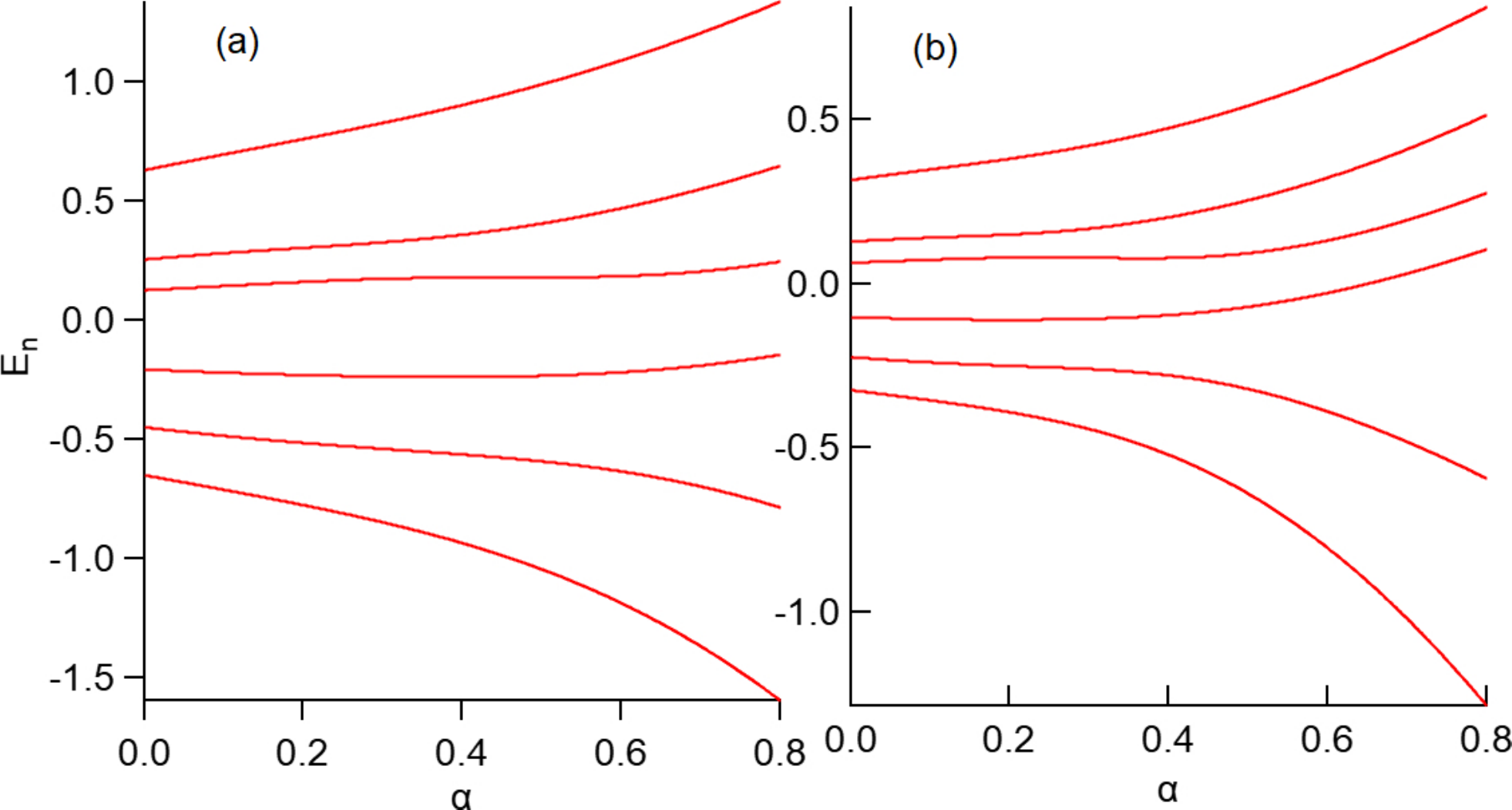}
  \caption{\label{Fig:correlations} Renormalization of the energy levels due to the bath, in the sample Hamiltonian. (a) For the sample system that we study. (b) For a system with half the level spacing in (a).}
\end{figure}

The energy levels of the system are $E=[-0.6490,-0.4476,-0.2045,0.1274,0.2561,0.6317]$. The frequency spread is $\omega_s=\text{std}[\vert\Gamma(\omega_k)\vert]/\alpha=0.70$.
The coupling operator is also drawn from the ensemble and normalized as $\Vert A\Vert_F=1$.

Fig.~\ref{Fig:correlations} displays the system energy levels versus the weak coupling constant. The
Hamiltonian of the system includes the Lamb-shift in second order in $\alpha$, obtained by the technique described
in appendix~\ref{appendix:Dissipator}. The level repulsion versus $\alpha$ is due to the correlations induced by the bath,
the characteristics of nonsecular dynamics. However, in our example [Fig.~\ref{Fig:correlations}(a)], the dynamics is only mildly nonsecular,
to keep the spread frequency much smaller than the relaxation rate, within the range of validity of the approximation.~\footnote{The bottleneck is the double integral
in the TCL4 in appendix~\ref{appendix:tcl}; this will be significantly improved in future work, as we recently learned how to reduce the double integral to a single integral.}

To determine CP-violations, we calculate the Choi matrix~\cite{CHOI1975285}, defined as
\begin{equation}
\mathcal{C}_t=\frac{1}{N}\sum_{i,j=1}^NE_{ij}\otimes T_\leftarrow\exp \left(\int_0^td\tau\mathcal{L}_\tau\right)E_{ij}.
\end{equation}
The dynamical map generated by $\mathcal{L}_t$ will be CP, iff $\mathcal{C}_t$ is PSD~\cite{CHOI1975285}.
\begin{figure}
\centering
\includegraphics[width=0.49\textwidth]{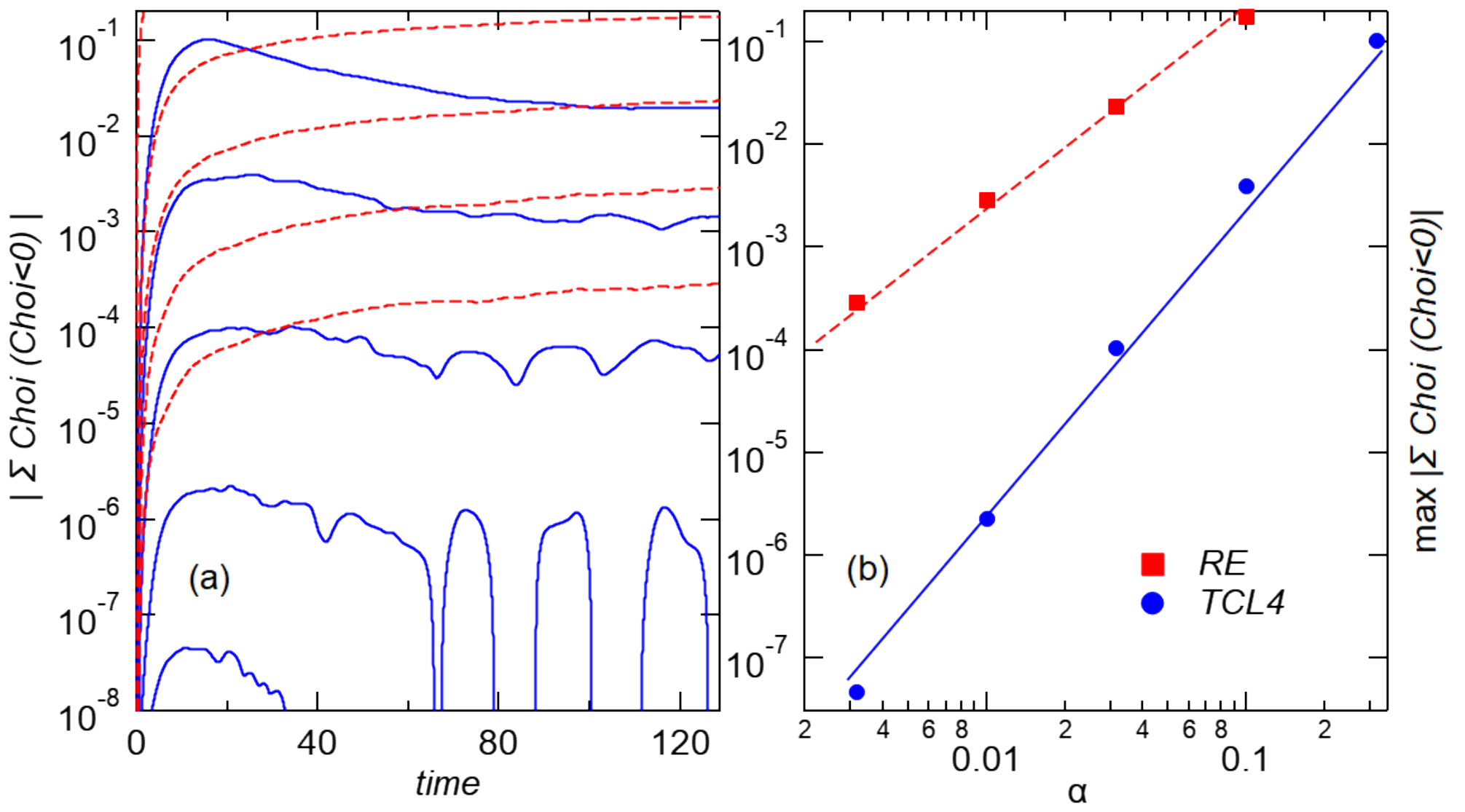}
  \caption{\label{Fig:Choi} Sum of the negative eigenvalues ($\Sigma$) of the Choi matrix versus time, determined from the Redfield (dashed red) and the TCL4 (full blue) master equations. (a) The weak coupling constant takes values $0.00316,0.01,0.0316,0.1$ and $0.315$, bottom to top. (b) The maximum in magnitude of $\Sigma$ versus the weak coupling constant. The lines between points are power laws $\alpha^2$ and $\alpha^3$, respectively. $\omega_c=10$ and $\omega_s=0.7$.}
\end{figure}

We numerically solve the Redfield and the TCL4 equation assuming $E_{ij}=\vert i\rangle\langle j\vert$ for the initial condition. Then we compute the Choi matrix. The CP-violation is
the sum ($\Sigma$) of the negative eigenvalues of the matrix.
Fig.~\ref{Fig:Choi}(a) displays $\vert\Sigma\vert$ versus time, for the Redfield and the TCL4 equation, at different weak-coupling constants. The violation is strongly suppressed in the TCL4 equation.  The maximum of $\vert\Sigma\vert$ versus $\alpha$ is a power law, as shown in Fig.~\ref{Fig:Choi}(b), with the exponents 2 and 3
for the Redfield and the TCL4, respectively. This cannot be by accident and assures that the cumbersome TCL4 coefficients are correctly calculated.
\begin{figure}
\centering
\includegraphics[width=0.49\textwidth]{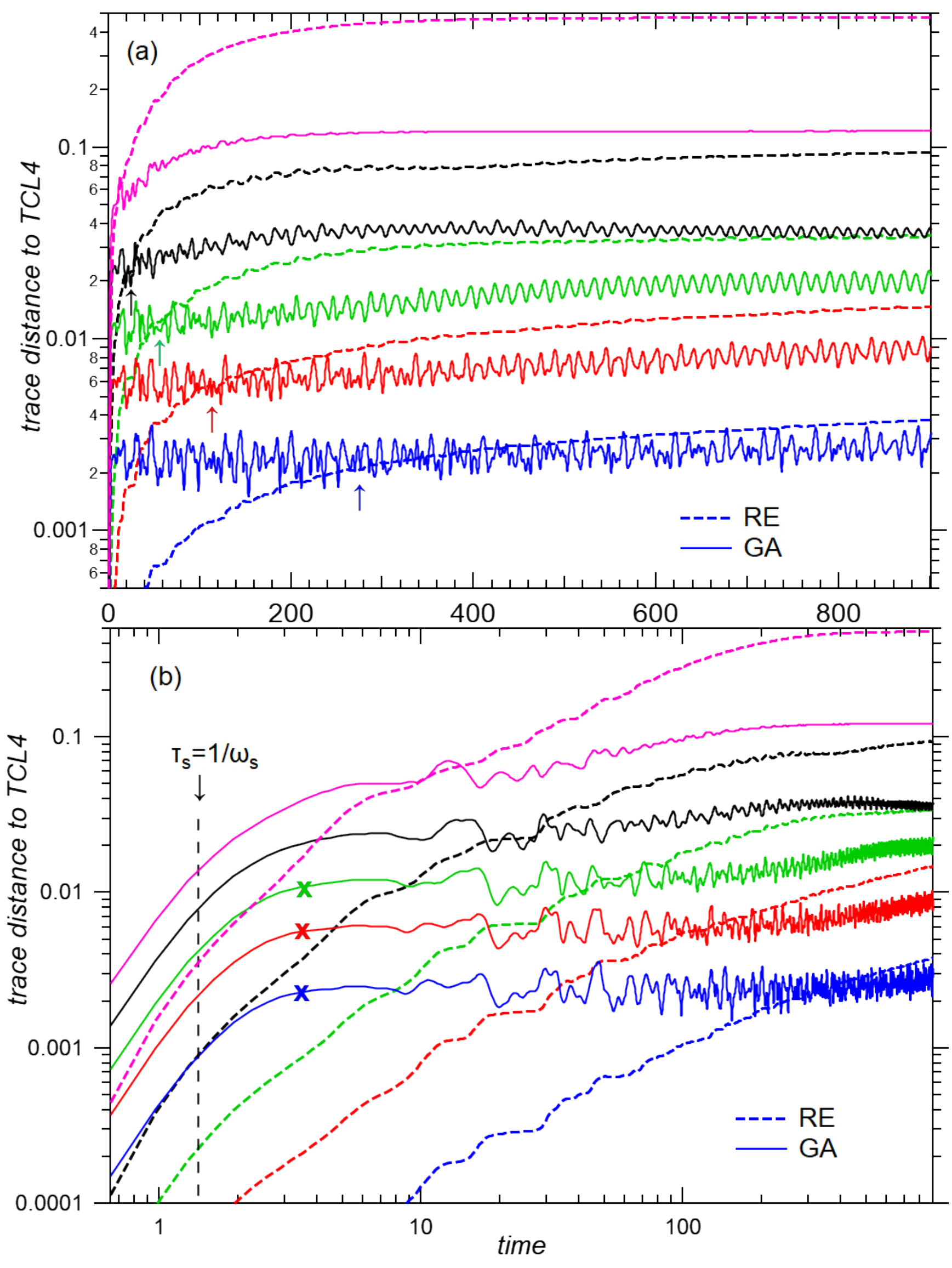}
  \caption{\label{Fig:traces} Trace distance between the states obtained by solving the Redfield (thick-dashed) and the geometric (thin-full) master equations, and that of the TCL4. $\alpha=0.0125,0.025,0.05,0.1$ and $0.2$, bottom to top.
The arrows in (a) indicate the time limit beyond which the GAME becomes more accurate than the RE.
$\omega_c=10$ and $\omega_s=0.7$.}
\end{figure}

For the initial state of the system, we assume $\vert\Psi(0)\rangle=[1 1 1 1 1 1]/\sqrt{6}$. We solve the three master equations (non-Markovian versions) using the RK4 method.
The observed energy relaxation time is $T_1=1/(0.175\alpha)$ (not shown).

Fig.~\ref{Fig:traces}(a) displays the trace distance versus time, measured between the two second order (in $\alpha$) master equation solutions and
the TCL4 solution. At small coupling to the bath, (the lowest three pairs of curves), the solution of the geometric equation increases in trace distance
very fast, but also rapidly stabilizes the distance. On the other hand, over a long time scale, the Redfield equation solution drifts, becoming factor of two to three less accurate than the GAME.

The arrows show the time above which the GAME becomes more accurate than the RE. This time is approximately $0.56T_1$, where $T_1$ is the energy relaxation time.
Thus, it is fair to
conclude that the stabilization imposed by enforcing the CP constraint, insulates the GAME from the problems of the RE at long time scale, thereby making it a better approximation
at long time scale.

Figs.~\ref{Fig:traces}(b) shows the same results on logarithmic time scale.
The system spread time is indicated by the vertical dashed line. We note that the Redfield solution is much more accurate at small time.
Thus, the trace-distance between the GAME and the RE solutions, slightly above $1/\omega_s$, (at the bold crosses),
 well approximates the trace-distance between the GAME and the exact quantum state,
especially so at weaker coupling to the bath. This can be verified for example, by using the triangle inequality for trace distance between the GAME and the exact states,
with the Redfield state as
intermediary, followed by neglecting the trace distance between the Redfield and the exact state, at those crosses. Thus, by solving the RE and the GAME, we
can estimate the error of the GAME relative to the exact quantum state.

\begin{figure}
\centering
\includegraphics[width=0.49\textwidth]{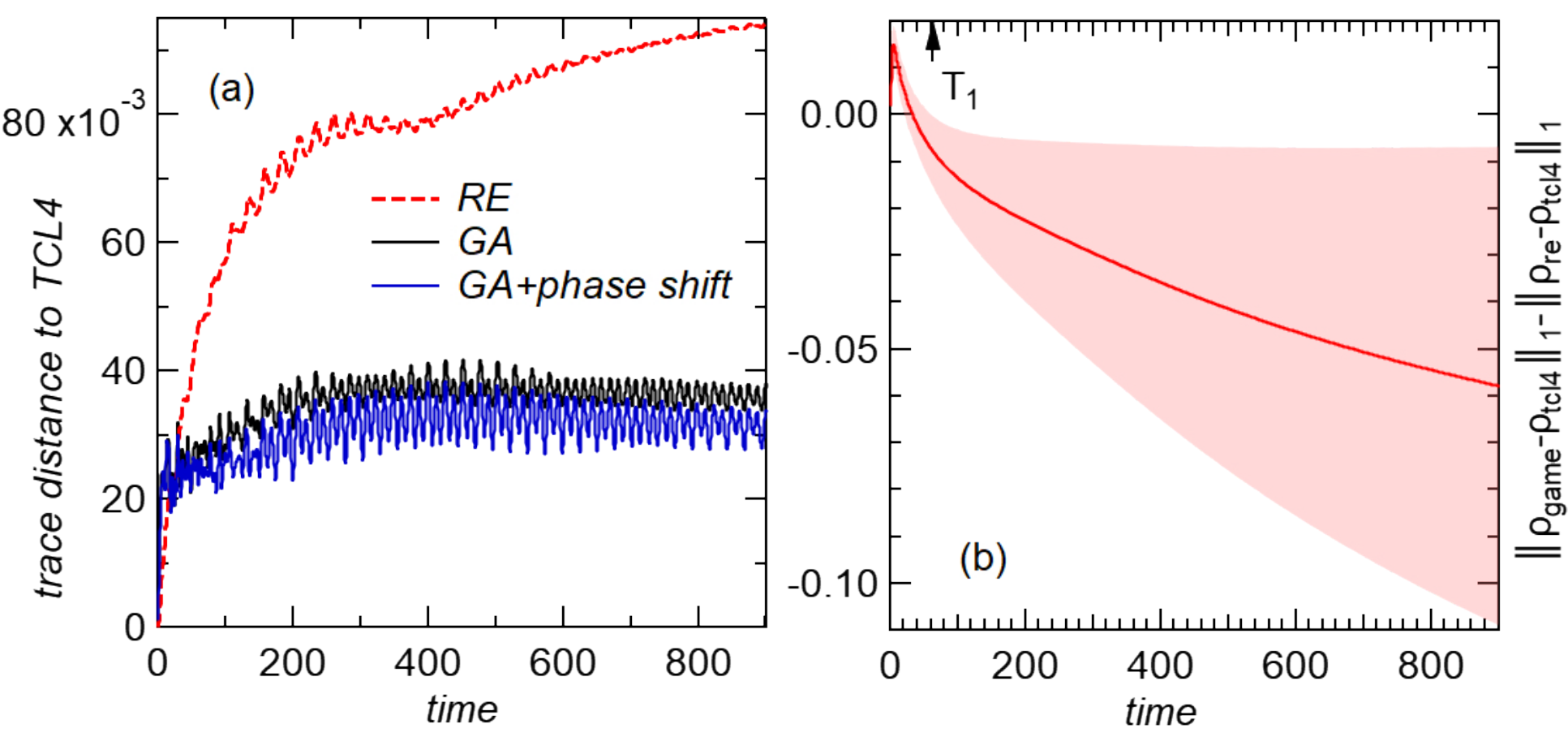}
  \caption{\label{Fig:gauge} Trace distance between the solutions of the Redfield, GAME, and phase-shifted GAME versus time, to the TCL4 solution, for the 6 by 6 random Hamiltonian.
$\omega_c=10$, $\omega_s=0.7$ and $\alpha=0.1$.
(b) Red line: The average difference between the trace distances, between the Redfield and TCL, and the GAME and the TCL, averaged over the ensemble of pure initial states and the ensemble of the coupling operators. Shaded area: $\pm$ the standard deviation.}
\end{figure}

While the logarithmic vertical scales in Fig.~\ref{Fig:traces} are encompassing, they may evade good first impressions.
Thus, in Fig.~\ref{Fig:gauge} we show the trace distances at $\alpha=0.1$, which is near the validity threshold.
We see that the trace distance of the Redfield trajectory goes through an inflection near $t=360$,
and starts to accelerate before undergoing yet another inflection around $t=480$. The solution of the GAME has no such latter inflections.

We have studied the trace distance versus time on a uniformly distribution of pure initial states, by randomly generating $250$ initial states
using the ground state of $250$ hermitian matrices in the GUE.
In addition, we studied the same over the GUE of coupling operators with unit Frobenius norm. Then we determined
the average difference between the trace distance to the TCL4 state and the standard deviation, and show the results in Fig.~\ref{Fig:gauge}(b).
The enhanced accuracy of the GAME, above the dissipation time scale, is robust, but there are significant fluctuations in the
accuracy. Similar results are obtained on the ensemble of classical random initial states (e.g., diagonal in the eigenbasis).
At small $t$, Fig.~\ref{Fig:gauge}(b)
shows that the RE is more accurate than the GAME. The arrow indicates that the accuracy crossover time is at about the energy relaxation time.

We have not addressed yet any role of the complex impedance $Z(\omega)$, which introduces a frequency dependent phase shift into the GAME, as a degree of freedom.
After realizing that
\begin{equation}
\Gamma(\omega)+\Gamma^\dagger(\omega')-2\sqrt{J(\omega')}\sqrt{J(\omega)}=i\frac{\partial S}{\partial\omega}(\omega-\omega')+O(\omega-\omega')^2,
\end{equation}
we choose a gauge $Z(\omega)=J(\omega)e^{i\phi(\omega)}$, where
\begin{equation}
\phi(\omega)=\int_{\omega_0}^{\omega} d\omega' \frac{1}{J(\omega')}\frac{\partial S}{\partial \omega'}.
\label{Eq:gauge}
\end{equation}
The gauge reduces the magnitude of the $X$-kernel near the frequency diagonal, from linear to nearly quadratic in frequency difference.
But the integral~\ref{Eq:gauge} diverges when $J(\omega)$ is zero, so there are some tradeoffs. In those and the very near frequency regions, we keep the phase constant.
After this transformation, we solve the geometric master equation and compare the distance to the TCL4. The result in Fig.~\ref{Fig:gauge} shows a modest reduction in the trace distance.
It will be interesting to study how various gauges displace the solutions around the TCL4 solution, to learn if this could possibly be a technique of triangulation of the exact state.

\section{Example 3. Dynamical Decoupling\label{Sec:DD}}

Here we show that enforcing complete positivity on the Redfield equation by the geometric-arithmetic approximation, to treat
the problem of dynamical decoupling (DD) directly in the interaction picture, eliminates most of the issues that render the Redfield equation very difficult to solve numerically.
The DD protocol involves applying a time-dependent system Hamiltonian on a quantum system, in the form of
a sequence of pulses~\cite{Viola,ZANARDI199977}. In Markovian dynamics, the DD has been studied in Refs.~\cite{Szczygielski,Gough,mozgunov}. If the time
between the pulses is shorter than the environmental correlation time, it can cause dephasing
suppression which may be important in quantum error correction~\cite{Suter,LidarCUP}.

Here we study an example of a
spin-1/2 boson problem in the regime of pure dephasing. For simplicity we assume that the free Hamiltonian is zero, so that
the system Hamiltonian is solely due to the external field,
\begin{equation}
H_S(t)=\frac{\pi}{2}\sum_{n=-\infty}^\infty\delta[t-(n+\nu)\tau]\sigma_y.
\end{equation}

The spin-1/2 is flipped by a $\pi$ pulse periodically with a period $\tau$, and
is coupled to Ohmic bath at zero temperature with spectral density given by Eq.~\ref{eq:spdensity}, with traceless operator $A=\sigma_x/2$.
Here, $\nu$ varies in the interval $(0,1)$, so the first pulse at $t>0$ occurs at $t=\nu\tau$.
In the interaction picture, $A(t)$ undergoes a series of discrete sign changes at times $t=(n+\nu)\tau$.
Namely, the unitary propagator at time $t$ is $U=\exp (-\int_0^t H_S(\tau)d\tau)=\exp (-ik\pi\sigma_y/2)=(-i\sigma_y)^k$, where $k=\text{floor}(t/\tau+\nu)$.
After some algebra, it follows that $A(t)=U(t)^\dagger A U(t)=(-1)^kA$.

In the following, we will let $\nu=1-0^+$. That is, we assume that the initial, factorized  system-environment state, is set at a time just after a pulse.
As a function of $\nu$, the following results do not change qualitatively.
However, we note that some memory of the initial time is preserved over the entire course of the dynamics,
which means that the dynamics is weakly non-Markovian, even in the asymptotic limit.
Since we find that these non-Markovian effects do not change the following graphs significantly,
nor they affect the dephasing time, we will not discuss them further.

\begin{figure}
\centering
\includegraphics[width=0.49\textwidth]{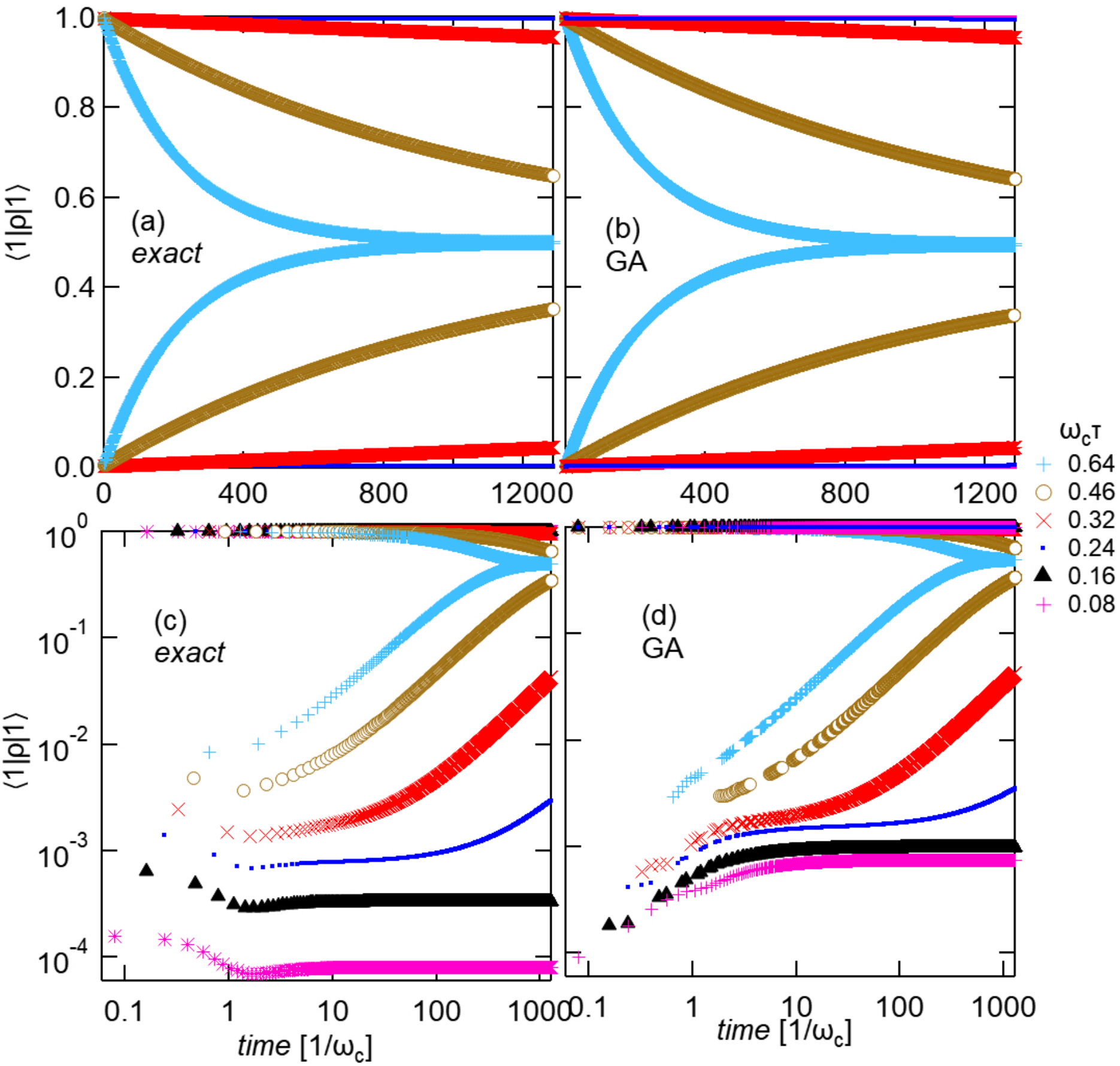}
  \caption{\label{Fig:DD1} Population versus time for pure dephasing of spin-1/2 in the presence of DD. The upper branches display dephasing of the initial polarization,
while the lower branches correspond to the flipped population due to the pulses. (a) and (b) correspond to the solutions of the exact TEMPO
algorithm and the GAME$_\text{t}$. The pulse length $\tau$ is shown in the legend.
$\alpha=0.2$.}
\end{figure}

\subsection{Path-Integral Solution}
First we solve the DD problem by the path integral technique known as Time Evolving Matrix Product Operator
(TEMPO). The introduction to the method and the python code we use
are available in Ref.~\cite{Strathearn2018}.
The TEMPO algorithm can also be modified to calculate the process tensor~\cite{Jorgensen2019}. This alternative formulation has been used in both optimizing quantum control procedures~\cite{Fux2021} and in calculating exact bath dynamics~\cite{Gribben2021}.
An open source python package~\cite{TimeEvolvingMPO} improves the performance and includes new approaches.

Here we work in the Schr\"odinger picture. In the DD, the unitary propagator imposed by a single pulse will be $U_0\otimes\mathbb{1}$, where $U_0=-i\sigma_y$. Let the unitary propagator
for the complete system, in time interval $\tau$
between the pulses, be $W_\tau$. Then, since the unitary dynamics is divisible, the unitary propagator
 at time $t=N\tau$ will be $U_0W_\tau U_0...U_0W_\tau=(U_0 W_\tau)^N.$ The corresponding Liouvillian is
$\mathcal{L}_0\mathcal{L}_\tau \mathcal{L}_0...\mathcal{L}_0\mathcal{L}_\tau=(\mathcal{L}_0\mathcal{L}_\tau)^N$. Here $\mathcal{L}_0=e^{-i(\pi/2)[\sigma_y,\bullet]}$
is the Liouvillian for a single pulse, while $\mathcal{L}_\tau=e^{-i[H_B+H_I,\bullet]\tau}$ is the Liouvillian for the continuous evolution of the system and bath
between consecutive pulses.

Then, the reduced state of the system after $N$ pulses in the Schr\"odinger picture
is
\begin{equation}
\rho(N\tau)=\text{tr}_B \mathcal{L}_0\mathcal{L}_\tau \mathcal{L}_0...\mathcal{L}_0\mathcal{L}_\tau\rho_S(0)\otimes\rho_B.
\end{equation}
Next, the state can be written as a path
sum over system states, by inserting identity resolutions between each
$\mathcal{L}_0$ and $\mathcal{L}_\tau$. Due to the form of the Liouvillian $\mathcal{L}_\tau$, tracing over the bath can be
accounted for by a discretized Feynman-Vernon functional with $N$ time steps~\cite{FEYNMAN1963118}.
An important simplification here is that there is no need to subdivide
the time step of length $\tau$ and use a Trotter splitting, because the system Hamiltonian (equal to zero) commutes with $H_B+H_I$ between the pulses.

TEMPO utilizes matrix product states to store the N-index tensor of the system, known as the augmented density tensor, which is the central entity in the exact Quasi Adiabatic Path Integral method (QUAPI)~\cite{Makri1,Makri2,THORWART2004333,Nalbach}. In TEMPO, the
tensor is efficiently compressed by singular value decomposition (SVD), thereby making the calculation of the path sum numerically tractable.
Out of the three errors in TEMPO, namely, the Trotter splitting error, the finite memory error,
and the SVD truncation error, our application only exhibits the last one. We calculate using the open source code available in Ref.~\cite{Strathearn2018}, except that we made a minor modification to take advantage of the absence of the Trotter splitting.

The degenerate spin-1/2 is initialized into $\vert 1\rangle$ with spin along $+z$ axis, and interacts with the bath
and the external pulses. In the asymptotic Markovian limit, the precise position of the pulse relative to
the initial time has a weak effect and we shall not discuss it here. The result for the population as a function of time is shown in Fig.~\ref{Fig:DD1}(a) and~(c),
for different pulse lengths. Each curve has two branches reflecting the fact that $s_z$ is flipped half of the time. Decreasing the pulse length
below the correlation time $1/\omega_c$ causes
a rapid increase in the dephasing time $T_2$.

\subsection{Master Equation Solutions}

\begin{figure}
\centering
\includegraphics[width=0.49\textwidth]{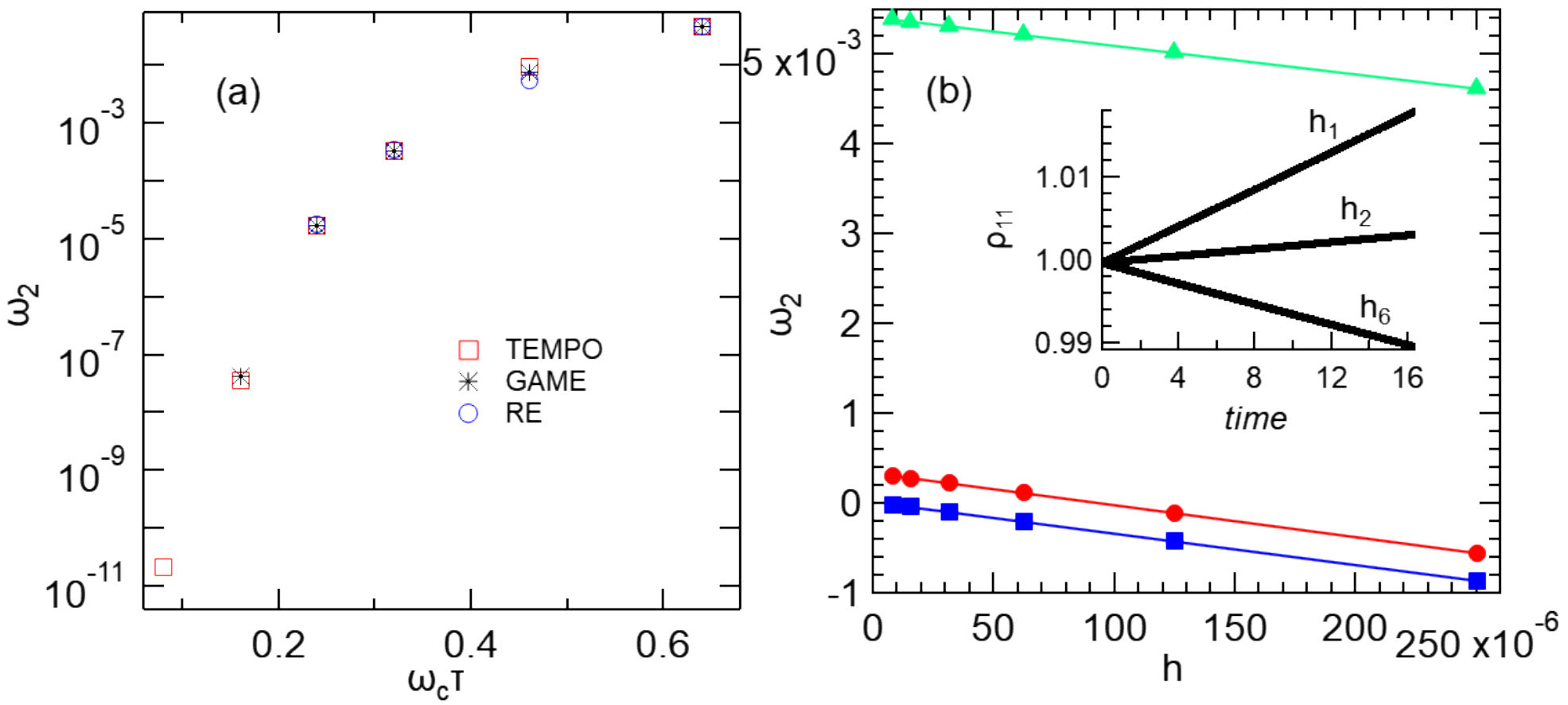}
  \caption{\label{Fig:DD2} (a) Dephasing rate of spin-1/2 versus time between $\pi$-pulses, obtained from the TEMPO, GAME, and the RE.
(b) Dephasing rate versus time step, in the Redfield equation. Triangles: $\tau=0.48/\omega_c$, circles: $\tau=0.32/\omega_c$, and squares: $\tau=0.24/\omega_c$.
The lines show best linear fits.
Inset: Population versus time in the Redfield equation, at pulsing period $\tau=0.32/\omega_c$. $h_1=2.5\times 10^{-4}$, $h_2=1.25\times 10^{-4}$, and
$h_6=7.81\times 10^{-6}$.
$\alpha=0.2$ and $\omega_c=10$.}
\end{figure}

Next we solve the dynamical decoupling using the Redfield and the GAME.
In the first case, we obtain $\Lambda(t)=\int_0^t d\tau C(\tau)A(t-\tau)$,
per Eq.~\ref{eq:lambdat0int}, by numerical integration using the Simpson's rule,
e.g., $\int_{0}^{t}f(\tau)d\tau=(h/3)\sum_{j=1}^{n/2}\{f[(2j-2)h]+4f[(2j-1)h]+f(2jh)\}$, where $h$ is the time step.
Similarly, we integrate over the non-markovian jump-correlator, according to Eq.~\ref{Eq:jumpBCFnonMarkov}, and get the time dependent generator of the
GAME.
The rotation to the
Schr\"odinger picture involves the free system propagator driven by the pulses.
Subsequently, the MEs are solved using the RK4 method with the step $2h$.

The populations versus time obtained by the GAME are shown in Fig.~\ref{Fig:DD1}(b) and~(d),
indicating very good agreement with TEMPO. By fitting a branch to an exponential, we obtain the dephasing rate and plot it versus $\tau$, showing good agreement
in Fig.~\ref{Fig:DD2}(a). The dephasing rate decays superexponentially with the pulsing frequency.

We also solve the dynamical decoupling using the RE, and find that the population versus time can be in good agreement with TEMPO, but
it requires significantly smaller time step, to reach the accuracy of the GAME.
Fig.~\ref{Fig:DD2}(b), inset, shows the population versus time, (upper branch), at pulsing period $\tau=0.32/\omega_c$, at three different time steps.
If the step is too large, the dephasing rate will be
negative, resulting in numerical instability.

The main part of Fig.~\ref{Fig:DD2}(b) shows the dephasing rate versus time step, obtained from the best linear fit of
$\rho_{11}$ versus time, at each $h$. The extrapolation to $h=0$ gives a positive dephasing rate, showing very
good agrement with the other two methods, as shown in Fig.~\ref{Fig:DD2}(a).
But it also shows how much more
difficult it is to calculate the dephasing rate, when the driving frequency is high.

The RE is sensitive to phase errors imposed by discretization, required to calculate the time integrals.
In the interaction picture, the RE has terms $\Lambda(t)\rho A(t)-A(t)\Lambda(t)\rho$ (Eq.~\ref{eq:useful}).
The integral of this expression over time is analogous to lock-in detection, with $A(t)$ as the reference signal and $\Lambda(t)$ as the measured
signal. Depending on $t$, $A(t)\Lambda(t)$ can be $<0$ (e.g., not PSD).
Then, if the time step is too large and
the frequency of the pulsing is high compared to the bath relaxation rate,
there will be a dominance of the terms with $A\Lambda(t)<0$ in this phase sensitive situation. This causes the instability of the RE, which can be
cured by reducing the time step.

The generator of the GAME, $M(t)$, also
exhibits similar phase errors, but the terms are now $M(t)\rho M(t)^\dagger-\{M(t)^\dagger M(t),\rho\}/2$, which are phase insensitive.
The result is that there is no instability and $h$ can be much larger. we could say in engineering language, if the RE is a lock-in amplifier, then the GAME will be
a linear amplifier~\cite{girvin}.

Fig.~\ref{Fig:DD2}(b) shows the best linear fits of the dephasing rate in the RE versus $h$, which is commensurate with the driving. We get
\begin{equation}
\omega_2(h)=\omega_2(0)-0.356h,
\label{Eq:DDaccuracy}
\end{equation}
with the slope approximately independent of the driving frequency.  Since the dephasing rate $\omega_2(0)$ is superexponentially reduced with frequency of the driving,
the required step $h$ also has to be superexponentially suppressed to yield the correct sign. All these issues go away
in the GAME, because of its insensitivity to the phase of the generator. This greatly simplifies the calculation of the dephasing rates,
and opens possibilities to study more complicated time dynamics based on direct calculation of the time-ordered exponential in the unitary propagator. This could allow one
to bypass additional assumptions that enhance the accuracy of the model, such as the Floquet formalism~\cite{AlickiPer,Szczygielski,szczygielski2020howland} or adiabatic
driving~\cite{Davies1978OpenQS,Albash_2012}.

\section{Discussion and Conclusion}

This paper reformulates the geometric-arithmetic master equation introduced in Ref.~\cite{Davidovic2020}, with emphasis on open quantum systems with
large numbers of Bohr frequencies, at or near zero temperature, and fast driving. The equation's accuracy comparable to the weak-coupling constant
is interpreted
in two distinct ways. In one point of view, the displacement
of the state, between the RE and the GAME,  is characterized by the coarse-grained derivative of the position. This derivative
drops as $1/t$ versus time $t$, above a corner time corresponding to the inverse frequency spread of the system and the environment. The latter is similar to the minimax frequency,
$\min[\max(\omega),\omega_c]$, where $\max(\omega)$ is the maximum system oscillation frequency and $\omega_c$ is the bath relaxation frequency. The displacement of the state
caused by the geometric-arithmetic approximation will be less than one, if the frequency spread is larger than the dissipation rate.

An alternative view point is based on limited ranks and the decay rates of the singular values, of various superoperator kernels
describing the system's displacement dynamics. We find that the high accuracy of the GAME requires that the s-values decay very rapidly.
Since the spectral density, which is not analytic at $\omega=0$ and $T=0$, is
contained within the rank-3 velocity kernel, it does not affect the decay-rates of the singular values.

It is not the pathology in the spectral density,
but the band limiting kernel, $\sin[(\omega-\omega')t]/[\pi(\omega-\omega')]$, which governs the decay rates.
The relatively large frequency spread of the GAME, compared to that in the secular approximation, opens a possibility to study rapid non-secular dynamics in
complex weakly coupled systems, such as light harvesting systems, quantum error correcting codes,
and many-body physics, consistent with the complete positivity requirement.

The applications discussed in this paper include, the Heisenberg spin-1/2 chain, which clearly demonstrates the
underpinnings of the geometric-arithmetic approximation in terms of the rapid decay of the singular-values of the position-displacement kernel matrices.
In a random system Hamiltonian, drawn from the Gaussian unitary ensemble,
we show that the GAME can be more accurate than the less approximated Redfield equation, at long time scales.
Thus, complete positivity is a practical constraint, by giving new insights into open quantum dynamics.

By introducing a gauge to the geometric mean of the decay rates, we have shown that the accuracy of the geometric master equation can be further improved.
A best case scenario will utilize multiple gauges as quantum state triangulation points, to examine if the exact quantum dynamics will be within reach.

In dynamical decoupling of a qubit, a similar scenario plays out. We find that the GAME becomes very simple to use, and it can be superexponentially accelerated,
relative to the perturbative Redfield equation,  particularly if the frequency of the driving is high compared to the bath relaxation rate.
This enhancement in efficiency of the GAME is due to the lack of phase sensitivity in the generator.

The author would like to thank Donghyun Jin, Grihith Manchanda, Huo Chen, Frederik Nathan, Mark Rudner, Nicholas Ezzell, Evgeny Mozgunov, Daniel Lidar, Archak Purkayastha, Gerald Fux and Brendon Lovett
for discussion; and Gerald Fux for adding a feature to the open source package on our request. This research was supported
by DOE contract DE-FG02-06ER46281. Additional support from the Georgia Tech Quantum Alliance (GTQA), a center
funded by the Georgia Tech Institute of Electronics and Nanotechnology was used to develop exact numerical
methods on entangled low-dimensional magnetic system.

\appendix

\section{Column Major Order\label{appendix:column}}
Some expressions in this paper depend on how double indices are translated into single indices. In the column major ordering that we utilize, if $A$ is a matrix with elements $A_{ij}$, it will be mapped to the column array $[A_{11} A_{21} ... A_{N1} A_{12} A_{22} ... A_{NN}]^T$. By inspection, one can show $(A\otimes B)_{(ij),(kl)}=A_{jl}B_{ik}$, where $(ij)$ and $(kl)$ on the LHS
are single indices in the column major order, e.g., $(ij)=i+(j-1)N$ and $(kl)=k+(l-1)N$. (In the row-major ordering, $(ij)=j+(i-1)N$ and $(kl)=l+(k-1)N$, so
the RHS would be $A_{ik}B_{jl}$). In the column-major order, $(A \otimes B)\rho$ translates, into the operator notation, as $B\rho A^T$, since $[(A \otimes B)\rho]_{nm}=\sum_{ij} [A \otimes B]_{nm,ij}\rho_{ij}=
\sum_{ij}A_{mj}B_{ni}\rho_{ij}=\sum_{ij}B_{ni}\rho_{ij}A_{jm}^T$.

As an example, we derive the dissipative part of the Redfield equation, (\ref{Eq:classicRE}), in the traceless case, i.e. $\text{Tr}A_\alpha=0$. The
 dissipative dynamics in Eq.~\ref{Eq:DisTensor} is
 \begin{equation}
\mathcal{D}(t)\varrho=D(t)\varrho-\frac{1}{2}\{\text{Tr}_1[G(t)]\otimes \mathbb{1}+\mathbb{1}\otimes \text{Tr}_1[G(t)]^T\}\varrho.
\label{eq:DDyn}
\end{equation}

Consider now a simplified dynamical matrix (Eq.~\ref{Eq:Dmat}): $D(t)=A^\star\otimes\Lambda+\Lambda^\star\otimes A$. $D(t)\rho$ translates, in the operator notation,
as $\Lambda\varrho A^\dagger+h.c.$, which accounts for the first term on the RHS of Eq.~\ref{Eq:classicRE}.
The middle term, on the RHS of Eq.~\ref{eq:DDyn}, has the superoperator with the following matrix elements:
\begin{eqnarray}
&-&\big[\frac{1}{2}(\text{Tr}_1 G)\otimes \mathbb{1}\big]_{nm,ij}=-\frac{1}{2}(\text{Tr}_1 G)_{mj}\mathbb{1}_{ni}\\
&=&-\frac{1}{2}\sum_k G_{km,kj}\delta_{ni}=-\frac{1}{2}\sum_k D_{kk,mj}\delta_{ni}\\
&=&-\frac{1}{2}\sum_k \big( A^\star\otimes\Lambda+\Lambda^\star\otimes A \big)_{kk,mj}\delta_{ni}\\
&=&-\frac{1}{2}\sum_k \big(A^\star_{kj}\Lambda_{km}+\Lambda^\star_{kj}A_{km}\big)\delta_{ni}\\
&=&-\frac{1}{2} \big(A^\dagger\Lambda+\Lambda^\dagger A\big)_{jm}\delta_{ni}.
\end{eqnarray}

The effect of the superoperator on $\rho$ is then
\begin{eqnarray}
\nonumber
&-&\sum_{ij}\big[\frac{1}{2}(\text{Tr}_1 G)\otimes \mathbb{1}\big]_{nm,ij}\rho_{ij}=-\frac{1}{2}\sum_j  \big(A^\dagger\Lambda+\Lambda^\dagger A\big)_{jm}\rho_{nj}\\
&=&-\frac{1}{2}\big[\rho(A^\dagger\Lambda+\Lambda^\dagger A)\big]_{nm}.
\end{eqnarray}
Similarly, the last term on the RHS of Eq.~\ref{eq:DDyn} maps to the hermitian conjugate of the middle term,
leading to
 \begin{equation}
\mathcal{D}(t)\varrho=\Lambda\varrho A+A\varrho \Lambda^\dagger-\frac{1}{2}\{\rho,A^\dagger\Lambda+\Lambda^\dagger A\}.
\label{eq:DDynb}
\end{equation}
Adding the Lamb shift component $-i[H_{ls},\varrho]$, with $H_{ls}$ given by Eq.~\ref{eq:ULS}, two terms in the anticommutator cancel out,
and one arrives at Eq.~\ref{Eq:classicRE}.

\section{Unique Hamiltonian and Dissipator in a Trace-Hermiticity Preserving Map\label{appendix:Dissipator}}

Any master equation can be expressed in terms of the time-dependent generator of quantum dynamics~\cite{Gorini}, $\mathcal{L}_t$,
which is a superoperator in $M(N)$,
\begin{equation}
\frac{d\varrho_S}{dt} = \mathcal{L}_t\varrho_S(t)=\sum\limits_{nimj}L_{ni,mj}(t)E_{ni}\varrho_SE_{mj}^\dagger.
\label{Eq:TdepGen}
\end{equation}
$E_{ij}=\vert i\rangle\langle j\vert$ make the complete orthonormal set (COS) in $M(N)$.
$L(t)$ is hermitian, [e.g., $L_{ni,mj}(t)=L_{mj,ni}^\star (t)$] and satisfies
$\sum_n L_{ni,nj}(t)=0,\,\forall i,j$, which assures trace preservation.
As a result, the RHS of Eq.~\ref{Eq:TdepGen} can also be written down as
\begin{equation}
\frac{d\varrho_S}{dt}=\frac{1}{2}\sum\limits_{nimj}L_{ni,mj}(t)\{[E_{ni}\varrho_S,E_{mj}^\dagger]+[E_{ni},\varrho_SE_{mj}^\dagger]\}.
\label{Eq:TdepGenLL}
\end{equation}

For a given hermiticity and trace-preserving map, the question arises
if the decomposition of the generator into the sum of the dissipative-like and the unitary-like terms, e.g., $[E_{ni}\varrho_S,E_{mj}^\dagger]+
[E_{ni},\varrho_S E_{mj}^\dagger]$ and $-i[H,\varrho_S]$, respectively, is unique. The answer is no, which we show by the next example.

{\bf Example}: In the Redfield master equation, expressing Eq.~\ref{eq:useful} in the basis directly leads to Eq.~\ref{Eq:TdepGen}, with no Hamiltonian term and
\begin{eqnarray}
\nonumber
&&L_{ni,mj}(t)=\sum_{\alpha\beta}\big\{\Lambda_{\alpha\beta}(t)_{ni}A_{\alpha}(t)_{mj}^\star+A_{\alpha}(t)_{ni}\Lambda_{\alpha\beta}(t)_{mj}^\star\\
&-&\sum_{k}[A_{\alpha}(t)_{nk}\Lambda_{\alpha\beta}(t)_{ki}\delta_{jm}
+\delta_{in}\Lambda_{\alpha\beta}(t)_{kj}^\star A_{\alpha}(t)_{km}]\big\}.
\label{Eq:TdepCoeffInt}
\end{eqnarray}

In contrast to the main text, here we do not assume that the coupling operators $A_\alpha$ are traceless. Nevertheless, Eq.~\ref{Eq:TdepCoeffInt} is
 the same as
\begin{eqnarray}
\nonumber
\frac{d\varrho_S}{dt}&=&\frac{1}{2}\sum\limits_{nimj}N_{ni,mj}(t)\{
[E_{ni}\varrho_S, E_{mj}^\dagger] + [E_{ni},\varrho_S E_{mj}^\dagger]\}\\
&-&i[H_{ls}(1),\varrho_S],
\label{Eq:Standard}
\end{eqnarray}
where
\begin{equation}
N_{ni,mj}(t)=\sum_{\alpha\beta}\big[\Lambda_{\alpha\beta}(t)_{ni}A_{\alpha}(t)_{mj}^\star+A_{\alpha}(t)_{ni}\Lambda_{\alpha\beta}(t)_{mj}^\star\big],
\end{equation}
and
\begin{equation}
H_{ls}(1)=\frac{1}{2i}\sum\limits_{\alpha\beta}\big[A_\alpha(t)\Lambda_{\alpha\beta}(t)-\Lambda_{\alpha\beta}^\dagger(t)A_\alpha(t)\big].
\end{equation}
The last two expressions are formally the same as the relaxation tensor and the Lamb-shift of the traceless case, e.g., $N_{ni,mj}(t)=D_{nm,ij}(t)$ where $D$ is given by Eq.~\ref{Eq:Dmat}, while $H_{ls}(1)$ is given by Eq.~\ref{eq:ULS}.

As we shall find below, the master equation~\ref{Eq:TdepCoeffInt} is also the same as
\begin{eqnarray}\label{Eq:TdepGenLLL}
\frac{d\varrho_S}{dt}&=&-i[H_{ls},\varrho_S]\\
\nonumber
&+&\frac{1}{2}\sum\limits_{nimj}G_{ni,mj}(t)\{[E_{ni}\varrho_S,E_{mj}^\dagger]+[E_{ni},\varrho_SE_{mj}^\dagger]\},
\end{eqnarray}
where, now, $G_{ni,mj}(t)=D_{nm,ij}(t)$, with $D$ given by Eq.~\ref{Eq:Dmatappendix}, and $H_{ls}$ is given by Eq.~\ref{eq:ULSappendix}.

Thus, in the same basis $\{E_{ij}\}$, there are at least three different ways, e.g., Eqs.~\ref{Eq:TdepGenLL},~\ref{Eq:Standard}, and~\ref{Eq:TdepGenLLL}, to split the RE between the dissipative-like  and the unitary-like terms.
These forms are all GKSL-like, which demonstrates non-uniqueness of the dissipative-like and the unitary-like components of the RE. This implies that the dissipative-like terms generally carry information about the unitary dynamics, and vice versa.

For the Redfield equation in particular, if the system Hamiltonian is time-independent, applying Eqs.~\ref{eq:REbreuer} and~\ref{eq:lambdat0sum} to Eq.~\ref{Eq:TdepCoeffInt},
in the eigenbasis of $H_S$, leads to
\begin{eqnarray}
\nonumber
L_{ni,mj}(t)&=&\sum_{\alpha\beta}\big\{A_{\beta}(t)_{ni}A_{\alpha}(t)_{mj}^\star\\
\nonumber
&\times&\big[\Gamma_{\alpha\beta}(\omega_{in},t)+\Gamma_{\beta\alpha}^\star(\omega_{jm},t)\big]\\
\nonumber
&-&\sum _{k}A_{\alpha}(t)_{nk}A_{\beta}(t)_{ik}^\star\Gamma_{\alpha\beta}(\omega_{ik},t)\delta_{jm}\\
&-&\sum _{k}\delta_{in}A_{\beta}(t)_{km}A_{\alpha}(t)_{kj}^\star\Gamma_{\beta\alpha}^\star(\omega_{jk},t)\big\}.
\label{Eq:TdepCoeff}
\end{eqnarray}
Here $\omega_{ij}\equiv E_i-E_j$ is a discrete Bohr frequency of the system.

The idea of the dissipator is it has all the information
about the dephasing and other nonunitary processes in the system, but is disentangled with the unitary dynamics completely.
To find the unique Hamiltonian and the unique dissipator, we need to identify which Hamiltonian and dissipator is invariant with changing the basis in $M(N)$.

\begin{figure}
\centering
\includegraphics[width=0.45\textwidth]{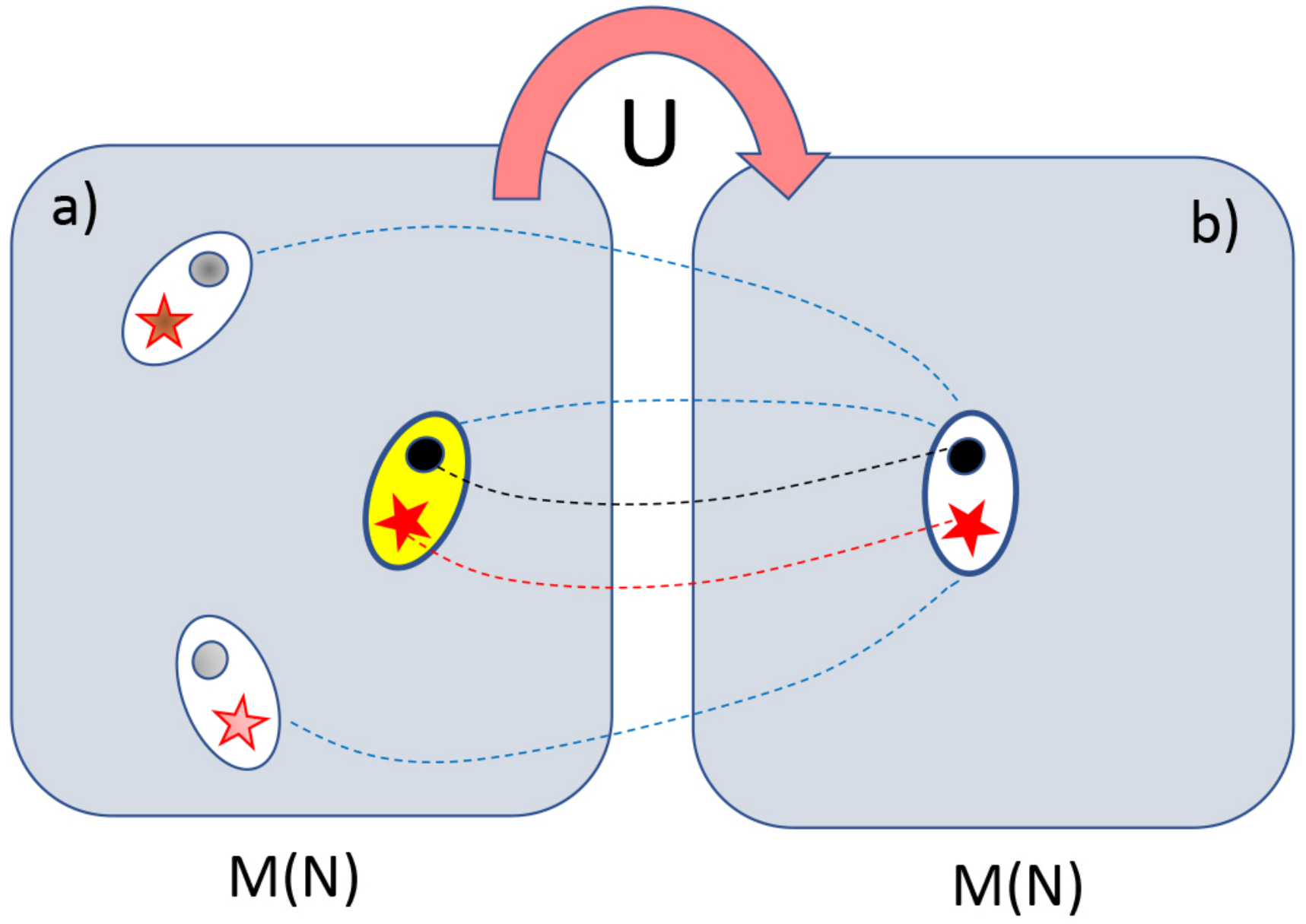}
  \caption{\label{Fig:figmap} Unitary transformation $U$ in the $C^\star$ algebra $M(N)$ of the $N\times N$ complex matrices. Each ellipse represents the same
 generator $\mathcal{L}_t$ of reduced dynamics of the open quantum system.
a) In the basis  $E_{ij}=\vert i\rangle\langle j\vert$, the splitting of $\mathcal{L}_t$ between the dissipative-like and the
unitary-like component (circle and star, respectively) is not unique.
b) The basis $F_{ij}$ includes the unity matrix, and
the splitting of $\mathcal{L}_t$ is unique. In the yellow ellipse in a), both the Hamiltonian and the dissipator are transformed by $U$
into the unique Hamiltonian and dissipator in b), respectively.}
\end{figure}

We sketch this idea in Fig.~\ref{Fig:figmap} to gain conceptual understanding.
Reference frame (a) represents $M(N)$ in the basis $\{E_{ij}\}$, which does not include the unity matrix.
Reference frame (b)
also represents $M(N)$, but the basis now includes the unity matrix. Reference frames (a) and (b) are connected by a unitary transformation $U$. The generator
of the RE can be split between the unitary-like and dissipative-like components, as we just discussed. The ellipses in (a)
represent different splits of the generator, with dissipative-like and unitary-like component implied by circles and stars, respectively.
In frame (b) the dissipative and unitary components are uniquely defined by the presence of the unity matrix in the basis~\cite[Lemma~2.3]{Gorini}.
In frame (a), on the other hand, out of the many ways to split the Redfield generator,
there can be only one with both the relaxation tensor and the Hamiltonian connected to the unique two in frame (b) via the unitary transformation.
At this splitting, the dissipative and the unitary
dynamics are disentangled since changing the basis in $M(N)$ changes neither the relaxation tensor nor the Hamiltonian.
Thus, our next goal is to transform the dissipative and the unitary generators from frame (b) to frame (a).

To this end, we apply the proof of~\cite[Lemma~2.3]{Gorini}.
Consider an operator in $M(N)$
\begin{equation}
\mathcal{L} (X)=\sum\limits_{nimj}^N L_{ni,mj}E_{ni}XE_{mj}^\dagger,
\end{equation}
with the properties tr$\mathcal{L}(X)=0$ and $[\mathcal{L} (X)]^\dagger=\mathcal{L} X^\dagger$.
As an example, $L_{ni,mj}$ could be the matrix $L(t)$ from Eq.~\ref{Eq:TdepGen} at fixed time, but the following procedure can be applied to any trace-preserving and
hermiticity-preserving linear map.

Now, switch the basis in $M(N)$ as follows:
\begin{equation}
F_{ij}=
\begin{cases}
E_{ij},\,\,i\neq j,\\
\frac{1}{\sqrt{N}}\sum\limits_{p=1}^N e^{2\pi i\frac{jp}{N}}E_{pp},\,\, \text{otherwise.}\\
\end{cases}
\label{eq:unitary}
\end{equation}
Importantly, $F_{NN}=(1/N)^{1/2}\mathbb{1}$, while the rest of the basis operators are traceless.
Substituting, we find
\begin{eqnarray}
\nonumber
\mathcal{L} (X)&=&\sideset{}{'}\sum\limits_{n,i=1}^N\sideset{}{'}\sum\limits_{m,j=1}^N  L_{ni,mj}F_{ni}X F_{mj}^\dagger\\
\nonumber
 &+&\sideset{}{'}\sum\limits_{n,i=1}^N\sum\limits_{p=1}^{N}c_{ni,p}
F_{ni}XF_{pp}^\dagger\\
\nonumber
 &+&\sum\limits_{p=1}^{N}\sideset{}{'}\sum\limits_{n,i=1}^N c_{ni,p}^\star
F_{pp}XF_{ni}^\dagger\\
 &+&\sum\limits_{p,q=1}^{N}c_{pq}
F_{pp}X F_{qq}^\dagger.
\label{eq:normalForm}
\end{eqnarray}
Here
\begin{equation}
c_{ni,p}=\frac{1}{\sqrt{N}}\sum\limits_{m=1}^NL_{ni,mm}e^{2\pi i\frac{mp}{N}},
\end{equation}
\begin{equation}
c_{pq}=\frac{1}{N}\sum\limits_{n,m=1}^N L_{nn,mm}e^{2\pi i\frac{mq-pn}{N}},
\end{equation}
with primes indicating unequal indices. Now,~\cite[Lemma~2.3]{Gorini} applies to Eq.~\ref{eq:normalForm} because the basis includes the unity matrix, leading to the unique
separation of the unitary and dissipative dynamics:
\begin{eqnarray}
\nonumber
&&\mathcal{L} (X)=-i[H,X]\\
\nonumber
 &+&\frac{1}{2}\sideset{}{'}\sum\limits_{n,i=1}^N\sideset{}{'}\sum\limits_{m,j=1}^N  L_{ni,mj}\{
[F_{ni}X, F_{mj}^\dagger] + [F_{ni},X F_{mj}^\dagger]\}\\
\nonumber
 &+&\frac{1}{2}\sideset{}{'}\sum\limits_{n,i=1}^N\sum\limits_{p=1}^{N-1}c_{ni,p}\{
[F_{ni}X, F_{pp}^\dagger] + [F_{ni},X F_{pp}^\dagger]\}\\
\nonumber
 &+&\frac{1}{2}\sum\limits_{p=1}^{N-1}\sideset{}{'}\sum\limits_{n,i=1}^N c_{ni,p}^\star\{
[F_{pp}X, F_{ni}^\dagger] + [F_{pp},X F_{ni}^\dagger]\}\\
 &+&\frac{1}{2}\sum\limits_{p,q=1}^{N-1}c_{pq}\{
[F_{pp}X, F_{qq}^\dagger] + [F_{pp},X F_{qq}^\dagger]\}.
\label{Eq:lemma}
\end{eqnarray}
The unique traceless Hamiltonian $H$ is determined using $H=(1/2i)(F^\dagger-F)$ (see~\cite[Eq.~(2.8)]{Gorini}), where
\begin{equation}
F=\frac{1}{\sqrt{N}}\big(\sideset{}{'}\sum\limits_{n,i=1}c_{ni,N}F_{ni}+\sum\limits_{p=1}^{N-1}c_{p,N}F_{pp}\big).
\end{equation}

Transforming back to the basis $\{E_{ij}\}$, after some algebra, we obtain
\begin{equation}
F=\frac{1}{N}\sum\limits_{inm=1}^NL_{ni,mm}E_{ni}-\frac{\mathbb{1}}{N^2}\sum\limits_{nm=1}^NL_{nn,mm},
\end{equation}
which corresponds to the traceless Lamb shift
\begin{equation}
\boxed{
H=\frac{1}{2Ni}\sum\limits_{inm=1}^N(L_{mm,in}-L_{ni,mm})E_{ni}.
}
\label{Eq:fullLS}
\end{equation}

The remaining terms, e.g.,
$L_{ni,mj}(t),\,n\neq i,\, m\neq j$; $c_{ni,p}(t),\, n\neq i, p=1,...,N-1$; $c_{p,ni}(t),\, n\neq i, p=1,...,N-1$; and $c_{pq}(t),\,p,q=1,...,N-1$
make the unique $(N^2-1)\times (N^2-1)$ Kossakowski matrix $K$. The interest in this matrix is that, for Markovian dynamics, its PSD is
equivalent to the condition of CP of the map~\cite{lindblad1976,Gorini}, while for non-Markovian dynamics, it is only a sufficient condition~\cite{Benatti2019}. But here, we
are primarily interested in its uniqueness property.

To determine the relaxation tensor in the frame $\{E_{ni}\}$, we again change the basis from $F_{ij}$ to $E_{ij}$. This is equivalent to
applying the unitary transformation $U^\dagger$ onto matrix $K\oplus 0$. [That is, the Kossakowski matrix $K$ is the top left $(N^2-1)\times(N^2-1)$ block of  $K\oplus0$, while all other
elements  outside the block are zero.] Thus the relaxation tensor (viewed as an $N^2\times N^2$ matrix) and the Kossakowski
matrix, respectively $G$ and $K$, share the same eigenvalues, $G$ has an extra eigenvalue of $0$, and the PSD of one matrix is equivalent to that of the other.
Since $K$ is uniquely defined~\cite[Lemma 2.3]{Gorini}, $G$ will also be unique.
After some algebra, the relaxation tensor can be shown to be
\begin{equation}
\label{eq:RM}
\boxed{
G_{ni,mj}=L_{ni,mj}-\frac{1}{N}\sum_{k=1}^N(\delta_{ni}L_{kk,mj}+L_{ni,kk}\delta_{jm}).}
\end{equation}

We define the dynamical tensor as realignment of $G$, e.g., $D_{nm,ij}\equiv G_{ni,mj}$.
It is easy to show that the realignment is basis change invariant. %That is, a unitary transformation of $G$
%can either be performed directly, or indirectly by realignment of $G$ to $D$, the unitary transformation of $D$, and realignment %from the transformed $D$ to the transformed $G$.
%One useful property is that the Frobenius norms of the relaxation and the dynamical tensors are identical.

In terms of the unique Hamiltonian $H$ (Eq.~\ref{Eq:fullLS}), the unique relaxation tensor $G$ (Eq.~\ref{eq:RM}), and the unique dynamical tensor $D$, the dynamical map is
\begin{equation}
\mathcal{L}(X)=-i[H,X]+\mathcal{D}X,
\label{Eq:dissipatorRE1}
\end{equation}
with the dissipator
\begin{equation}
\mathcal{D}=D-\frac{1}{2}[\text{Tr}_1(G)\otimes \mathbb{1}+\mathbb{1}\otimes \text{Tr}_1(G^T)].
\label{Eq:DisTensorG1}
\end{equation}

Now, let us insert the coefficients from Eq.~\ref{Eq:TdepCoeffInt} into Eqs.~\ref{Eq:fullLS} and~\ref{eq:RM}.
After tedious algebra, we obtain the traceless Hamiltonian, or the Lamb shift,
\begin{eqnarray}
\nonumber
H_{ls}(t)&=&\frac{1}{2i}\sum\limits_{\alpha\beta}\big\{[A_\alpha(t)\Lambda_{\alpha\beta}(t)-\Lambda_{\alpha\beta}^\dagger(t)A_\alpha(t)]\\
\nonumber
&-&\text{tr}[A_\alpha(t)]\frac{\Lambda_{\alpha\beta}(t)-\Lambda_{\alpha\beta}^\dagger(t)}{2Ni}\\
\nonumber
&+&\text{tr}\big[\frac{\Lambda_{\alpha\beta}(t)-\Lambda_{\alpha\beta}^\dagger(t)}{2Ni}\big]A_\alpha(t)\\
&-&\text{tr}[\frac{\Lambda_{\alpha\beta}(t)-\Lambda_{\alpha\beta}^\dagger (t)}{2Ni}A_\alpha(t)]\mathbb{1}\big\},
\label{eq:ULSappendix}
\end{eqnarray}
where $N$ is the dimension of the system Hamiltonian Hilbert space.

The dynamical tensor $D(t)$, also in the interaction picture, is
\begin{eqnarray}
\nonumber
&&D(t)=\sum\limits_{\alpha\beta}\big(A_\alpha^\star(t)\otimes\Lambda_{\alpha\beta}(t)+\Lambda_{\alpha\beta}^\star(t)\otimes A_{\alpha}(t)\\
\nonumber
&-&\frac{1}{N}\{\text{tr}[\Lambda_{\alpha\beta}(t)]A_\alpha^\star(t)\otimes\mathbb{1}+\text{tr}[\Lambda_{\alpha\beta}^\star(t)] \mathbb{1}\otimes A_\alpha(t)\}\\
\nonumber
&-&\frac{\text{tr}[A_\alpha(t)]}{N}[\Lambda_{\alpha\beta}^\star(t)\otimes\mathbb{1}+\mathbb{1}\otimes\Lambda_{\alpha\beta}(t)]\\
\label{Eq:Dmatappendix}
&+&\frac{\text{tr}\{[\Lambda_{\alpha\beta}(t)+\Lambda_{\alpha\beta}^\dagger(t)]A_\alpha(t)\}}{N}\mathbb{1}\otimes\mathbb{1}\big).
\end{eqnarray}

As discussed in the traceless case in the main text, we can rewrite Eq.~\ref{eq:useful} in terms of matrices only. In
the general case with a nonzero trace of the coupling operators, one can show
\begin{eqnarray}
\nonumber
\frac{d\varrho_S}{dt}&=&-i[H_{ls}(t),\varrho_S]\\
\nonumber
&+&\sum_{\alpha\beta}\Big(\Lambda_{\alpha\beta}(t)\varrho_S A_\alpha (t)+A_\alpha(t)\varrho\Lambda_{\alpha\beta}^\dagger(t)\\
\nonumber
&-&\frac{1}{2}\big\{A_\alpha(t)\Lambda_{\alpha\beta}(t)+\Lambda_{\alpha\beta}^\dagger(t) A_\alpha(t),\varrho\big\}\\
\nonumber
&-&\{\frac{\text{tr}[\Lambda_{\alpha\beta}(t)]}{N}\varrho_S A_\alpha(t)+\frac{\text{tr}[\Lambda_{\alpha\beta}^\dagger(t)]}{N}\varrho_S A_\alpha(t)\}\\
\label{eq:UD}
&-&\frac{\text{tr}[A_\alpha(t)]}{N}[\Lambda_{\alpha\beta}(t)\varrho_S+\varrho_S\Lambda_{\alpha\beta}^\dagger(t)]\Big).
\end{eqnarray}

\section{Matrix Geometric-Arithmetic Mean Approximation\label{appendix:MGAM}}
\begin{figure}
\centering
\includegraphics[width=0.4\textwidth]{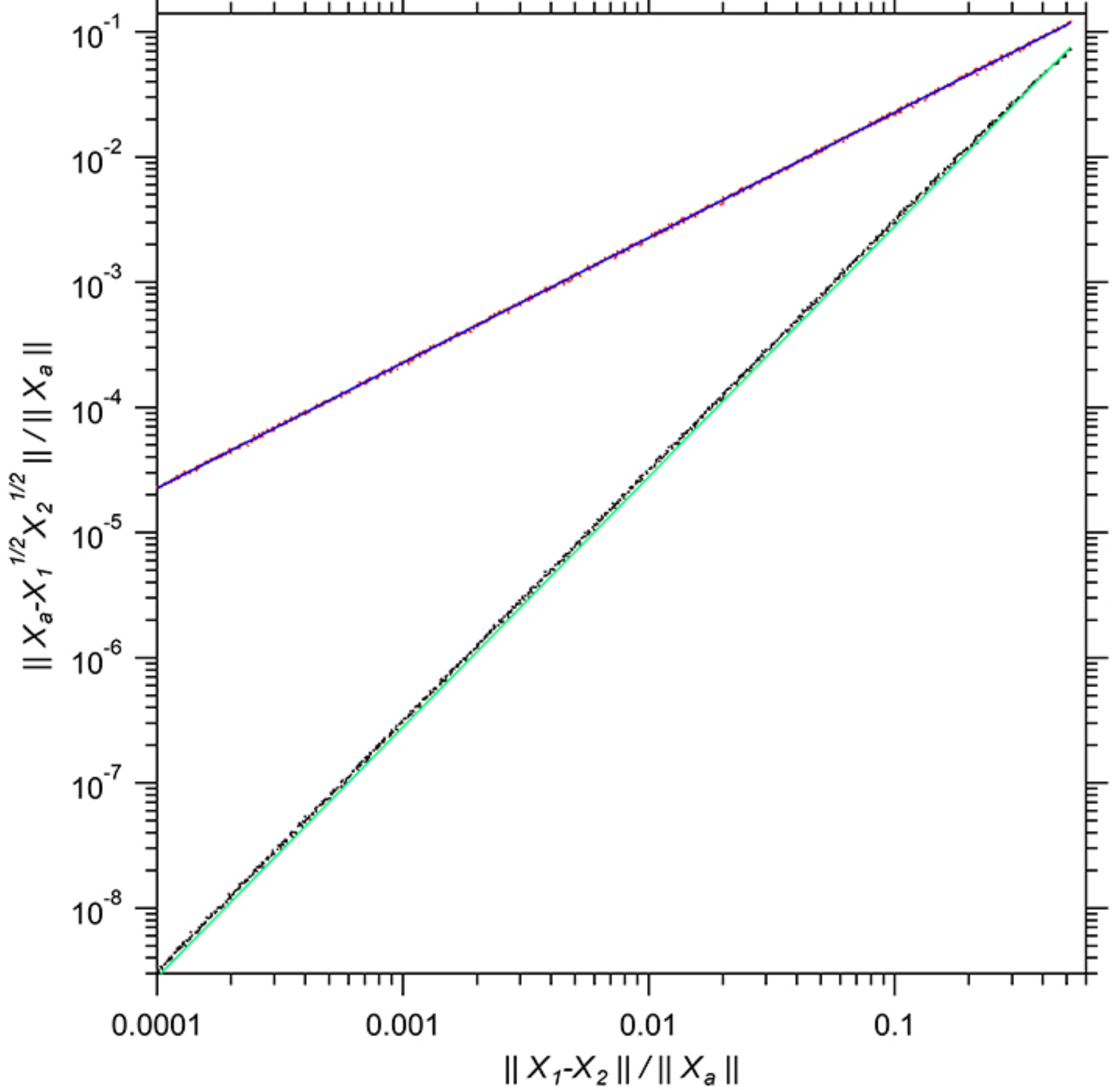}
  \caption{\label{Fig:MatrixGAME} Matrix geometric-arithmetic mean approximation: Relative Frobenius norm-distance between the arithmetic and geometric means of two matrices, as a function of the relative
  norm-distance between the matrices. $X_a=(X_1+X_2)/2$. The top and the bottom lines
  correspond the approximation used in Eq.~\ref{eq:ansatz} ($Z-J$) and the symmetrized geometric mean, respectively. Blue and green lines are best fits to power law.}
\end{figure}

Matrix geometric-arithmetic mean inequalities have been discussed in Ref.~\cite{BHATIA20082177} and references therein.
Here we examine the geometric-arithmetic approximation numerically, by creating two independent ensembles $X$ and $Y$, of $50\times 50$ matrices.
Each matrix element is drawn from a gaussian distribution of real and imaginary parts with variance $1$. Then, two ensembles of random PSD matrices are calculated
as $X_1=\sqrt{X^\dagger X}$ and $X_2=\sqrt{(X+\epsilon Y)^\dagger (X+\epsilon Y)}$, where $\epsilon$ varies in the interval $[0.0001,0.4]$ through the ensemble.
Fig.~\ref{Fig:MatrixGAME} displays $\Vert (X_1+X_2)/2-\sqrt{X_1}\sqrt{X_2}\Vert/\Vert (X_1+X_2)/2\Vert$ versus $\Vert X_1-X_2\Vert/\Vert (X_1+X_2)/2\Vert$. The best fit to power law, indicated by the blue line, leads to $\Vert (X_1+X_2)/2-\sqrt{X_1}\sqrt{X_2}\Vert=0.228\Vert X_1-X_2\Vert$. So the error is linear with the norm-distance.

An alternative, symmetrized, geometric-arithmetic mean approximation is more accurate, as shown by the plot of $\Vert (X_1+X_2)/2-(\sqrt{X_1}\sqrt{X_2}+\sqrt{X_2}\sqrt{X_1})/2\Vert/\\Vert (X_1+X_2)/2\Vert$ in Fig.~\ref{Fig:MatrixGAME}. In the latter case the error is quadratic in the  norm distance, e.g., $\Vert (X_1+X_2)/2-(\sqrt{X_1}\sqrt{X_2}+\sqrt{X_2}\sqrt{X_1})/2\Vert=0.280\Vert X_1-X_2\Vert^2/\Vert (X_1+X_2)/2\Vert$, analogous to the usual geometric-arithmetic mean approximation between positive scalars.
If the baths are independent, then the matrices $J$ will commute and the error due to the geometric-arithmetic mean approximation will be quadratic with the frequency difference, at small frequency difference.

\section{Proof of Inequality~\ref{Eq:boundExplBY}~\label{appendix:normIneq}}

Let us express the superoperator $\mathbb{Y}$ in Lindblad-like form.
In terms of the solutions of the eigenvalue equation~\ref{eq:decompY},
the spectral decomposition of the kernel is
\begin{equation}
Y_{\alpha\beta}(\omega_k,\omega_q)=\sum_n y_nU_{\alpha k,n}U_{\beta q,n}^\star.
\end{equation}
Inserting this into Eq.~\ref{Eq:Y-operator}, we  have
\begin{equation}
\mathbb{Y}=\sum_n y_n\sum\limits_{\alpha\beta kq}
U_{\alpha k,n} U_{\beta q,n}^\star\mathbb{A}_{\alpha\beta}(\omega_k,\omega_q).
\end{equation}
Furthermore, renormalizing the system operators as
\begin{equation}
a_\alpha(\omega_k)=\sum_{n}U_{\alpha k,n} d_{n},
\label{eq:reno}
\end{equation}
we arrive at the form we wanted,
\begin{equation}
\mathbb{Y}=\sum_n y_n\mathbb{Y}_n,
\label{Eq:LiuvD1}
\end{equation}
with generators
\begin{equation}
\mathbb{Y}_{n}= d_n^\star\otimes d_n-\frac{1}{2}\{[d_n^\dagger d_n]^T\otimes \mathbb{1}+\mathbb{1}\otimes d_n^\dagger d_n\}.
\label{Eq:LiuvD2}
\end{equation}

Apply the operator norm and the triangle inequality on Eq.~\ref{Eq:LiuvD1}. We have
\begin{equation}
\Vert\mathbb{Y}\Vert_\infty\leq\sum_n\vert y_n\vert \Vert\mathbb{Y}_{n}\Vert_\infty.
\end{equation}
%Under the premise that the eigenvalue magnitudes decay exponentially or super-exponentially,
%the cost of the triangle equality will not be high.
Next we apply the triangle inequality on Eq.~\ref{Eq:LiuvD2} and the norm multiplicativity under the tensor product, leading to
\begin{equation}
\Vert\mathbb{Y}\Vert_\infty\leq\sum_n\vert y_{n}\vert\left[\Vert d_n\Vert_\infty^2+\Vert d_n^\dagger d_n\Vert_\infty\right].
\end{equation}
Then we apply submultiplicativity on the operator norm on the second term on the RHS, and bound the operator norm by the
Frobenius norm. This results in
\begin{equation}
\Vert\mathbb{Y}\Vert_\infty\leq 2\sum_n\vert y_{n}\vert\Vert d_n\Vert_F^2.
\end{equation}
Inserting the inverse of the unitary transformation in Eq.~\ref{eq:reno} and applying the orthogonality relation~\ref{Eq:orthogonality},
we arrive at Eq.~\ref{Eq:boundExplBY}.

\section{TCL$4$\label{appendix:tcl}}
Terms at second order in $\alpha$ are given by~\cite[Eq.~29]{Breuer_1999}:
\begin{widetext}
\begin{eqnarray}
  \label{eq:K_4_exp}
    \left(\frac{d\varrho}{dt}\right)_2&=&\int_0^tdt_1\int_0^{t_1}dt_2\int_0^{t_2}
  dt_3\Big\{  \langle 20\rangle\langle 31\rangle\left[\hat 0,\varrho \hat 3\left[\hat 1,\hat 2\right]\right]- \langle 02\rangle\langle 31\rangle\left[\hat 0,\left[\hat 1,\hat 2\right]\varrho \hat 3\right]\nonumber\\
 &+& \langle 30\rangle\langle 21\rangle\left[\hat 0,\hat 1\varrho\left[\hat 3,\hat 2\right]+\varrho\left[\hat 2\hat 1,\hat 3\right]\right]
   - \langle 30 \rangle\langle12\rangle\left[\hat 0,\hat 2\varrho\left[\hat 1,\hat 3\right]\right]\Big\}+\text{h.c.}
   \label{eq:monstercorrelation}
\end{eqnarray}
\end{widetext}
Here the short-hand notion means
\begin{eqnarray}
\nonumber
\hat{i}&=&A(t_i),\,\,i\doteq 0,1,2,3,\,t_0=t\\
\nonumber
\langle ij\rangle&=&C(t_i-t_j),\,\,i,j\doteq 0,1,2,3.
\end{eqnarray}

The TCL4$_\text{t}$ master equation
is obtained by replacing $L_{ni,mj}(t)$ in Eqs.~\ref{Eq:TdepGen} or~\ref{Eq:TdepCoeff} with $L_{ni,mj}(t)+\delta L_{ni,mj}(t)+\delta L_{mj,ni}^\star(t)$. Evaluating the expression~\ref{eq:monstercorrelation} is very complicated.
After moving to the Schr\"odinger picture, the five commutators, on the RHS of Eq.~\ref{eq:monstercorrelation}, are expanded into pairs of terms, respectively as
\begin{widetext}
\begin{eqnarray}
\delta L_{ni,mj}(t)&=&\sum\limits_{a,b}A_{ni}A_{ja}A_{ab}A_{bm}\mathcal{H}_{jmab}(t)
-\sum\limits_{a,b,k}\delta_{ni}A_{ja}A_{ab}A_{bk}A_{km}\mathcal{H}_{jkab}(t)\\
&-&\sum\limits_{a,b}A_{na}A_{ab}A_{bi}A_{jm}\mathcal{F}_{ijmab}(t)
+\sum\limits_{a,b}A_{nb}A_{bi}A_{ja}A_{am}\mathcal{F}_{ijanb}(t)\\
\label{Eq:dI}
&+&\sum\limits_{a,b}A_{na}A_{ai}A_{jb}A_{bm}\mathcal{I}_{ijmab}(t)
-\sum\limits_{a,b}A_{ni}A_{jb}A_{ba}A_{am}\mathcal{I}_{ijanb}(t)\\
&+&\sum\limits_{a,b}A_{ni}A_{jb}A_{ba}A_{am}\mathcal{P}_{jmab}(t)
-\sum\limits_{a,b,k}\delta_{ni}A_{jk}A_{ka}A_{ab}A_{bm}\mathcal{P}_{jbak}(t)\\
\label{Eq:dY}
&-&\sum\limits_{a,b}A_{na}A_{ai}A_{jb}A_{bm}\mathcal{Y}_{ijmab}(t)
+\sum\limits_{a,b}A_{ni}A_{jb}A_{ba}A_{am}\mathcal{Y}_{ijanb}(t),
\label{eq:monsterformula}
\end{eqnarray}
where
\begin{eqnarray}
\label{corH1}
\mathcal{H}_{jmab}(t)&=&\int\limits_0^td\tau\big[e^{i\omega_{bj}(t-\tau)}H_{bm,ja}(t,\tau)-
e^{i(\omega_{ab}+\omega_{mj})(t-\tau)}H_{ab,ja}(t,\tau)\big]\\
\label{corF1}
\mathcal{F}_{ijmab}(t)&=&\int\limits_0^td\tau\big[ e^{i(\omega_{mj}+\omega_{ba})(t-\tau)}
F_{bi,jm}(t,\tau)-e^{i(\omega_{mj}+\omega_{ib})(t-\tau)}F_{ab,jm}(t,\tau)\big]\\
\label{corI1}
\mathcal{I}_{ijmab}(t)&=&\int\limits_0^td\tau\big[
e^{i(\omega_{ia}+\omega_{mb})(t-\tau)}I_{bm,jb}(t,\tau)-
e^{i(\omega_{ia}+\omega_{bj})(t-\tau)}I_{jb,bm}(t,\tau)\big]\\
\label{corP1}
\mathcal{P}_{jmab}(t)&=&\int\limits_0^td\tau
\big[e^{i\omega_{aj}(t-\tau)}I_{jb,am}(t,\tau)
  -e^{i\omega_{mb}(t-\tau)}I_{ba,jb}(t,\tau)\big]\\
\label{corY1}
\mathcal{Y}_{ijmab}(t)&=&\int\limits_0^td\tau
\big [e^{i(\omega_{bj}+\omega_{ia})(t-\tau)}Y_{ai,bm}(t,\tau)-
e^{i(\omega_{mb}+\omega_{ia})(t-\tau)}Y_{ai,jb}(t,\tau)\big].
\end{eqnarray}

Here we define four bath correlation functions

\begin{eqnarray}
\label{corF}
F_{ab,jm}(t,\tau)&=&\int\limits_0^{\tau}du C(t-u)e^{i\omega_{ab}(u-t)}
\times\big[\Gamma_{jm}(\tau)-\Gamma_{jm}(\tau-u)\big]^\star
\label{corH}
\\
H_{ab,jm}(t,\tau)&=&\int\limits_0^{\tau}du C(u-t)e^{i\omega_{ab}(u-t)}
\times\big[\Gamma_{jm}(\tau)-\Gamma_{jm}(\tau-u)\big]^\star\\
\label{corI}
I_{ab,jm}(t,\tau)&=&\int\limits_0^{\tau}du C(u-\tau)e^{i\omega_{ab}(u-\tau)}
\times\big[\Gamma_{jm}(t)-\Gamma_{jm}(t-u)\big]^\star\\
Y_{ab,jm}(t,\tau)&=&\int\limits_0^{\tau}du C(\tau-u)e^{i\omega_{ab}(u-\tau)}
\label{corY}\times\big[\Gamma_{jm}(t)-\Gamma_{jm}(t-u)\big]^\star,
\end{eqnarray}
\end{widetext}
where $\Gamma_{jm}(t)=\Gamma(\omega_{jm},t)$ is the timed spectral density (See Eq.~\ref{eq:sdtOhmic}).
Since the four correlation functions embody four point correlations in time, there is no escape from rank-4 tensors, in contrast to the RE which can be studied entirely in terms of
(rank-2) matrices. Note that if the spectral density is flat or $C(t)=c \delta(t)$, then $\Gamma(\omega,t)=\int_0^t \,d\tau C(\tau)e^{i\omega\tau}=c/2$ is also flat, and the four-point correlation functions in Eqs.~\ref{corF}-\ref{corY} are all zero, in agreement with Ref.~\cite[Eqs.~B3 and~B4]{JangS}.
We determine the integrals using the Simpson rule. The smallest sub-step in time is $0.0025/\Delta$,
and increases by factor of two in each consecutive integral.

Fig.~\ref{Fig:circled} displays a typical time dependence of the matrix elements of $D(t)$, showing the imaginary versus the real part. The convergence to Markovianity is excruciatingly slow for the Ohmic bath at $T=0$,
making it virtually impossible to write down the Markovian equation.

\bibliographystyle{apsrev4-1}
\bibliography{master}
\end{document}